\documentclass[trackchanges, twocolumn]{aastex701}

\usepackage{booktabs}
\usepackage{graphicx}
\usepackage{caption}
\usepackage{rotating}
\usepackage{placeins}
\usepackage{amsmath}

\newcommand{\PHOEBE}{\texttt{PHOEBE}}
\newcommand{\ellc}{\texttt{ellc}}
\newcommand{\Korg}{\texttt{Korg}}
\newcommand{\iSpec}{\texttt{iSpec}}
\newcommand{\Gaia}{\textit{Gaia}}
\newcommand{\TESS}{\textit{TESS}}

\newcommand{\feh}{$[\rm{Fe/H}]$}
\defcitealias{Torres10}{T10}

\newcommand{\alpham}{\ensuremath{[\alpha/\mathrm{M}]}}
\newcommand{\mh}{\ensuremath{[\mathrm{M/H}]}}

\newcommand{\setval}[3]{%
  \expandafter\gdef\csname paperval@#1@#2\endcsname{#3}%
}

\newcommand{\papervalmissing}[2]{%
  \PackageWarning{papervalues}{No value for '#2' of star '#1'}%
  \mbox{\textbf{??}}%
}

\newcommand{\getval}[2]{%
  \ifcsname paperval@#1@#2\endcsname
    \csname paperval@#1@#2\endcsname
  \else
    \papervalmissing{#1}{#2}%
  \fi
}

\newcommand{\defineinfixval}[3]{% #1 = macro base name, #2 = key prefix, #3 = key suffix
  \expandafter\newcommand\csname #1\endcsname[2]{\ensuremath{\getval{##1}{#2##2#3}}}%
  \expandafter\newcommand\csname #1bare\endcsname[2]{\getval{##1}{#2##2#3}}%
  \expandafter\newcommand\csname #1par\endcsname[2]{\ensuremath{\getval{##1}{#2##2#3@paren}}}%
  \expandafter\newcommand\csname #1parbare\endcsname[2]{\getval{##1}{#2##2#3@paren}}%
}

\newcommand{\definecompval}[2]{% #1 = macro base name, #2 = value key prefix
  \defineinfixval{#1}{#2}{}%
}

\newcommand{\definesysval}[2]{% #1 = macro base name, #2 = full value key
  \expandafter\newcommand\csname #1\endcsname[1]{\ensuremath{\getval{##1}{#2}}}%
  \expandafter\newcommand\csname #1bare\endcsname[1]{\getval{##1}{#2}}%
  \expandafter\newcommand\csname #1par\endcsname[1]{\ensuremath{\getval{##1}{#2@paren}}}%
  \expandafter\newcommand\csname #1parbare\endcsname[1]{\getval{##1}{#2@paren}}%
}

\newcommand{\definepointinfixval}[3]{% #1 = macro base name, #2 = key prefix, #3 = key suffix
  \expandafter\newcommand\csname #1\endcsname[2]{\ensuremath{\getval{##1}{#2##2#3}}}%
  \expandafter\newcommand\csname #1bare\endcsname[2]{\getval{##1}{#2##2#3}}%
}

\newcommand{\definepointcompval}[2]{% #1 = macro base name, #2 = value key prefix
  \definepointinfixval{#1}{#2}{}%
}

\newcommand{\definepointsysval}[2]{% #1 = macro base name, #2 = full value key
  \expandafter\newcommand\csname #1\endcsname[1]{\ensuremath{\getval{##1}{#2}}}%
  \expandafter\newcommand\csname #1bare\endcsname[1]{\getval{##1}{#2}}%
}

\definecompval{mass}{m}
\definecompval{radius}{r}
\definecompval{teff}{teff}
\definecompval{logjitter}{logs}

\definesysval{period}{period}
\definesysval{ecc}{ecc}
\definesysval{inclination}{incl_deg}
\definesysval{argperi}{omega_deg}
\definesysval{sysvel}{vgamma}
\definesysval{massratio}{q}
\definesysval{sma}{sma}
\definesysval{teffratio}{teffratio}
\definesysval{thirdlight}{l3_frac}
\definesysval{radiussum}{requivsum}
\definesysval{phasezero}{phi0}
\definesysval{rvoffset}{rv_offset_0}

\definesysval{impactprimary}{b_primary}
\definesysval{impactsecondary}{b_secondary}

\definepointcompval{specteff}{dis.teff}
\definepointcompval{speclogg}{dis.logg}
\definepointcompval{specvsini}{dis.vsini}
\definepointcompval{specmicro}{dis.vmic}

\definepointsysval{specmet}{dis.met}
\definepointsysval{specalpha}{dis.alpha_fe}
\definepointsysval{speclightratio}{dis.l2}

\defineinfixval{age}{age.}{.coeval}
\defineinfixval{ageprimary}{age.}{.primary}
\defineinfixval{agesecondary}{age.}{.secondary}
\defineinfixval{logage}{logage.}{.coeval}

\definepointinfixval{agedeltachisq}{age.}{.dchi2}
\definepointinfixval{agepvalue}{age.}{.pval}

\setval{J1556+1218}{age.basti.coeval}{11.8\pm0.5}%
\setval{J1556+1218}{age.basti.coeval@paren}{11.84(47)}%
\setval{J1556+1218}{age.basti.dchi2}{0.48}%
\setval{J1556+1218}{age.basti.primary}{11.9\pm0.6}%
\setval{J1556+1218}{age.basti.primary@paren}{11.92(57)}%
\setval{J1556+1218}{age.basti.pval}{0.421}%
\setval{J1556+1218}{age.basti.secondary}{11.7\pm0.9}%
\setval{J1556+1218}{age.basti.secondary@paren}{11.67(89)}%
\setval{J1556+1218}{age.mist.coeval}{11.8\pm0.5}%
\setval{J1556+1218}{age.mist.coeval@paren}{11.84(49)}%
\setval{J1556+1218}{age.mist.dchi2}{0.48}%
\setval{J1556+1218}{age.mist.primary}{11.9\pm0.6}%
\setval{J1556+1218}{age.mist.primary@paren}{11.90(58)}%
\setval{J1556+1218}{age.mist.pval}{0.424}%
\setval{J1556+1218}{age.mist.secondary}{11.7^{+1.0}_{-0.9}}%
\setval{J1556+1218}{age.mist.secondary@paren}{11.65(95)}%
\setval{J1556+1218}{age.padova.coeval}{11.7^{+0.5}_{-0.4}}%
\setval{J1556+1218}{age.padova.coeval@paren}{11.66(45)}%
\setval{J1556+1218}{age.padova.dchi2}{0.50}%
\setval{J1556+1218}{age.padova.primary}{11.7\pm0.6}%
\setval{J1556+1218}{age.padova.primary@paren}{11.71(57)}%
\setval{J1556+1218}{age.padova.pval}{0.422}%
\setval{J1556+1218}{age.padova.secondary}{11.6\pm0.9}%
\setval{J1556+1218}{age.padova.secondary@paren}{11.60(86)}%
\setval{J1556+1218}{b_primary}{0.258\pm0.001}%
\setval{J1556+1218}{b_primary@paren}{0.2576(13)}%
\setval{J1556+1218}{b_primary_frac}{0.1502\pm0.0007}%
\setval{J1556+1218}{b_primary_frac@paren}{0.15023(73)}%
\setval{J1556+1218}{b_secondary}{0.258\pm0.001}%
\setval{J1556+1218}{b_secondary@paren}{0.2576(13)}%
\setval{J1556+1218}{b_secondary_frac}{0.1502\pm0.0007}%
\setval{J1556+1218}{b_secondary_frac@paren}{0.15021(73)}%
\setval{J1556+1218}{cosi}{0.01641^{+0.00008}_{-0.00009}}%
\setval{J1556+1218}{cosi@paren}{0.016414(86)}%
\setval{J1556+1218}{dis.alpha}{-1.74}%
\setval{J1556+1218}{dis.alpha_fe}{0.32}%
\setval{J1556+1218}{dis.l2}{0.27}%
\setval{J1556+1218}{dis.logg1}{4.3}%
\setval{J1556+1218}{dis.logg2}{4.5}%
\setval{J1556+1218}{dis.met}{-2.06}%
\setval{J1556+1218}{dis.rv1}{0}%
\setval{J1556+1218}{dis.rv2}{0}%
\setval{J1556+1218}{dis.teff1}{6000}%
\setval{J1556+1218}{dis.teff2}{5600}%
\setval{J1556+1218}{dis.vmic1}{1.1}%
\setval{J1556+1218}{dis.vmic2}{0.9}%
\setval{J1556+1218}{dis.vsini1}{8.5}%
\setval{J1556+1218}{dis.vsini2}{5.0}%
\setval{J1556+1218}{ecc}{0.00004^{+0.0002}_{-0.00004}}%
\setval{J1556+1218}{ecc@paren}{0.00004(13)}%
\setval{J1556+1218}{incl}{1.55438^{+0.00009}_{-0.00008}}%
\setval{J1556+1218}{incl@paren}{1.554382(86)}%
\setval{J1556+1218}{incl_deg}{89.060\pm0.005}%
\setval{J1556+1218}{incl_deg@paren}{89.0595(49)}%
\setval{J1556+1218}{k}{0.7147\pm0.0004}%
\setval{J1556+1218}{k@paren}{0.71472(44)}%
\setval{J1556+1218}{logage.basti.coeval}{10.07\pm0.02}%
\setval{J1556+1218}{logage.basti.coeval@paren}{10.073(17)}%
\setval{J1556+1218}{logage.basti.primary}{10.08\pm0.02}%
\setval{J1556+1218}{logage.basti.primary@paren}{10.076(21)}%
\setval{J1556+1218}{logage.basti.secondary}{10.07\pm0.03}%
\setval{J1556+1218}{logage.basti.secondary@paren}{10.067(33)}%
\setval{J1556+1218}{logage.mist.coeval}{10.07\pm0.02}%
\setval{J1556+1218}{logage.mist.coeval@paren}{10.074(18)}%
\setval{J1556+1218}{logage.mist.primary}{10.08\pm0.02}%
\setval{J1556+1218}{logage.mist.primary@paren}{10.076(21)}%
\setval{J1556+1218}{logage.mist.secondary}{10.07^{+0.03}_{-0.04}}%
\setval{J1556+1218}{logage.mist.secondary@paren}{10.066(35)}%
\setval{J1556+1218}{logage.padova.coeval}{10.07\pm0.02}%
\setval{J1556+1218}{logage.padova.coeval@paren}{10.067(17)}%
\setval{J1556+1218}{logage.padova.primary}{10.07\pm0.02}%
\setval{J1556+1218}{logage.padova.primary@paren}{10.069(21)}%
\setval{J1556+1218}{logage.padova.secondary}{10.06\pm0.03}%
\setval{J1556+1218}{logage.padova.secondary@paren}{10.064(32)}%
\setval{J1556+1218}{logs1}{-1.1^{+0.3}_{-0.4}}%
\setval{J1556+1218}{logs1@paren}{-1.12(33)}%
\setval{J1556+1218}{logs2}{-6.0\pm3.0}%
\setval{J1556+1218}{logs2@paren}{-5.6(30)}%
\setval{J1556+1218}{m1}{0.759\pm0.004}%
\setval{J1556+1218}{m1@paren}{0.7593(39)}%
\setval{J1556+1218}{m2}{0.696\pm0.003}%
\setval{J1556+1218}{m2@paren}{0.6956(34)}%
\setval{J1556+1218}{omega}{-1.1^{+3}_{-0.5}}%
\setval{J1556+1218}{omega@paren}{-1.1(16)}%
\setval{J1556+1218}{omega_deg}{-60^{+150}_{-30}}%
\setval{J1556+1218}{omega_deg@paren}{-63(90)}%
\setval{J1556+1218}{pblum_primary}{8.869\pm0.003}%
\setval{J1556+1218}{pblum_primary@paren}{8.8692(27)}%
\setval{J1556+1218}{period}{6.26817\pm0.00001}%
\setval{J1556+1218}{period@paren}{6.268171(10)}%
\setval{J1556+1218}{phi0}{0.73962^{+0.00008}_{-0.00007}}%
\setval{J1556+1218}{phi0@paren}{0.739618(76)}%
\setval{J1556+1218}{q}{0.916\pm0.003}%
\setval{J1556+1218}{q@paren}{0.9162(30)}%
\setval{J1556+1218}{r1}{1.033\pm0.002}%
\setval{J1556+1218}{r1@paren}{1.0325(17)}%
\setval{J1556+1218}{r2}{0.738\pm0.001}%
\setval{J1556+1218}{r2@paren}{0.7380(13)}%
\setval{J1556+1218}{requivsum}{1.771\pm0.003}%
\setval{J1556+1218}{requivsum@paren}{1.7705(30)}%
\setval{J1556+1218}{rv_offset_0}{0.7\pm0.2}%
\setval{J1556+1218}{rv_offset_0@paren}{0.68(16)}%
\setval{J1556+1218}{sma}{16.20\pm0.03}%
\setval{J1556+1218}{sma@paren}{16.205(26)}%
\setval{J1556+1218}{teff1}{6500\pm200}%
\setval{J1556+1218}{teff1@paren}{6490(210)}%
\setval{J1556+1218}{teff2}{6100\pm200}%
\setval{J1556+1218}{teff2@paren}{6090(200)}%
\setval{J1556+1218}{teffratio}{0.9386\pm0.0004}%
\setval{J1556+1218}{teffratio@paren}{0.93856(42)}%
\setval{J1556+1218}{vgamma}{-15.14\pm0.10}%
\setval{J1556+1218}{vgamma@paren}{-15.145(98)}%
\setval{J1556+1218}{x}{0.000\pm0.001}%
\setval{J1556+1218}{x@paren}{0.0004(12)}%
\setval{J1556+1218}{y}{-0.002^{+0.008}_{-0.01}}%
\setval{J1556+1218}{y@paren}{-0.002(11)}%
\setval{J1909+4025}{age.basti.coeval}{7.1\pm0.3}%
\setval{J1909+4025}{age.basti.coeval@paren}{7.07(26)}%
\setval{J1909+4025}{age.basti.dchi2}{0.43}%
\setval{J1909+4025}{age.basti.primary}{7.0^{+0.4}_{-0.3}}%
\setval{J1909+4025}{age.basti.primary@paren}{7.04(36)}%
\setval{J1909+4025}{age.basti.pval}{0.458}%
\setval{J1909+4025}{age.basti.secondary}{7.1^{+0.4}_{-0.3}}%
\setval{J1909+4025}{age.basti.secondary@paren}{7.10(36)}%
\setval{J1909+4025}{age.mist.coeval}{7.1\pm0.3}%
\setval{J1909+4025}{age.mist.coeval@paren}{7.08(29)}%
\setval{J1909+4025}{age.mist.dchi2}{0.45}%
\setval{J1909+4025}{age.mist.primary}{7.1\pm0.4}%
\setval{J1909+4025}{age.mist.primary@paren}{7.05(42)}%
\setval{J1909+4025}{age.mist.pval}{0.428}%
\setval{J1909+4025}{age.mist.secondary}{7.1\pm0.4}%
\setval{J1909+4025}{age.mist.secondary@paren}{7.12(41)}%
\setval{J1909+4025}{age.padova.coeval}{6.9\pm0.3}%
\setval{J1909+4025}{age.padova.coeval@paren}{6.94(27)}%
\setval{J1909+4025}{age.padova.dchi2}{0.42}%
\setval{J1909+4025}{age.padova.primary}{6.9\pm0.4}%
\setval{J1909+4025}{age.padova.primary@paren}{6.91(37)}%
\setval{J1909+4025}{age.padova.pval}{0.458}%
\setval{J1909+4025}{age.padova.secondary}{7.0\pm0.4}%
\setval{J1909+4025}{age.padova.secondary@paren}{6.96(38)}%
\setval{J1909+4025}{b_primary}{0.568^{+0.006}_{-0.007}}%
\setval{J1909+4025}{b_primary@paren}{0.5684(68)}%
\setval{J1909+4025}{b_primary_frac}{0.284\pm0.002}%
\setval{J1909+4025}{b_primary_frac@paren}{0.2839(19)}%
\setval{J1909+4025}{b_secondary}{0.570^{+0.007}_{-0.008}}%
\setval{J1909+4025}{b_secondary@paren}{0.5695(73)}%
\setval{J1909+4025}{b_secondary_frac}{0.284\pm0.002}%
\setval{J1909+4025}{b_secondary_frac@paren}{0.2844(20)}%
\setval{J1909+4025}{cosi}{0.0580\pm0.0004}%
\setval{J1909+4025}{cosi@paren}{0.05796(40)}%
\setval{J1909+4025}{dis.alpha}{-0.94}%
\setval{J1909+4025}{dis.alpha_fe}{0.22}%
\setval{J1909+4025}{dis.l2}{0.50}%
\setval{J1909+4025}{dis.logg1}{4.2}%
\setval{J1909+4025}{dis.logg2}{4.2}%
\setval{J1909+4025}{dis.met}{-1.16}%
\setval{J1909+4025}{dis.rv1}{0}%
\setval{J1909+4025}{dis.rv2}{0}%
\setval{J1909+4025}{dis.teff1}{6000}%
\setval{J1909+4025}{dis.teff2}{6000}%
\setval{J1909+4025}{dis.vmic1}{1.1}%
\setval{J1909+4025}{dis.vmic2}{1.1}%
\setval{J1909+4025}{dis.vsini1}{17.8}%
\setval{J1909+4025}{dis.vsini2}{17.7}%
\setval{J1909+4025}{ecc}{0.0009^{+0.0005}_{-0.0007}}%
\setval{J1909+4025}{ecc@paren}{0.00095(60)}%
\setval{J1909+4025}{incl}{1.5128\pm0.0004}%
\setval{J1909+4025}{incl@paren}{1.51280(40)}%
\setval{J1909+4025}{incl_deg}{86.68\pm0.02}%
\setval{J1909+4025}{incl_deg@paren}{86.677(23)}%
\setval{J1909+4025}{k}{1.00\pm0.02}%
\setval{J1909+4025}{k@paren}{1.002(20)}%
\setval{J1909+4025}{l3_frac}{0.038\pm0.003}%
\setval{J1909+4025}{l3_frac@paren}{0.0377(33)}%
\setval{J1909+4025}{logage.basti.coeval}{9.85\pm0.02}%
\setval{J1909+4025}{logage.basti.coeval@paren}{9.850(16)}%
\setval{J1909+4025}{logage.basti.primary}{9.85\pm0.02}%
\setval{J1909+4025}{logage.basti.primary@paren}{9.847(22)}%
\setval{J1909+4025}{logage.basti.secondary}{9.85\pm0.02}%
\setval{J1909+4025}{logage.basti.secondary@paren}{9.851(22)}%
\setval{J1909+4025}{logage.mist.coeval}{9.85\pm0.02}%
\setval{J1909+4025}{logage.mist.coeval@paren}{9.850(18)}%
\setval{J1909+4025}{logage.mist.primary}{9.85\pm0.03}%
\setval{J1909+4025}{logage.mist.primary@paren}{9.848(26)}%
\setval{J1909+4025}{logage.mist.secondary}{9.85^{+0.03}_{-0.02}}%
\setval{J1909+4025}{logage.mist.secondary@paren}{9.852(25)}%
\setval{J1909+4025}{logage.padova.coeval}{9.84\pm0.02}%
\setval{J1909+4025}{logage.padova.coeval@paren}{9.841(17)}%
\setval{J1909+4025}{logage.padova.primary}{9.84\pm0.02}%
\setval{J1909+4025}{logage.padova.primary@paren}{9.840(23)}%
\setval{J1909+4025}{logage.padova.secondary}{9.84\pm0.02}%
\setval{J1909+4025}{logage.padova.secondary@paren}{9.843(24)}%
\setval{J1909+4025}{logs1}{-5.0\pm3.0}%
\setval{J1909+4025}{logs1@paren}{-5.3(32)}%
\setval{J1909+4025}{logs2}{-5^{+3}_{-4}}%
\setval{J1909+4025}{logs2@paren}{-4.8(35)}%
\setval{J1909+4025}{m1}{0.895\pm0.006}%
\setval{J1909+4025}{m1@paren}{0.8951(56)}%
\setval{J1909+4025}{m2}{0.894\pm0.005}%
\setval{J1909+4025}{m2@paren}{0.8941(49)}%
\setval{J1909+4025}{omega}{1.43^{+0.05}_{-0.4}}%
\setval{J1909+4025}{omega@paren}{1.43(23)}%
\setval{J1909+4025}{omega_deg}{82^{+3}_{-23}}%
\setval{J1909+4025}{omega_deg@paren}{82(13)}%
\setval{J1909+4025}{pblum_primary}{6.0\pm0.1}%
\setval{J1909+4025}{pblum_primary@paren}{6.03(12)}%
\setval{J1909+4025}{period}{3.475545\pm0.000006}%
\setval{J1909+4025}{period@paren}{3.4755451(61)}%
\setval{J1909+4025}{phi0}{0.6579\pm0.0003}%
\setval{J1909+4025}{phi0@paren}{0.65788(26)}%
\setval{J1909+4025}{q}{0.999\pm0.004}%
\setval{J1909+4025}{q@paren}{0.9989(38)}%
\setval{J1909+4025}{r1}{1.19\pm0.01}%
\setval{J1909+4025}{r1@paren}{1.194(12)}%
\setval{J1909+4025}{r2}{1.20\pm0.01}%
\setval{J1909+4025}{r2@paren}{1.196(12)}%
\setval{J1909+4025}{requivsum}{2.390\pm0.005}%
\setval{J1909+4025}{requivsum@paren}{2.3902(46)}%
\setval{J1909+4025}{rv_offset_0}{0.9\pm0.3}%
\setval{J1909+4025}{rv_offset_0@paren}{0.90(27)}%
\setval{J1909+4025}{sma}{11.72\pm0.02}%
\setval{J1909+4025}{sma@paren}{11.717(22)}%
\setval{J1909+4025}{teff1}{6100\pm200}%
\setval{J1909+4025}{teff1@paren}{6140(170)}%
\setval{J1909+4025}{teff2}{6100\pm200}%
\setval{J1909+4025}{teff2@paren}{6140(170)}%
\setval{J1909+4025}{teffratio}{0.9993\pm0.0002}%
\setval{J1909+4025}{teffratio@paren}{0.99928(16)}%
\setval{J1909+4025}{vgamma}{-92.8\pm0.1}%
\setval{J1909+4025}{vgamma@paren}{-92.85(14)}%
\setval{J1909+4025}{x}{0.004^{+0.003}_{-0.001}}%
\setval{J1909+4025}{x@paren}{0.0044(22)}%
\setval{J1909+4025}{y}{0.030^{+0.008}_{-0.02}}%
\setval{J1909+4025}{y@paren}{0.030(12)}%
\setval{J0001+2156}{age.basti.coeval}{10.3^{+0.6}_{-0.5}}%
\setval{J0001+2156}{age.basti.coeval@paren}{10.32(59)}%
\setval{J0001+2156}{age.basti.dchi2}{1.07}%
\setval{J0001+2156}{age.basti.primary}{10.5\pm0.6}%
\setval{J0001+2156}{age.basti.primary@paren}{10.55(57)}%
\setval{J0001+2156}{age.basti.pval}{0.226}%
\setval{J0001+2156}{age.basti.secondary}{9^{+1}_{-2}}%
\setval{J0001+2156}{age.basti.secondary@paren}{9.0(15)}%
\setval{J0001+2156}{age.mist.coeval}{10.2^{+0.7}_{-0.5}}%
\setval{J0001+2156}{age.mist.coeval@paren}{10.20(61)}%
\setval{J0001+2156}{age.mist.dchi2}{1.07}%
\setval{J0001+2156}{age.mist.primary}{10.4^{+0.7}_{-0.6}}%
\setval{J0001+2156}{age.mist.primary@paren}{10.40(68)}%
\setval{J0001+2156}{age.mist.pval}{0.227}%
\setval{J0001+2156}{age.mist.secondary}{9^{+1}_{-2}}%
\setval{J0001+2156}{age.mist.secondary@paren}{8.9(15)}%
\setval{J0001+2156}{age.padova.coeval}{10.2^{+0.4}_{-0.5}}%
\setval{J0001+2156}{age.padova.coeval@paren}{10.20(46)}%
\setval{J0001+2156}{age.padova.dchi2}{1.13}%
\setval{J0001+2156}{age.padova.primary}{10.3^{+0.7}_{-0.5}}%
\setval{J0001+2156}{age.padova.primary@paren}{10.29(59)}%
\setval{J0001+2156}{age.padova.pval}{0.226}%
\setval{J0001+2156}{age.padova.secondary}{9.0\pm1.0}%
\setval{J0001+2156}{age.padova.secondary@paren}{8.8(14)}%
\setval{J0001+2156}{b_primary}{0.03^{+0.03}_{-0.02}}%
\setval{J0001+2156}{b_primary@paren}{0.028(25)}%
\setval{J0001+2156}{b_primary_frac}{0.02\pm0.02}%
\setval{J0001+2156}{b_primary_frac@paren}{0.022(19)}%
\setval{J0001+2156}{b_secondary}{0.03^{+0.03}_{-0.02}}%
\setval{J0001+2156}{b_secondary@paren}{0.028(25)}%
\setval{J0001+2156}{b_secondary_frac}{0.02\pm0.02}%
\setval{J0001+2156}{b_secondary_frac@paren}{0.021(19)}%
\setval{J0001+2156}{cosi}{0.003^{+0.003}_{-0.002}}%
\setval{J0001+2156}{cosi@paren}{0.0026(24)}%
\setval{J0001+2156}{dis.alpha}{-2.12}%
\setval{J0001+2156}{dis.alpha_fe}{0.40}%
\setval{J0001+2156}{dis.l2}{0.08}%
\setval{J0001+2156}{dis.logg1}{3.7}%
\setval{J0001+2156}{dis.logg2}{4.7}%
\setval{J0001+2156}{dis.met}{-2.51}%
\setval{J0001+2156}{dis.rv1}{0}%
\setval{J0001+2156}{dis.rv2}{0}%
\setval{J0001+2156}{dis.teff1}{5700}%
\setval{J0001+2156}{dis.teff2}{5500}%
\setval{J0001+2156}{dis.vmic1}{1.1}%
\setval{J0001+2156}{dis.vmic2}{0.9}%
\setval{J0001+2156}{dis.vsini1}{10.4}%
\setval{J0001+2156}{dis.vsini2}{5.0}%
\setval{J0001+2156}{ecc}{0.0005^{+0.002}_{-0.0004}}%
\setval{J0001+2156}{ecc@paren}{0.0005(11)}%
\setval{J0001+2156}{incl}{1.568^{+0.002}_{-0.003}}%
\setval{J0001+2156}{incl@paren}{1.5682(24)}%
\setval{J0001+2156}{incl_deg}{89.8^{+0.1}_{-0.2}}%
\setval{J0001+2156}{incl_deg@paren}{89.85(14)}%
\setval{J0001+2156}{k}{0.3074\pm0.0005}%
\setval{J0001+2156}{k@paren}{0.30735(51)}%
\setval{J0001+2156}{logage.basti.coeval}{10.01^{+0.03}_{-0.02}}%
\setval{J0001+2156}{logage.basti.coeval@paren}{10.014(25)}%
\setval{J0001+2156}{logage.basti.primary}{10.02\pm0.02}%
\setval{J0001+2156}{logage.basti.primary@paren}{10.023(23)}%
\setval{J0001+2156}{logage.basti.secondary}{9.96^{+0.06}_{-0.08}}%
\setval{J0001+2156}{logage.basti.secondary@paren}{9.955(73)}%
\setval{J0001+2156}{logage.mist.coeval}{10.01^{+0.03}_{-0.02}}%
\setval{J0001+2156}{logage.mist.coeval@paren}{10.009(26)}%
\setval{J0001+2156}{logage.mist.primary}{10.02\pm0.03}%
\setval{J0001+2156}{logage.mist.primary@paren}{10.017(28)}%
\setval{J0001+2156}{logage.mist.secondary}{9.95^{+0.06}_{-0.08}}%
\setval{J0001+2156}{logage.mist.secondary@paren}{9.950(73)}%
\setval{J0001+2156}{logage.padova.coeval}{10.01\pm0.02}%
\setval{J0001+2156}{logage.padova.coeval@paren}{10.008(20)}%
\setval{J0001+2156}{logage.padova.primary}{10.01^{+0.03}_{-0.02}}%
\setval{J0001+2156}{logage.padova.primary@paren}{10.012(25)}%
\setval{J0001+2156}{logage.padova.secondary}{9.95^{+0.06}_{-0.08}}%
\setval{J0001+2156}{logage.padova.secondary@paren}{9.945(71)}%
\setval{J0001+2156}{logs1}{-6.0\pm3.0}%
\setval{J0001+2156}{logs1@paren}{-5.7(30)}%
\setval{J0001+2156}{logs2}{-0.1^{+0.3}_{-0.4}}%
\setval{J0001+2156}{logs2@paren}{-0.08(34)}%
\setval{J0001+2156}{m1}{0.821\pm0.010}%
\setval{J0001+2156}{m1@paren}{0.821(10)}%
\setval{J0001+2156}{m2}{0.658^{+0.006}_{-0.005}}%
\setval{J0001+2156}{m2@paren}{0.6583(55)}%
\setval{J0001+2156}{omega}{-1.6^{+3}_{-0.2}}%
\setval{J0001+2156}{omega@paren}{-1.6(18)}%
\setval{J0001+2156}{omega_deg}{-90^{+190}_{-10}}%
\setval{J0001+2156}{omega_deg@paren}{-90(100)}%
\setval{J0001+2156}{pblum_primary}{11.587^{+0.004}_{-0.003}}%
\setval{J0001+2156}{pblum_primary@paren}{11.5870(35)}%
\setval{J0001+2156}{period}{9.5858\pm0.0001}%
\setval{J0001+2156}{period@paren}{9.58581(15)}%
\setval{J0001+2156}{phi0}{0.7011\pm0.0009}%
\setval{J0001+2156}{phi0@paren}{0.70114(94)}%
\setval{J0001+2156}{q}{0.802\pm0.005}%
\setval{J0001+2156}{q@paren}{0.8020(53)}%
\setval{J0001+2156}{r1}{2.031\pm0.008}%
\setval{J0001+2156}{r1@paren}{2.0311(81)}%
\setval{J0001+2156}{r2}{0.624\pm0.003}%
\setval{J0001+2156}{r2@paren}{0.6243(26)}%
\setval{J0001+2156}{requivsum}{2.66\pm0.01}%
\setval{J0001+2156}{requivsum@paren}{2.655(11)}%
\setval{J0001+2156}{rv_offset_0}{0.6\pm0.2}%
\setval{J0001+2156}{rv_offset_0@paren}{0.63(22)}%
\setval{J0001+2156}{sma}{21.63\pm0.07}%
\setval{J0001+2156}{sma@paren}{21.627(73)}%
\setval{J0001+2156}{teff1}{5900\pm200}%
\setval{J0001+2156}{teff1@paren}{5900(190)}%
\setval{J0001+2156}{teff2}{5700\pm200}%
\setval{J0001+2156}{teff2@paren}{5680(190)}%
\setval{J0001+2156}{teffratio}{0.963^{+0.002}_{-0.001}}%
\setval{J0001+2156}{teffratio@paren}{0.9633(15)}%
\setval{J0001+2156}{vgamma}{-228.0\pm0.1}%
\setval{J0001+2156}{vgamma@paren}{-228.03(14)}%
\setval{J0001+2156}{x}{-0.002^{+0.003}_{-0.004}}%
\setval{J0001+2156}{x@paren}{-0.0023(36)}%
\setval{J0001+2156}{y}{-0.02\pm0.03}%
\setval{J0001+2156}{y@paren}{-0.019(27)}%
\setval{J0043+3505}{age.basti.coeval}{7.7\pm0.3}%
\setval{J0043+3505}{age.basti.coeval@paren}{7.68(29)}%
\setval{J0043+3505}{age.basti.dchi2}{0.60}%
\setval{J0043+3505}{age.basti.primary}{7.5\pm0.4}%
\setval{J0043+3505}{age.basti.primary@paren}{7.54(42)}%
\setval{J0043+3505}{age.basti.pval}{0.370}%
\setval{J0043+3505}{age.basti.secondary}{7.8\pm0.4}%
\setval{J0043+3505}{age.basti.secondary@paren}{7.83(40)}%
\setval{J0043+3505}{age.mist.coeval}{8.2\pm0.3}%
\setval{J0043+3505}{age.mist.coeval@paren}{8.18(32)}%
\setval{J0043+3505}{age.mist.dchi2}{0.62}%
\setval{J0043+3505}{age.mist.primary}{8.0\pm0.5}%
\setval{J0043+3505}{age.mist.primary@paren}{7.99(47)}%
\setval{J0043+3505}{age.mist.pval}{0.365}%
\setval{J0043+3505}{age.mist.secondary}{8.3\pm0.5}%
\setval{J0043+3505}{age.mist.secondary@paren}{8.34(46)}%
\setval{J0043+3505}{age.padova.coeval}{7.7\pm0.3}%
\setval{J0043+3505}{age.padova.coeval@paren}{7.67(28)}%
\setval{J0043+3505}{age.padova.dchi2}{0.58}%
\setval{J0043+3505}{age.padova.primary}{7.5\pm0.4}%
\setval{J0043+3505}{age.padova.primary@paren}{7.51(41)}%
\setval{J0043+3505}{age.padova.pval}{0.372}%
\setval{J0043+3505}{age.padova.secondary}{7.8\pm0.4}%
\setval{J0043+3505}{age.padova.secondary@paren}{7.81(42)}%
\setval{J0043+3505}{b_primary}{0.686^{+0.005}_{-0.006}}%
\setval{J0043+3505}{b_primary@paren}{0.6863(53)}%
\setval{J0043+3505}{b_primary_frac}{0.3484\pm0.0006}%
\setval{J0043+3505}{b_primary_frac@paren}{0.34836(63)}%
\setval{J0043+3505}{b_secondary}{0.685^{+0.005}_{-0.006}}%
\setval{J0043+3505}{b_secondary@paren}{0.6852(56)}%
\setval{J0043+3505}{b_secondary_frac}{0.3478\pm0.0006}%
\setval{J0043+3505}{b_secondary_frac@paren}{0.34778(61)}%
\setval{J0043+3505}{cosi}{0.0959\pm0.0002}%
\setval{J0043+3505}{cosi@paren}{0.09594(22)}%
\setval{J0043+3505}{dis.alpha}{-0.54}%
\setval{J0043+3505}{dis.alpha_fe}{0.22}%
\setval{J0043+3505}{dis.l2}{0.52}%
\setval{J0043+3505}{dis.logg1}{4.0}%
\setval{J0043+3505}{dis.logg2}{4.0}%
\setval{J0043+3505}{dis.met}{-0.76}%
\setval{J0043+3505}{dis.rv1}{0}%
\setval{J0043+3505}{dis.rv2}{0}%
\setval{J0043+3505}{dis.teff1}{5600}%
\setval{J0043+3505}{dis.teff2}{5800}%
\setval{J0043+3505}{dis.vmic1}{1.1}%
\setval{J0043+3505}{dis.vmic2}{1.1}%
\setval{J0043+3505}{dis.vsini1}{26.9}%
\setval{J0043+3505}{dis.vsini2}{26.6}%
\setval{J0043+3505}{ecc}{0.0006^{+0.001}_{-0.0005}}%
\setval{J0043+3505}{ecc@paren}{0.00060(92)}%
\setval{J0043+3505}{incl}{1.4747\pm0.0002}%
\setval{J0043+3505}{incl@paren}{1.47470(22)}%
\setval{J0043+3505}{incl_deg}{84.49\pm0.01}%
\setval{J0043+3505}{incl_deg@paren}{84.494(13)}%
\setval{J0043+3505}{k}{0.97\pm0.02}%
\setval{J0043+3505}{k@paren}{0.970(16)}%
\setval{J0043+3505}{logage.basti.coeval}{9.89\pm0.02}%
\setval{J0043+3505}{logage.basti.coeval@paren}{9.885(16)}%
\setval{J0043+3505}{logage.basti.primary}{9.88\pm0.02}%
\setval{J0043+3505}{logage.basti.primary@paren}{9.877(24)}%
\setval{J0043+3505}{logage.basti.secondary}{9.89\pm0.02}%
\setval{J0043+3505}{logage.basti.secondary@paren}{9.894(22)}%
\setval{J0043+3505}{logage.mist.coeval}{9.91\pm0.02}%
\setval{J0043+3505}{logage.mist.coeval@paren}{9.913(17)}%
\setval{J0043+3505}{logage.mist.primary}{9.90\pm0.03}%
\setval{J0043+3505}{logage.mist.primary@paren}{9.903(25)}%
\setval{J0043+3505}{logage.mist.secondary}{9.92\pm0.02}%
\setval{J0043+3505}{logage.mist.secondary@paren}{9.921(24)}%
\setval{J0043+3505}{logage.padova.coeval}{9.88\pm0.02}%
\setval{J0043+3505}{logage.padova.coeval@paren}{9.885(16)}%
\setval{J0043+3505}{logage.padova.primary}{9.88\pm0.02}%
\setval{J0043+3505}{logage.padova.primary@paren}{9.876(24)}%
\setval{J0043+3505}{logage.padova.secondary}{9.89\pm0.02}%
\setval{J0043+3505}{logage.padova.secondary@paren}{9.892(23)}%
\setval{J0043+3505}{logs1}{-4^{+3}_{-4}}%
\setval{J0043+3505}{logs1@paren}{-3.9(38)}%
\setval{J0043+3505}{logs2}{-0.3^{+0.4}_{-4}}%
\setval{J0043+3505}{logs2@paren}{-0.3(24)}%
\setval{J0043+3505}{m1}{0.938^{+0.010}_{-0.009}}%
\setval{J0043+3505}{m1@paren}{0.9381(95)}%
\setval{J0043+3505}{m2}{0.926\pm0.008}%
\setval{J0043+3505}{m2@paren}{0.9262(83)}%
\setval{J0043+3505}{omega}{-1.56^{+2}_{-0.08}}%
\setval{J0043+3505}{omega@paren}{-1.6(12)}%
\setval{J0043+3505}{omega_deg}{-89^{+137}_{-4}}%
\setval{J0043+3505}{omega_deg@paren}{-89(71)}%
\setval{J0043+3505}{pblum_primary}{6.18\pm0.10}%
\setval{J0043+3505}{pblum_primary@paren}{6.18(10)}%
\setval{J0043+3505}{period}{3.202755\pm0.000007}%
\setval{J0043+3505}{period@paren}{3.2027554(73)}%
\setval{J0043+3505}{phi0}{0.75493\pm0.00010}%
\setval{J0043+3505}{phi0@paren}{0.754934(97)}%
\setval{J0043+3505}{q}{0.987\pm0.006}%
\setval{J0043+3505}{q@paren}{0.9874(61)}%
\setval{J0043+3505}{r1}{1.57\pm0.01}%
\setval{J0043+3505}{r1@paren}{1.575(13)}%
\setval{J0043+3505}{r2}{1.53\pm0.01}%
\setval{J0043+3505}{r2@paren}{1.527(13)}%
\setval{J0043+3505}{requivsum}{3.101\pm0.010}%
\setval{J0043+3505}{requivsum@paren}{3.1008(99)}%
\setval{J0043+3505}{rv_offset_0}{1.0\pm0.4}%
\setval{J0043+3505}{rv_offset_0@paren}{0.99(42)}%
\setval{J0043+3505}{sma}{11.25\pm0.03}%
\setval{J0043+3505}{sma@paren}{11.249(34)}%
\setval{J0043+3505}{teff1}{6900\pm200}%
\setval{J0043+3505}{teff1@paren}{6910(190)}%
\setval{J0043+3505}{teff2}{7100\pm200}%
\setval{J0043+3505}{teff2@paren}{7110(200)}%
\setval{J0043+3505}{teffratio}{1.0284^{+0.0007}_{-0.0006}}%
\setval{J0043+3505}{teffratio@paren}{1.02843(66)}%
\setval{J0043+3505}{vgamma}{-157.9^{+0.3}_{-0.2}}%
\setval{J0043+3505}{vgamma@paren}{-157.93(25)}%
\setval{J0043+3505}{x}{-0.000\pm0.002}%
\setval{J0043+3505}{x@paren}{-0.0000(23)}%
\setval{J0043+3505}{y}{-0.02^{+0.03}_{-0.02}}%
\setval{J0043+3505}{y@paren}{-0.024(23)}%
\setval{J1905+6233}{age.basti.coeval}{5.6\pm0.2}%
\setval{J1905+6233}{age.basti.coeval@paren}{5.57(20)}%
\setval{J1905+6233}{age.basti.dchi2}{4.57}%
\setval{J1905+6233}{age.basti.primary}{5.2^{+0.3}_{-0.2}}%
\setval{J1905+6233}{age.basti.primary@paren}{5.16(25)}%
\setval{J1905+6233}{age.basti.pval}{0.024}%
\setval{J1905+6233}{age.basti.secondary}{6.0\pm0.3}%
\setval{J1905+6233}{age.basti.secondary@paren}{6.01(30)}%
\setval{J1905+6233}{age.mist.coeval}{5.6\pm0.2}%
\setval{J1905+6233}{age.mist.coeval@paren}{5.55(22)}%
\setval{J1905+6233}{age.mist.dchi2}{4.27}%
\setval{J1905+6233}{age.mist.primary}{5.1\pm0.3}%
\setval{J1905+6233}{age.mist.primary@paren}{5.12(29)}%
\setval{J1905+6233}{age.mist.pval}{0.026}%
\setval{J1905+6233}{age.mist.secondary}{6.0^{+0.4}_{-0.3}}%
\setval{J1905+6233}{age.mist.secondary@paren}{6.04(34)}%
\setval{J1905+6233}{age.padova.coeval}{5.5\pm0.2}%
\setval{J1905+6233}{age.padova.coeval@paren}{5.52(19)}%
\setval{J1905+6233}{age.padova.dchi2}{4.24}%
\setval{J1905+6233}{age.padova.primary}{5.1\pm0.3}%
\setval{J1905+6233}{age.padova.primary@paren}{5.11(26)}%
\setval{J1905+6233}{age.padova.pval}{0.035}%
\setval{J1905+6233}{age.padova.secondary}{6.0\pm0.3}%
\setval{J1905+6233}{age.padova.secondary@paren}{5.96(31)}%
\setval{J1905+6233}{b_primary}{1.366^{+0.007}_{-0.008}}%
\setval{J1905+6233}{b_primary@paren}{1.3659(76)}%
\setval{J1905+6233}{b_primary_frac}{0.6966^{+0.0006}_{-0.0005}}%
\setval{J1905+6233}{b_primary_frac@paren}{0.69657(54)}%
\setval{J1905+6233}{b_secondary}{1.365^{+0.007}_{-0.008}}%
\setval{J1905+6233}{b_secondary@paren}{1.3655(75)}%
\setval{J1905+6233}{b_secondary_frac}{0.6965^{+0.0005}_{-0.0007}}%
\setval{J1905+6233}{b_secondary_frac@paren}{0.69646(59)}%
\setval{J1905+6233}{cosi}{0.1953\pm0.0002}%
\setval{J1905+6233}{cosi@paren}{0.19530(23)}%
\setval{J1905+6233}{dis.alpha}{-0.78}%
\setval{J1905+6233}{dis.alpha_fe}{0.22}%
\setval{J1905+6233}{dis.l2}{0.48}%
\setval{J1905+6233}{dis.logg1}{4.2}%
\setval{J1905+6233}{dis.logg2}{4.2}%
\setval{J1905+6233}{dis.met}{-1.00}%
\setval{J1905+6233}{dis.rv1}{0}%
\setval{J1905+6233}{dis.rv2}{0}%
\setval{J1905+6233}{dis.teff1}{6400}%
\setval{J1905+6233}{dis.teff2}{6400}%
\setval{J1905+6233}{dis.vmic1}{1.4}%
\setval{J1905+6233}{dis.vmic2}{1.3}%
\setval{J1905+6233}{dis.vsini1}{26.5}%
\setval{J1905+6233}{dis.vsini2}{27.9}%
\setval{J1905+6233}{ecc}{0.0004^{+0.0009}_{-0.0002}}%
\setval{J1905+6233}{ecc@paren}{0.00042(54)}%
\setval{J1905+6233}{incl}{1.3742\pm0.0002}%
\setval{J1905+6233}{incl@paren}{1.37424(23)}%
\setval{J1905+6233}{incl_deg}{78.74\pm0.01}%
\setval{J1905+6233}{incl_deg@paren}{78.738(13)}%
\setval{J1905+6233}{k}{0.96\pm0.01}%
\setval{J1905+6233}{k@paren}{0.961(11)}%
\setval{J1905+6233}{logage.basti.coeval}{9.75\pm0.02}%
\setval{J1905+6233}{logage.basti.coeval@paren}{9.746(16)}%
\setval{J1905+6233}{logage.basti.primary}{9.71\pm0.02}%
\setval{J1905+6233}{logage.basti.primary@paren}{9.713(21)}%
\setval{J1905+6233}{logage.basti.secondary}{9.78\pm0.02}%
\setval{J1905+6233}{logage.basti.secondary@paren}{9.779(22)}%
\setval{J1905+6233}{logage.mist.coeval}{9.74\pm0.02}%
\setval{J1905+6233}{logage.mist.coeval@paren}{9.745(17)}%
\setval{J1905+6233}{logage.mist.primary}{9.71^{+0.02}_{-0.03}}%
\setval{J1905+6233}{logage.mist.primary@paren}{9.709(25)}%
\setval{J1905+6233}{logage.mist.secondary}{9.78\pm0.02}%
\setval{J1905+6233}{logage.mist.secondary@paren}{9.781(24)}%
\setval{J1905+6233}{logage.padova.coeval}{9.74^{+0.01}_{-0.02}}%
\setval{J1905+6233}{logage.padova.coeval@paren}{9.742(15)}%
\setval{J1905+6233}{logage.padova.primary}{9.71\pm0.02}%
\setval{J1905+6233}{logage.padova.primary@paren}{9.708(22)}%
\setval{J1905+6233}{logage.padova.secondary}{9.78\pm0.02}%
\setval{J1905+6233}{logage.padova.secondary@paren}{9.775(23)}%
\setval{J1905+6233}{logs1}{-5.0\pm3.0}%
\setval{J1905+6233}{logs1@paren}{-5.1(32)}%
\setval{J1905+6233}{logs2}{-4^{+3}_{-4}}%
\setval{J1905+6233}{logs2@paren}{-4.2(36)}%
\setval{J1905+6233}{m1}{0.999\pm0.007}%
\setval{J1905+6233}{m1@paren}{0.9992(67)}%
\setval{J1905+6233}{m2}{0.956\pm0.006}%
\setval{J1905+6233}{m2@paren}{0.9560(64)}%
\setval{J1905+6233}{omega}{0^{+2}_{-1}}%
\setval{J1905+6233}{omega@paren}{-0.3(13)}%
\setval{J1905+6233}{omega_deg}{-20^{+90}_{-60}}%
\setval{J1905+6233}{omega_deg@paren}{-19(76)}%
\setval{J1905+6233}{pblum_primary}{6.55^{+0.08}_{-0.07}}%
\setval{J1905+6233}{pblum_primary@paren}{6.546(73)}%
\setval{J1905+6233}{period}{2.503678\pm0.000006}%
\setval{J1905+6233}{period@paren}{2.5036783(62)}%
\setval{J1905+6233}{phi0}{0.9833\pm0.0005}%
\setval{J1905+6233}{phi0@paren}{0.98332(46)}%
\setval{J1905+6233}{q}{0.957\pm0.004}%
\setval{J1905+6233}{q@paren}{0.9567(42)}%
\setval{J1905+6233}{r1}{1.387\pm0.008}%
\setval{J1905+6233}{r1@paren}{1.3870(81)}%
\setval{J1905+6233}{r2}{1.332\pm0.008}%
\setval{J1905+6233}{r2@paren}{1.3324(81)}%
\setval{J1905+6233}{requivsum}{2.719\pm0.006}%
\setval{J1905+6233}{requivsum@paren}{2.7195(63)}%
\setval{J1905+6233}{rv_offset_0}{0.8\pm0.4}%
\setval{J1905+6233}{rv_offset_0@paren}{0.75(43)}%
\setval{J1905+6233}{sma}{9.70\pm0.02}%
\setval{J1905+6233}{sma@paren}{9.699(21)}%
\setval{J1905+6233}{teff1}{7460^{+90}_{-160}}%
\setval{J1905+6233}{teff1@paren}{7460(130)}%
\setval{J1905+6233}{teff2}{7430^{+90}_{-160}}%
\setval{J1905+6233}{teff2@paren}{7430(130)}%
\setval{J1905+6233}{teffratio}{0.997\pm0.002}%
\setval{J1905+6233}{teffratio@paren}{0.9966(18)}%
\setval{J1905+6233}{vgamma}{29.3^{+0.2}_{-0.1}}%
\setval{J1905+6233}{vgamma@paren}{29.26(15)}%
\setval{J1905+6233}{x}{0.008\pm0.004}%
\setval{J1905+6233}{x@paren}{0.0083(40)}%
\setval{J1905+6233}{y}{-0.00\pm0.03}%
\setval{J1905+6233}{y@paren}{-0.004(26)}%
\setval{J0807+6945}{age.basti.coeval}{6.0\pm0.4}%
\setval{J0807+6945}{age.basti.coeval@paren}{6.02(37)}%
\setval{J0807+6945}{age.basti.dchi2}{0.53}%
\setval{J0807+6945}{age.basti.primary}{6.1\pm0.5}%
\setval{J0807+6945}{age.basti.primary@paren}{6.13(52)}%
\setval{J0807+6945}{age.basti.pval}{0.425}%
\setval{J0807+6945}{age.basti.secondary}{5.9\pm0.5}%
\setval{J0807+6945}{age.basti.secondary@paren}{5.91(53)}%
\setval{J0807+6945}{age.mist.coeval}{7.2^{+0.5}_{-0.4}}%
\setval{J0807+6945}{age.mist.coeval@paren}{7.25(45)}%
\setval{J0807+6945}{age.mist.dchi2}{0.49}%
\setval{J0807+6945}{age.mist.primary}{7.4^{+0.6}_{-0.7}}%
\setval{J0807+6945}{age.mist.primary@paren}{7.39(64)}%
\setval{J0807+6945}{age.mist.pval}{0.438}%
\setval{J0807+6945}{age.mist.secondary}{7.1\pm0.6}%
\setval{J0807+6945}{age.mist.secondary@paren}{7.11(64)}%
\setval{J0807+6945}{age.padova.coeval}{6.1\pm0.4}%
\setval{J0807+6945}{age.padova.coeval@paren}{6.11(37)}%
\setval{J0807+6945}{age.padova.dchi2}{0.47}%
\setval{J0807+6945}{age.padova.primary}{6.2\pm0.5}%
\setval{J0807+6945}{age.padova.primary@paren}{6.22(54)}%
\setval{J0807+6945}{age.padova.pval}{0.431}%
\setval{J0807+6945}{age.padova.secondary}{6.0\pm0.5}%
\setval{J0807+6945}{age.padova.secondary@paren}{6.00(51)}%
\setval{J0807+6945}{b_primary}{0.185\pm0.002}%
\setval{J0807+6945}{b_primary@paren}{0.1850(22)}%
\setval{J0807+6945}{b_primary_frac}{0.093\pm0.001}%
\setval{J0807+6945}{b_primary_frac@paren}{0.0929(10)}%
\setval{J0807+6945}{b_secondary}{0.185\pm0.002}%
\setval{J0807+6945}{b_secondary@paren}{0.1850(23)}%
\setval{J0807+6945}{b_secondary_frac}{0.093\pm0.001}%
\setval{J0807+6945}{b_secondary_frac@paren}{0.0929(10)}%
\setval{J0807+6945}{cosi}{0.00619\pm0.00007}%
\setval{J0807+6945}{cosi@paren}{0.006193(72)}%
\setval{J0807+6945}{dis.alpha}{-0.38}%
\setval{J0807+6945}{dis.alpha_fe}{0.35}%
\setval{J0807+6945}{dis.l2}{0.50}%
\setval{J0807+6945}{dis.logg1}{4.3}%
\setval{J0807+6945}{dis.logg2}{4.4}%
\setval{J0807+6945}{dis.met}{-0.73}%
\setval{J0807+6945}{dis.rv1}{0}%
\setval{J0807+6945}{dis.rv2}{0}%
\setval{J0807+6945}{dis.teff1}{5700}%
\setval{J0807+6945}{dis.teff2}{5700}%
\setval{J0807+6945}{dis.vmic1}{1.0}%
\setval{J0807+6945}{dis.vmic2}{1.0}%
\setval{J0807+6945}{dis.vsini1}{5.1}%
\setval{J0807+6945}{dis.vsini2}{5.2}%
\setval{J0807+6945}{ecc}{0.00007^{+0.0004}_{-0.00006}}%
\setval{J0807+6945}{ecc@paren}{0.00007(24)}%
\setval{J0807+6945}{incl}{1.56460\pm0.00007}%
\setval{J0807+6945}{incl@paren}{1.564603(72)}%
\setval{J0807+6945}{incl_deg}{89.645\pm0.004}%
\setval{J0807+6945}{incl_deg@paren}{89.6452(41)}%
\setval{J0807+6945}{k}{0.991\pm0.007}%
\setval{J0807+6945}{k@paren}{0.9913(71)}%
\setval{J0807+6945}{logage.basti.coeval}{9.78\pm0.03}%
\setval{J0807+6945}{logage.basti.coeval@paren}{9.780(27)}%
\setval{J0807+6945}{logage.basti.primary}{9.79\pm0.04}%
\setval{J0807+6945}{logage.basti.primary@paren}{9.787(37)}%
\setval{J0807+6945}{logage.basti.secondary}{9.77\pm0.04}%
\setval{J0807+6945}{logage.basti.secondary@paren}{9.772(39)}%
\setval{J0807+6945}{logage.mist.coeval}{9.86\pm0.03}%
\setval{J0807+6945}{logage.mist.coeval@paren}{9.860(27)}%
\setval{J0807+6945}{logage.mist.primary}{9.87\pm0.04}%
\setval{J0807+6945}{logage.mist.primary@paren}{9.869(38)}%
\setval{J0807+6945}{logage.mist.secondary}{9.85\pm0.04}%
\setval{J0807+6945}{logage.mist.secondary@paren}{9.852(39)}%
\setval{J0807+6945}{logage.padova.coeval}{9.79\pm0.03}%
\setval{J0807+6945}{logage.padova.coeval@paren}{9.786(26)}%
\setval{J0807+6945}{logage.padova.primary}{9.79\pm0.04}%
\setval{J0807+6945}{logage.padova.primary@paren}{9.794(38)}%
\setval{J0807+6945}{logage.padova.secondary}{9.78\pm0.04}%
\setval{J0807+6945}{logage.padova.secondary@paren}{9.778(37)}%
\setval{J0807+6945}{logs1}{-5.0\pm3.0}%
\setval{J0807+6945}{logs1@paren}{-5.2(32)}%
\setval{J0807+6945}{logs2}{-5^{+3}_{-4}}%
\setval{J0807+6945}{logs2@paren}{-4.5(34)}%
\setval{J0807+6945}{m1}{0.904\pm0.007}%
\setval{J0807+6945}{m1@paren}{0.9036(72)}%
\setval{J0807+6945}{m2}{0.906^{+0.007}_{-0.008}}%
\setval{J0807+6945}{m2@paren}{0.9058(75)}%
\setval{J0807+6945}{omega}{1.6^{+0.3}_{-3}}%
\setval{J0807+6945}{omega@paren}{1.6(18)}%
\setval{J0807+6945}{omega_deg}{90^{+20}_{-190}}%
\setval{J0807+6945}{omega_deg@paren}{90(100)}%
\setval{J0807+6945}{pblum_primary}{6.34\pm0.04}%
\setval{J0807+6945}{pblum_primary@paren}{6.339(41)}%
\setval{J0807+6945}{period}{15.19652\pm0.00006}%
\setval{J0807+6945}{period@paren}{15.196523(57)}%
\setval{J0807+6945}{phi0}{0.6163\pm0.0001}%
\setval{J0807+6945}{phi0@paren}{0.61627(15)}%
\setval{J0807+6945}{q}{1.002\pm0.005}%
\setval{J0807+6945}{q@paren}{1.0024(54)}%
\setval{J0807+6945}{r1}{1.053\pm0.005}%
\setval{J0807+6945}{r1@paren}{1.0530(47)}%
\setval{J0807+6945}{r2}{1.044^{+0.005}_{-0.004}}%
\setval{J0807+6945}{r2@paren}{1.0439(46)}%
\setval{J0807+6945}{requivsum}{2.097^{+0.005}_{-0.006}}%
\setval{J0807+6945}{requivsum@paren}{2.0969(55)}%
\setval{J0807+6945}{rv_offset_0}{0.7\pm0.2}%
\setval{J0807+6945}{rv_offset_0@paren}{0.70(21)}%
\setval{J0807+6945}{sma}{31.45\pm0.08}%
\setval{J0807+6945}{sma@paren}{31.448(80)}%
\setval{J0807+6945}{teff1}{4820\pm80}%
\setval{J0807+6945}{teff1@paren}{4817(83)}%
\setval{J0807+6945}{teff2}{4820\pm80}%
\setval{J0807+6945}{teff2@paren}{4816(83)}%
\setval{J0807+6945}{teffratio}{0.9999^{+0.0003}_{-0.0004}}%
\setval{J0807+6945}{teffratio@paren}{0.99990(35)}%
\setval{J0807+6945}{vgamma}{-23.0\pm0.1}%
\setval{J0807+6945}{vgamma@paren}{-23.05(14)}%
\setval{J0807+6945}{x}{-0.0009^{+0.0008}_{-0.002}}%
\setval{J0807+6945}{x@paren}{-0.0009(12)}%
\setval{J0807+6945}{y}{0.00^{+0.02}_{-0.01}}%
\setval{J0807+6945}{y@paren}{0.002(14)}%
\setval{J0157+2928}{age.basti.coeval}{7.4\pm0.4}%
\setval{J0157+2928}{age.basti.coeval@paren}{7.40(41)}%
\setval{J0157+2928}{age.basti.dchi2}{0.48}%
\setval{J0157+2928}{age.basti.primary}{7.4\pm0.6}%
\setval{J0157+2928}{age.basti.primary@paren}{7.40(58)}%
\setval{J0157+2928}{age.basti.pval}{0.438}%
\setval{J0157+2928}{age.basti.secondary}{7.4^{+0.5}_{-0.6}}%
\setval{J0157+2928}{age.basti.secondary@paren}{7.42(57)}%
\setval{J0157+2928}{age.mist.coeval}{8.5^{+0.5}_{-0.4}}%
\setval{J0157+2928}{age.mist.coeval@paren}{8.52(45)}%
\setval{J0157+2928}{age.mist.dchi2}{0.48}%
\setval{J0157+2928}{age.mist.primary}{8.5\pm0.6}%
\setval{J0157+2928}{age.mist.primary@paren}{8.53(64)}%
\setval{J0157+2928}{age.mist.pval}{0.435}%
\setval{J0157+2928}{age.mist.secondary}{8.5^{+0.6}_{-0.7}}%
\setval{J0157+2928}{age.mist.secondary@paren}{8.54(63)}%
\setval{J0157+2928}{age.padova.coeval}{7.5\pm0.4}%
\setval{J0157+2928}{age.padova.coeval@paren}{7.50(42)}%
\setval{J0157+2928}{age.padova.dchi2}{0.48}%
\setval{J0157+2928}{age.padova.primary}{7.5\pm0.6}%
\setval{J0157+2928}{age.padova.primary@paren}{7.50(58)}%
\setval{J0157+2928}{age.padova.pval}{0.444}%
\setval{J0157+2928}{age.padova.secondary}{7.5^{+0.5}_{-0.6}}%
\setval{J0157+2928}{age.padova.secondary@paren}{7.51(58)}%
\setval{J0157+2928}{b_primary}{1.17\pm0.04}%
\setval{J0157+2928}{b_primary@paren}{1.168(38)}%
\setval{J0157+2928}{b_primary_frac}{0.48^{+0.01}_{-0.02}}%
\setval{J0157+2928}{b_primary_frac@paren}{0.483(15)}%
\setval{J0157+2928}{b_secondary}{1.17\pm0.04}%
\setval{J0157+2928}{b_secondary@paren}{1.168(38)}%
\setval{J0157+2928}{b_secondary_frac}{0.48^{+0.01}_{-0.02}}%
\setval{J0157+2928}{b_secondary_frac@paren}{0.483(15)}%
\setval{J0157+2928}{cosi}{0.044\pm0.002}%
\setval{J0157+2928}{cosi@paren}{0.0436(17)}%
\setval{J0157+2928}{dis.alpha}{-0.73}%
\setval{J0157+2928}{dis.alpha_fe}{0.38}%
\setval{J0157+2928}{dis.l2}{0.50}%
\setval{J0157+2928}{dis.logg1}{4.4}%
\setval{J0157+2928}{dis.logg2}{4.4}%
\setval{J0157+2928}{dis.met}{-1.11}%
\setval{J0157+2928}{dis.rv1}{0}%
\setval{J0157+2928}{dis.rv2}{0}%
\setval{J0157+2928}{dis.teff1}{5700}%
\setval{J0157+2928}{dis.teff2}{5700}%
\setval{J0157+2928}{dis.vmic1}{1.0}%
\setval{J0157+2928}{dis.vmic2}{1.0}%
\setval{J0157+2928}{dis.vsini1}{3.6}%
\setval{J0157+2928}{dis.vsini2}{3.8}%
\setval{J0157+2928}{ecc}{0.00011^{+0.0004}_{-0.00010}}%
\setval{J0157+2928}{ecc@paren}{0.00011(27)}%
\setval{J0157+2928}{incl}{1.527\pm0.002}%
\setval{J0157+2928}{incl@paren}{1.5272(17)}%
\setval{J0157+2928}{incl_deg}{87.50^{+0.1}_{-0.09}}%
\setval{J0157+2928}{incl_deg@paren}{87.503(98)}%
\setval{J0157+2928}{k}{1.41^{+0.09}_{-0.06}}%
\setval{J0157+2928}{k@paren}{1.413(74)}%
\setval{J0157+2928}{l3_frac}{0.09^{+0.03}_{-0.04}}%
\setval{J0157+2928}{l3_frac@paren}{0.087(33)}%
\setval{J0157+2928}{logage.basti.coeval}{9.87\pm0.02}%
\setval{J0157+2928}{logage.basti.coeval@paren}{9.869(24)}%
\setval{J0157+2928}{logage.basti.primary}{9.87\pm0.03}%
\setval{J0157+2928}{logage.basti.primary@paren}{9.869(34)}%
\setval{J0157+2928}{logage.basti.secondary}{9.87^{+0.03}_{-0.04}}%
\setval{J0157+2928}{logage.basti.secondary@paren}{9.871(33)}%
\setval{J0157+2928}{logage.mist.coeval}{9.93\pm0.02}%
\setval{J0157+2928}{logage.mist.coeval@paren}{9.930(23)}%
\setval{J0157+2928}{logage.mist.primary}{9.93\pm0.03}%
\setval{J0157+2928}{logage.mist.primary@paren}{9.931(33)}%
\setval{J0157+2928}{logage.mist.secondary}{9.93\pm0.03}%
\setval{J0157+2928}{logage.mist.secondary@paren}{9.932(32)}%
\setval{J0157+2928}{logage.padova.coeval}{9.87^{+0.02}_{-0.03}}%
\setval{J0157+2928}{logage.padova.coeval@paren}{9.875(24)}%
\setval{J0157+2928}{logage.padova.primary}{9.88\pm0.03}%
\setval{J0157+2928}{logage.padova.primary@paren}{9.875(34)}%
\setval{J0157+2928}{logage.padova.secondary}{9.88^{+0.03}_{-0.04}}%
\setval{J0157+2928}{logage.padova.secondary@paren}{9.876(33)}%
\setval{J0157+2928}{logs1}{-4^{+3}_{-4}}%
\setval{J0157+2928}{logs1@paren}{-4.5(32)}%
\setval{J0157+2928}{logs2}{-5.0\pm3.0}%
\setval{J0157+2928}{logs2@paren}{-5.2(30)}%
\setval{J0157+2928}{m1}{0.819\pm0.003}%
\setval{J0157+2928}{m1@paren}{0.8190(33)}%
\setval{J0157+2928}{m2}{0.819\pm0.003}%
\setval{J0157+2928}{m2@paren}{0.8194(27)}%
\setval{J0157+2928}{omega}{1.3^{+0.3}_{-3}}%
\setval{J0157+2928}{omega@paren}{1.3(17)}%
\setval{J0157+2928}{omega_deg}{80^{+20}_{-170}}%
\setval{J0157+2928}{omega_deg@paren}{75(95)}%
\setval{J0157+2928}{pblum_primary}{3.8^{+0.2}_{-0.3}}%
\setval{J0157+2928}{pblum_primary@paren}{3.82(29)}%
\setval{J0157+2928}{period}{8.19949\pm0.00002}%
\setval{J0157+2928}{period@paren}{8.199489(16)}%
\setval{J0157+2928}{phi0}{0.12871^{+0.00002}_{-0.00003}}%
\setval{J0157+2928}{phi0@paren}{0.128710(26)}%
\setval{J0157+2928}{q}{1.000\pm0.003}%
\setval{J0157+2928}{q@paren}{1.0004(25)}%
\setval{J0157+2928}{r1}{0.76^{+0.02}_{-0.03}}%
\setval{J0157+2928}{r1@paren}{0.756(26)}%
\setval{J0157+2928}{r2}{1.07\pm0.02}%
\setval{J0157+2928}{r2@paren}{1.068(19)}%
\setval{J0157+2928}{requivsum}{1.82^{+0.01}_{-0.02}}%
\setval{J0157+2928}{requivsum@paren}{1.821(15)}%
\setval{J0157+2928}{requivsum_over_2}{0.910^{+0.007}_{-0.008}}%
\setval{J0157+2928}{requivsum_over_2@paren}{0.9103(74)}%
\setval{J0157+2928}{rv_offset_0}{0.7\pm0.1}%
\setval{J0157+2928}{rv_offset_0@paren}{0.70(12)}%
\setval{J0157+2928}{sma}{20.16^{+0.03}_{-0.02}}%
\setval{J0157+2928}{sma@paren}{20.163(23)}%
\setval{J0157+2928}{teff1}{5700\pm200}%
\setval{J0157+2928}{teff1@paren}{5710(200)}%
\setval{J0157+2928}{teff2}{5700\pm200}%
\setval{J0157+2928}{teff2@paren}{5720(200)}%
\setval{J0157+2928}{teffratio}{1.002\pm0.001}%
\setval{J0157+2928}{teffratio@paren}{1.0019(10)}%
\setval{J0157+2928}{vgamma}{-134.62^{+0.06}_{-0.07}}%
\setval{J0157+2928}{vgamma@paren}{-134.616(66)}%
\setval{J0157+2928}{x}{-0.000\pm0.002}%
\setval{J0157+2928}{x@paren}{-0.0002(19)}%
\setval{J0157+2928}{y}{0.00^{+0.02}_{-0.01}}%
\setval{J0157+2928}{y@paren}{0.003(16)}%
\setval{J0345-0407}{age.basti.coeval}{3.0\pm2.0}%
\setval{J0345-0407}{age.basti.coeval@paren}{3.2(18)}%
\setval{J0345-0407}{age.basti.dchi2}{0.77}%
\setval{J0345-0407}{age.basti.primary}{2.0\pm2.0}%
\setval{J0345-0407}{age.basti.primary@paren}{1.9(21)}%
\setval{J0345-0407}{age.basti.pval}{0.305}%
\setval{J0345-0407}{age.basti.secondary}{5^{+2}_{-3}}%
\setval{J0345-0407}{age.basti.secondary@paren}{4.6(24)}%
\setval{J0345-0407}{age.mist.coeval}{4.0\pm2.0}%
\setval{J0345-0407}{age.mist.coeval@paren}{3.6(21)}%
\setval{J0345-0407}{age.mist.dchi2}{0.72}%
\setval{J0345-0407}{age.mist.primary}{2^{+3}_{-2}}%
\setval{J0345-0407}{age.mist.primary@paren}{2.1(24)}%
\setval{J0345-0407}{age.mist.pval}{0.319}%
\setval{J0345-0407}{age.mist.secondary}{5.0\pm3.0}%
\setval{J0345-0407}{age.mist.secondary@paren}{5.1(27)}%
\setval{J0345-0407}{age.padova.coeval}{4.0\pm2.0}%
\setval{J0345-0407}{age.padova.coeval@paren}{4.1(17)}%
\setval{J0345-0407}{age.padova.dchi2}{0.77}%
\setval{J0345-0407}{age.padova.primary}{3.0\pm2.0}%
\setval{J0345-0407}{age.padova.primary@paren}{2.8(23)}%
\setval{J0345-0407}{age.padova.pval}{0.304}%
\setval{J0345-0407}{age.padova.secondary}{5.0\pm2.0}%
\setval{J0345-0407}{age.padova.secondary@paren}{5.4(22)}%
\setval{J0345-0407}{b_primary}{1.39\pm0.05}%
\setval{J0345-0407}{b_primary@paren}{1.390(50)}%
\setval{J0345-0407}{b_primary_frac}{0.659\pm0.001}%
\setval{J0345-0407}{b_primary_frac@paren}{0.6591(13)}%
\setval{J0345-0407}{b_secondary}{1.35\pm0.05}%
\setval{J0345-0407}{b_secondary@paren}{1.351(49)}%
\setval{J0345-0407}{b_secondary_frac}{0.641\pm0.001}%
\setval{J0345-0407}{b_secondary_frac@paren}{0.6406(13)}%
\setval{J0345-0407}{cosi}{0.0434\pm0.0001}%
\setval{J0345-0407}{cosi@paren}{0.04338(12)}%
\setval{J0345-0407}{dis.alpha}{-0.25}%
\setval{J0345-0407}{dis.alpha_fe}{0.23}%
\setval{J0345-0407}{dis.k}{0.9884}%
\setval{J0345-0407}{dis.l2}{0.45}%
\setval{J0345-0407}{dis.logg1}{4.6}%
\setval{J0345-0407}{dis.logg2}{4.6}%
\setval{J0345-0407}{dis.met}{-0.48}%
\setval{J0345-0407}{dis.r1_rsun}{0.749}%
\setval{J0345-0407}{dis.r2_rsun}{0.7402}%
\setval{J0345-0407}{dis.rv1}{0}%
\setval{J0345-0407}{dis.rv2}{0}%
\setval{J0345-0407}{dis.teff1}{5000}%
\setval{J0345-0407}{dis.teff2}{4900}%
\setval{J0345-0407}{dis.vmic1}{0.9}%
\setval{J0345-0407}{dis.vmic2}{0.9}%
\setval{J0345-0407}{dis.vsini1}{4.1}%
\setval{J0345-0407}{dis.vsini2}{4.9}%
\setval{J0345-0407}{ecc}{0.015\pm0.002}%
\setval{J0345-0407}{ecc@paren}{0.0149(17)}%
\setval{J0345-0407}{incl}{1.5274\pm0.0001}%
\setval{J0345-0407}{incl@paren}{1.52741(13)}%
\setval{J0345-0407}{incl_deg}{87.514^{+0.008}_{-0.006}}%
\setval{J0345-0407}{incl_deg@paren}{87.5140(72)}%
\setval{J0345-0407}{k}{1.11^{+0.07}_{-0.08}}%
\setval{J0345-0407}{k@paren}{1.108(77)}%
\setval{J0345-0407}{logage.basti.coeval}{9.5^{+0.2}_{-0.4}}%
\setval{J0345-0407}{logage.basti.coeval@paren}{9.51(28)}%
\setval{J0345-0407}{logage.basti.primary}{9.3^{+0.4}_{-1}}%
\setval{J0345-0407}{logage.basti.primary@paren}{9.27(81)}%
\setval{J0345-0407}{logage.basti.secondary}{9.7^{+0.2}_{-0.4}}%
\setval{J0345-0407}{logage.basti.secondary@paren}{9.66(26)}%
\setval{J0345-0407}{logage.mist.coeval}{9.6^{+0.2}_{-0.4}}%
\setval{J0345-0407}{logage.mist.coeval@paren}{9.55(30)}%
\setval{J0345-0407}{logage.mist.primary}{9.3^{+0.4}_{-1}}%
\setval{J0345-0407}{logage.mist.primary@paren}{9.33(84)}%
\setval{J0345-0407}{logage.mist.secondary}{9.7^{+0.2}_{-0.4}}%
\setval{J0345-0407}{logage.mist.secondary@paren}{9.70(28)}%
\setval{J0345-0407}{logage.padova.coeval}{9.6\pm0.2}%
\setval{J0345-0407}{logage.padova.coeval@paren}{9.61(19)}%
\setval{J0345-0407}{logage.padova.primary}{9.4^{+0.3}_{-0.8}}%
\setval{J0345-0407}{logage.padova.primary@paren}{9.44(51)}%
\setval{J0345-0407}{logage.padova.secondary}{9.7^{+0.1}_{-0.2}}%
\setval{J0345-0407}{logage.padova.secondary@paren}{9.74(19)}%
\setval{J0345-0407}{logs1}{-3^{+2}_{-5}}%
\setval{J0345-0407}{logs1@paren}{-3.4(33)}%
\setval{J0345-0407}{logs2}{-5.0\pm3.0}%
\setval{J0345-0407}{logs2@paren}{-5.4(32)}%
\setval{J0345-0407}{m1}{0.82\pm0.02}%
\setval{J0345-0407}{m1@paren}{0.819(16)}%
\setval{J0345-0407}{m2}{0.783\pm0.009}%
\setval{J0345-0407}{m2@paren}{0.7831(89)}%
\setval{J0345-0407}{omega}{-1.87^{+0.03}_{-0.04}}%
\setval{J0345-0407}{omega@paren}{-1.874(35)}%
\setval{J0345-0407}{omega_deg}{-107.0\pm2.0}%
\setval{J0345-0407}{omega_deg@paren}{-107.4(20)}%
\setval{J0345-0407}{pblum_primary}{6.1^{+0.5}_{-0.4}}%
\setval{J0345-0407}{pblum_primary@paren}{6.10(47)}%
\setval{J0345-0407}{period}{9.65322\pm0.00001}%
\setval{J0345-0407}{period@paren}{9.653224(14)}%
\setval{J0345-0407}{phi0}{0.6091\pm0.0001}%
\setval{J0345-0407}{phi0@paren}{0.60911(13)}%
\setval{J0345-0407}{q}{0.956^{+0.009}_{-0.010}}%
\setval{J0345-0407}{q@paren}{0.9563(93)}%
\setval{J0345-0407}{r1}{0.71\pm0.03}%
\setval{J0345-0407}{r1@paren}{0.707(27)}%
\setval{J0345-0407}{r2}{0.78^{+0.02}_{-0.03}}%
\setval{J0345-0407}{r2@paren}{0.783(25)}%
\setval{J0345-0407}{requivsum}{1.489^{+0.009}_{-0.008}}%
\setval{J0345-0407}{requivsum@paren}{1.4892(82)}%
\setval{J0345-0407}{sma}{22.3\pm0.1}%
\setval{J0345-0407}{sma@paren}{22.32(12)}%
\setval{J0345-0407}{teff1}{4600\pm200}%
\setval{J0345-0407}{teff1@paren}{4560(190)}%
\setval{J0345-0407}{teff2}{4400\pm200}%
\setval{J0345-0407}{teff2@paren}{4420(190)}%
\setval{J0345-0407}{teffratio}{0.969^{+0.003}_{-0.004}}%
\setval{J0345-0407}{teffratio@paren}{0.9685(35)}%
\setval{J0345-0407}{vgamma}{-6.06^{+0.08}_{-0.09}}%
\setval{J0345-0407}{vgamma@paren}{-6.063(86)}%
\setval{J0345-0407}{x}{-0.036\pm0.002}%
\setval{J0345-0407}{x@paren}{-0.0364(21)}%
\setval{J0345-0407}{y}{-0.116^{+0.008}_{-0.007}}%
\setval{J0345-0407}{y@paren}{-0.1164(78)}%
\setval{J0727+2018}{age.basti.coeval}{4.2\pm0.5}%
\setval{J0727+2018}{age.basti.coeval@paren}{4.19(46)}%
\setval{J0727+2018}{age.basti.dchi2}{4.10}%
\setval{J0727+2018}{age.basti.primary}{3.7^{+0.5}_{-0.6}}%
\setval{J0727+2018}{age.basti.primary@paren}{3.65(55)}%
\setval{J0727+2018}{age.basti.pval}{0.018}%
\setval{J0727+2018}{age.basti.secondary}{5.7^{+0.7}_{-0.9}}%
\setval{J0727+2018}{age.basti.secondary@paren}{5.72(80)}%
\setval{J0727+2018}{age.mist.coeval}{4.3\pm0.5}%
\setval{J0727+2018}{age.mist.coeval@paren}{4.26(51)}%
\setval{J0727+2018}{age.mist.dchi2}{4.96}%
\setval{J0727+2018}{age.mist.primary}{3.6\pm0.6}%
\setval{J0727+2018}{age.mist.primary@paren}{3.62(59)}%
\setval{J0727+2018}{age.mist.pval}{0.009}%
\setval{J0727+2018}{age.mist.secondary}{6.1^{+0.8}_{-0.9}}%
\setval{J0727+2018}{age.mist.secondary@paren}{6.07(85)}%
\setval{J0727+2018}{age.padova.coeval}{4.3^{+0.4}_{-0.5}}%
\setval{J0727+2018}{age.padova.coeval@paren}{4.25(46)}%
\setval{J0727+2018}{age.padova.dchi2}{5.01}%
\setval{J0727+2018}{age.padova.primary}{3.6\pm0.5}%
\setval{J0727+2018}{age.padova.primary@paren}{3.60(53)}%
\setval{J0727+2018}{age.padova.pval}{0.009}%
\setval{J0727+2018}{age.padova.secondary}{5.8^{+0.7}_{-0.8}}%
\setval{J0727+2018}{age.padova.secondary@paren}{5.82(77)}%
\setval{J0727+2018}{b_primary}{1.36\pm0.06}%
\setval{J0727+2018}{b_primary@paren}{1.358(62)}%
\setval{J0727+2018}{b_primary_frac}{0.701\pm0.007}%
\setval{J0727+2018}{b_primary_frac@paren}{0.7009(69)}%
\setval{J0727+2018}{b_secondary}{1.38\pm0.06}%
\setval{J0727+2018}{b_secondary@paren}{1.379(63)}%
\setval{J0727+2018}{b_secondary_frac}{0.712^{+0.006}_{-0.007}}%
\setval{J0727+2018}{b_secondary_frac@paren}{0.7120(67)}%
\setval{J0727+2018}{cosi}{0.105\pm0.002}%
\setval{J0727+2018}{cosi@paren}{0.1054(18)}%
\setval{J0727+2018}{dis.alpha}{-0.63}%
\setval{J0727+2018}{dis.alpha_fe}{0.23}%
\setval{J0727+2018}{dis.k}{0.8814}%
\setval{J0727+2018}{dis.l2}{0.39}%
\setval{J0727+2018}{dis.logg1}{4.4}%
\setval{J0727+2018}{dis.logg2}{4.4}%
\setval{J0727+2018}{dis.met}{-0.85}%
\setval{J0727+2018}{dis.r1_rsun}{1.067}%
\setval{J0727+2018}{dis.r2_rsun}{0.9405}%
\setval{J0727+2018}{dis.rv1}{0}%
\setval{J0727+2018}{dis.rv2}{0}%
\setval{J0727+2018}{dis.teff1}{6300}%
\setval{J0727+2018}{dis.teff2}{6000}%
\setval{J0727+2018}{dis.vmic1}{1.2}%
\setval{J0727+2018}{dis.vmic2}{1.1}%
\setval{J0727+2018}{dis.vsini1}{13.0}%
\setval{J0727+2018}{dis.vsini2}{13.6}%
\setval{J0727+2018}{ecc}{0.008\pm0.003}%
\setval{J0727+2018}{ecc@paren}{0.0080(28)}%
\setval{J0727+2018}{incl}{1.465\pm0.002}%
\setval{J0727+2018}{incl@paren}{1.4652(18)}%
\setval{J0727+2018}{incl_deg}{83.95\pm0.10}%
\setval{J0727+2018}{incl_deg@paren}{83.95(11)}%
\setval{J0727+2018}{k}{0.93^{+0.09}_{-0.08}}%
\setval{J0727+2018}{k@paren}{0.934(83)}%
\setval{J0727+2018}{l3_frac}{0.13\pm0.03}%
\setval{J0727+2018}{l3_frac@paren}{0.126(27)}%
\setval{J0727+2018}{logage.basti.coeval}{9.62^{+0.04}_{-0.05}}%
\setval{J0727+2018}{logage.basti.coeval@paren}{9.622(48)}%
\setval{J0727+2018}{logage.basti.primary}{9.56^{+0.06}_{-0.07}}%
\setval{J0727+2018}{logage.basti.primary@paren}{9.563(66)}%
\setval{J0727+2018}{logage.basti.secondary}{9.76^{+0.05}_{-0.07}}%
\setval{J0727+2018}{logage.basti.secondary@paren}{9.758(62)}%
\setval{J0727+2018}{logage.mist.coeval}{9.63^{+0.05}_{-0.06}}%
\setval{J0727+2018}{logage.mist.coeval@paren}{9.629(53)}%
\setval{J0727+2018}{logage.mist.primary}{9.56^{+0.06}_{-0.08}}%
\setval{J0727+2018}{logage.mist.primary@paren}{9.559(71)}%
\setval{J0727+2018}{logage.mist.secondary}{9.78^{+0.05}_{-0.07}}%
\setval{J0727+2018}{logage.mist.secondary@paren}{9.783(62)}%
\setval{J0727+2018}{logage.padova.coeval}{9.63^{+0.04}_{-0.06}}%
\setval{J0727+2018}{logage.padova.coeval@paren}{9.629(48)}%
\setval{J0727+2018}{logage.padova.primary}{9.56^{+0.06}_{-0.07}}%
\setval{J0727+2018}{logage.padova.primary@paren}{9.557(65)}%
\setval{J0727+2018}{logage.padova.secondary}{9.77^{+0.05}_{-0.06}}%
\setval{J0727+2018}{logage.padova.secondary@paren}{9.765(58)}%
\setval{J0727+2018}{logs1}{-1.8^{+1.0}_{-5}}%
\setval{J0727+2018}{logs1@paren}{-1.8(31)}%
\setval{J0727+2018}{logs2}{-5.0\pm3.0}%
\setval{J0727+2018}{logs2@paren}{-5.3(33)}%
\setval{J0727+2018}{m1}{0.976^{+0.008}_{-0.007}}%
\setval{J0727+2018}{m1@paren}{0.9759(75)}%
\setval{J0727+2018}{m2}{0.868\pm0.005}%
\setval{J0727+2018}{m2@paren}{0.8675(51)}%
\setval{J0727+2018}{omega}{1.58\pm0.02}%
\setval{J0727+2018}{omega@paren}{1.580(18)}%
\setval{J0727+2018}{omega_deg}{90.5^{+1}_{-0.9}}%
\setval{J0727+2018}{omega_deg@paren}{90.5(10)}%
\setval{J0727+2018}{pblum_primary}{6.3\pm0.6}%
\setval{J0727+2018}{pblum_primary@paren}{6.28(56)}%
\setval{J0727+2018}{period}{4.216143\pm0.000007}%
\setval{J0727+2018}{period@paren}{4.2161435(71)}%
\setval{J0727+2018}{phi0}{0.9034\pm0.0002}%
\setval{J0727+2018}{phi0@paren}{0.90341(18)}%
\setval{J0727+2018}{q}{0.889\pm0.005}%
\setval{J0727+2018}{q@paren}{0.8889(46)}%
\setval{J0727+2018}{r1}{1.04\pm0.04}%
\setval{J0727+2018}{r1@paren}{1.038(41)}%
\setval{J0727+2018}{r2}{0.97\pm0.05}%
\setval{J0727+2018}{r2@paren}{0.971(47)}%
\setval{J0727+2018}{requivsum}{2.01\pm0.02}%
\setval{J0727+2018}{requivsum@paren}{2.008(19)}%
\setval{J0727+2018}{requivsum_over_2}{1.004^{+0.009}_{-0.010}}%
\setval{J0727+2018}{requivsum_over_2@paren}{1.0038(93)}%
\setval{J0727+2018}{rv_offset_0}{1.4\pm0.3}%
\setval{J0727+2018}{rv_offset_0@paren}{1.37(27)}%
\setval{J0727+2018}{sma}{13.46\pm0.03}%
\setval{J0727+2018}{sma@paren}{13.461(29)}%
\setval{J0727+2018}{teff1}{6200\pm200}%
\setval{J0727+2018}{teff1@paren}{6240(180)}%
\setval{J0727+2018}{teff2}{5900\pm200}%
\setval{J0727+2018}{teff2@paren}{5940(170)}%
\setval{J0727+2018}{teffratio}{0.953\pm0.007}%
\setval{J0727+2018}{teffratio@paren}{0.9526(69)}%
\setval{J0727+2018}{vgamma}{22.7\pm0.2}%
\setval{J0727+2018}{vgamma@paren}{22.74(18)}%
\setval{J0727+2018}{x}{-0.001^{+0.001}_{-0.002}}%
\setval{J0727+2018}{x@paren}{-0.0009(15)}%
\setval{J0727+2018}{y}{0.09^{+0.01}_{-0.02}}%
\setval{J0727+2018}{y@paren}{0.089(16)}%
\setval{J1047+4815}{age.basti.coeval}{6.1\pm0.3}%
\setval{J1047+4815}{age.basti.coeval@paren}{6.11(30)}%
\setval{J1047+4815}{age.basti.dchi2}{0.92}%
\setval{J1047+4815}{age.basti.primary}{6.3^{+0.4}_{-0.3}}%
\setval{J1047+4815}{age.basti.primary@paren}{6.27(35)}%
\setval{J1047+4815}{age.basti.pval}{0.288}%
\setval{J1047+4815}{age.basti.secondary}{5.7\pm0.6}%
\setval{J1047+4815}{age.basti.secondary@paren}{5.69(57)}%
\setval{J1047+4815}{age.mist.coeval}{7.5\pm0.4}%
\setval{J1047+4815}{age.mist.coeval@paren}{7.47(38)}%
\setval{J1047+4815}{age.mist.dchi2}{0.52}%
\setval{J1047+4815}{age.mist.primary}{7.6^{+0.5}_{-0.4}}%
\setval{J1047+4815}{age.mist.primary@paren}{7.56(44)}%
\setval{J1047+4815}{age.mist.pval}{0.435}%
\setval{J1047+4815}{age.mist.secondary}{7.2\pm0.7}%
\setval{J1047+4815}{age.mist.secondary@paren}{7.21(70)}%
\setval{J1047+4815}{age.padova.coeval}{6.2\pm0.3}%
\setval{J1047+4815}{age.padova.coeval@paren}{6.16(29)}%
\setval{J1047+4815}{age.padova.dchi2}{0.54}%
\setval{J1047+4815}{age.padova.primary}{6.3\pm0.3}%
\setval{J1047+4815}{age.padova.primary@paren}{6.25(34)}%
\setval{J1047+4815}{age.padova.pval}{0.408}%
\setval{J1047+4815}{age.padova.secondary}{5.9\pm0.5}%
\setval{J1047+4815}{age.padova.secondary@paren}{5.88(53)}%
\setval{J1047+4815}{b_primary}{0.109\pm0.005}%
\setval{J1047+4815}{b_primary@paren}{0.1087(49)}%
\setval{J1047+4815}{b_primary_frac}{0.062\pm0.003}%
\setval{J1047+4815}{b_primary_frac@paren}{0.0617(28)}%
\setval{J1047+4815}{b_secondary}{0.074\pm0.003}%
\setval{J1047+4815}{b_secondary@paren}{0.0745(33)}%
\setval{J1047+4815}{b_secondary_frac}{0.042\pm0.002}%
\setval{J1047+4815}{b_secondary_frac@paren}{0.0423(19)}%
\setval{J1047+4815}{cosi}{0.0049\pm0.0002}%
\setval{J1047+4815}{cosi@paren}{0.00486(22)}%
\setval{J1047+4815}{dis.alpha}{-0.31}%
\setval{J1047+4815}{dis.alpha_fe}{0.32}%
\setval{J1047+4815}{dis.l2}{0.33}%
\setval{J1047+4815}{dis.logg1}{4.2}%
\setval{J1047+4815}{dis.logg2}{4.4}%
\setval{J1047+4815}{dis.met}{-0.63}%
\setval{J1047+4815}{dis.rv1}{0}%
\setval{J1047+4815}{dis.rv2}{0}%
\setval{J1047+4815}{dis.teff1}{5700}%
\setval{J1047+4815}{dis.teff2}{5500}%
\setval{J1047+4815}{dis.vmic1}{1.0}%
\setval{J1047+4815}{dis.vmic2}{1.0}%
\setval{J1047+4815}{dis.vsini1}{11.7}%
\setval{J1047+4815}{dis.vsini2}{7.0}%
\setval{J1047+4815}{ecc}{0.4281\pm0.0002}%
\setval{J1047+4815}{ecc@paren}{0.42806(19)}%
\setval{J1047+4815}{incl}{1.5659\pm0.0002}%
\setval{J1047+4815}{incl@paren}{1.56593(22)}%
\setval{J1047+4815}{incl_deg}{89.72\pm0.01}%
\setval{J1047+4815}{incl_deg@paren}{89.721(13)}%
\setval{J1047+4815}{k}{0.7605\pm0.0006}%
\setval{J1047+4815}{k@paren}{0.76052(60)}%
\setval{J1047+4815}{logage.basti.coeval}{9.79\pm0.02}%
\setval{J1047+4815}{logage.basti.coeval@paren}{9.786(21)}%
\setval{J1047+4815}{logage.basti.primary}{9.80\pm0.02}%
\setval{J1047+4815}{logage.basti.primary@paren}{9.797(24)}%
\setval{J1047+4815}{logage.basti.secondary}{9.76^{+0.04}_{-0.05}}%
\setval{J1047+4815}{logage.basti.secondary@paren}{9.755(44)}%
\setval{J1047+4815}{logage.mist.coeval}{9.87\pm0.02}%
\setval{J1047+4815}{logage.mist.coeval@paren}{9.873(22)}%
\setval{J1047+4815}{logage.mist.primary}{9.88\pm0.03}%
\setval{J1047+4815}{logage.mist.primary@paren}{9.878(25)}%
\setval{J1047+4815}{logage.mist.secondary}{9.86\pm0.04}%
\setval{J1047+4815}{logage.mist.secondary@paren}{9.858(42)}%
\setval{J1047+4815}{logage.padova.coeval}{9.79\pm0.02}%
\setval{J1047+4815}{logage.padova.coeval@paren}{9.789(21)}%
\setval{J1047+4815}{logage.padova.primary}{9.80\pm0.02}%
\setval{J1047+4815}{logage.padova.primary@paren}{9.796(24)}%
\setval{J1047+4815}{logage.padova.secondary}{9.77\pm0.04}%
\setval{J1047+4815}{logage.padova.secondary@paren}{9.769(39)}%
\setval{J1047+4815}{logs1}{-6.0\pm3.0}%
\setval{J1047+4815}{logs1@paren}{-5.6(31)}%
\setval{J1047+4815}{logs2}{-1.5^{+0.4}_{-1}}%
\setval{J1047+4815}{logs2@paren}{-1.51(77)}%
\setval{J1047+4815}{m1}{0.961\pm0.005}%
\setval{J1047+4815}{m1@paren}{0.9610(46)}%
\setval{J1047+4815}{m2}{0.890\pm0.005}%
\setval{J1047+4815}{m2@paren}{0.8897(53)}%
\setval{J1047+4815}{omega}{-2.690\pm0.001}%
\setval{J1047+4815}{omega@paren}{-2.6903(11)}%
\setval{J1047+4815}{omega_deg}{-154.14\pm0.07}%
\setval{J1047+4815}{omega_deg@paren}{-154.145(65)}%
\setval{J1047+4815}{pblum_primary}{8.251\pm0.004}%
\setval{J1047+4815}{pblum_primary@paren}{8.2513(42)}%
\setval{J1047+4815}{period}{12.8894512\pm0.0000008}%
\setval{J1047+4815}{period@paren}{12.88945117(78)}%
\setval{J1047+4815}{phi0}{0.489307^{+0.000004}_{-0.000003}}%
\setval{J1047+4815}{phi0@paren}{0.4893067(34)}%
\setval{J1047+4815}{q}{0.926\pm0.003}%
\setval{J1047+4815}{q@paren}{0.9259(33)}%
\setval{J1047+4815}{r1}{1.276\pm0.002}%
\setval{J1047+4815}{r1@paren}{1.2761(23)}%
\setval{J1047+4815}{r2}{0.970\pm0.002}%
\setval{J1047+4815}{r2@paren}{0.9705(18)}%
\setval{J1047+4815}{requivsum}{2.247\pm0.004}%
\setval{J1047+4815}{requivsum@paren}{2.2466(41)}%
\setval{J1047+4815}{rv_offset_0}{0.7\pm0.1}%
\setval{J1047+4815}{rv_offset_0@paren}{0.73(14)}%
\setval{J1047+4815}{sma}{28.39\pm0.05}%
\setval{J1047+4815}{sma@paren}{28.391(49)}%
\setval{J1047+4815}{teff1}{6300\pm100}%
\setval{J1047+4815}{teff1@paren}{6330(120)}%
\setval{J1047+4815}{teff2}{6140\pm100}%
\setval{J1047+4815}{teff2@paren}{6140(110)}%
\setval{J1047+4815}{teffratio}{0.9693^{+0.0006}_{-0.0005}}%
\setval{J1047+4815}{teffratio@paren}{0.96926(55)}%
\setval{J1047+4815}{vgamma}{34.87^{+0.09}_{-0.08}}%
\setval{J1047+4815}{vgamma@paren}{34.871(83)}%
\setval{J1047+4815}{x}{-0.5888\pm0.0002}%
\setval{J1047+4815}{x@paren}{-0.58877(19)}%
\setval{J1047+4815}{y}{-0.2853\pm0.0007}%
\setval{J1047+4815}{y@paren}{-0.28532(73)}%
\setval{J1916+3828}{age.basti.coeval}{13.9^{+0.8}_{-0.6}}%
\setval{J1916+3828}{age.basti.coeval@paren}{13.93(73)}%
\setval{J1916+3828}{age.basti.dchi2}{1.51}%
\setval{J1916+3828}{age.basti.primary}{13.4\pm0.8}%
\setval{J1916+3828}{age.basti.primary@paren}{13.39(82)}%
\setval{J1916+3828}{age.basti.pval}{0.161}%
\setval{J1916+3828}{age.basti.secondary}{14.9^{+0.9}_{-1}}%
\setval{J1916+3828}{age.basti.secondary@paren}{14.9(10)}%
\setval{J1916+3828}{age.mist.coeval}{14.0\pm0.7}%
\setval{J1916+3828}{age.mist.coeval@paren}{14.02(72)}%
\setval{J1916+3828}{age.mist.dchi2}{1.44}%
\setval{J1916+3828}{age.mist.primary}{13.4^{+1.0}_{-0.8}}%
\setval{J1916+3828}{age.mist.primary@paren}{13.39(91)}%
\setval{J1916+3828}{age.mist.pval}{0.198}%
\setval{J1916+3828}{age.mist.secondary}{15.0\pm1.0}%
\setval{J1916+3828}{age.mist.secondary@paren}{14.9(11)}%
\setval{J1916+3828}{age.padova.coeval}{13.8^{+0.8}_{-0.7}}%
\setval{J1916+3828}{age.padova.coeval@paren}{13.78(74)}%
\setval{J1916+3828}{age.padova.dchi2}{1.40}%
\setval{J1916+3828}{age.padova.primary}{13.2\pm0.8}%
\setval{J1916+3828}{age.padova.primary@paren}{13.24(82)}%
\setval{J1916+3828}{age.padova.pval}{0.181}%
\setval{J1916+3828}{age.padova.secondary}{15.0\pm1.0}%
\setval{J1916+3828}{age.padova.secondary@paren}{14.7(11)}%
\setval{J1916+3828}{b_primary}{0.49\pm0.02}%
\setval{J1916+3828}{b_primary@paren}{0.490(17)}%
\setval{J1916+3828}{b_primary_frac}{0.30\pm0.01}%
\setval{J1916+3828}{b_primary_frac@paren}{0.302(11)}%
\setval{J1916+3828}{b_secondary}{0.47\pm0.02}%
\setval{J1916+3828}{b_secondary@paren}{0.475(17)}%
\setval{J1916+3828}{b_secondary_frac}{0.292\pm0.010}%
\setval{J1916+3828}{b_secondary_frac@paren}{0.292(11)}%
\setval{J1916+3828}{cosi}{0.045\pm0.002}%
\setval{J1916+3828}{cosi@paren}{0.0454(18)}%
\setval{J1916+3828}{dis.alpha}{-1.74}%
\setval{J1916+3828}{dis.alpha_fe}{0.42}%
\setval{J1916+3828}{dis.l2}{0.36}%
\setval{J1916+3828}{dis.logg1}{3.8}%
\setval{J1916+3828}{dis.logg2}{4.2}%
\setval{J1916+3828}{dis.met}{-2.16}%
\setval{J1916+3828}{dis.rv1}{0}%
\setval{J1916+3828}{dis.rv2}{0}%
\setval{J1916+3828}{dis.teff1}{6000}%
\setval{J1916+3828}{dis.teff2}{6400}%
\setval{J1916+3828}{dis.vmic1}{1.2}%
\setval{J1916+3828}{dis.vmic2}{1.3}%
\setval{J1916+3828}{dis.vsini1}{11.9}%
\setval{J1916+3828}{dis.vsini2}{8.0}%
\setval{J1916+3828}{ecc}{0.0161\pm0.0010}%
\setval{J1916+3828}{ecc@paren}{0.0161(10)}%
\setval{J1916+3828}{incl}{1.525\pm0.002}%
\setval{J1916+3828}{incl@paren}{1.5254(18)}%
\setval{J1916+3828}{incl_deg}{87.40^{+0.09}_{-0.1}}%
\setval{J1916+3828}{incl_deg@paren}{87.40(10)}%
\setval{J1916+3828}{k}{0.624\pm0.004}%
\setval{J1916+3828}{k@paren}{0.6238(42)}%
\setval{J1916+3828}{l3_frac}{0.07^{+0.01}_{-0.02}}%
\setval{J1916+3828}{l3_frac@paren}{0.075(16)}%
\setval{J1916+3828}{logage.basti.coeval}{10.14\pm0.02}%
\setval{J1916+3828}{logage.basti.coeval@paren}{10.144(23)}%
\setval{J1916+3828}{logage.basti.primary}{10.13\pm0.03}%
\setval{J1916+3828}{logage.basti.primary@paren}{10.127(27)}%
\setval{J1916+3828}{logage.basti.secondary}{10.17\pm0.03}%
\setval{J1916+3828}{logage.basti.secondary@paren}{10.175(29)}%
\setval{J1916+3828}{logage.mist.coeval}{10.15\pm0.02}%
\setval{J1916+3828}{logage.mist.coeval@paren}{10.147(22)}%
\setval{J1916+3828}{logage.mist.primary}{10.13\pm0.03}%
\setval{J1916+3828}{logage.mist.primary@paren}{10.127(30)}%
\setval{J1916+3828}{logage.mist.secondary}{10.17\pm0.03}%
\setval{J1916+3828}{logage.mist.secondary@paren}{10.174(33)}%
\setval{J1916+3828}{logage.padova.coeval}{10.14^{+0.03}_{-0.02}}%
\setval{J1916+3828}{logage.padova.coeval@paren}{10.139(23)}%
\setval{J1916+3828}{logage.padova.primary}{10.12\pm0.03}%
\setval{J1916+3828}{logage.padova.primary@paren}{10.122(27)}%
\setval{J1916+3828}{logage.padova.secondary}{10.17\pm0.03}%
\setval{J1916+3828}{logage.padova.secondary@paren}{10.167(33)}%
\setval{J1916+3828}{logs1}{-0.3^{+0.3}_{-0.8}}%
\setval{J1916+3828}{logs1@paren}{-0.34(58)}%
\setval{J1916+3828}{logs2}{-5^{+3}_{-4}}%
\setval{J1916+3828}{logs2@paren}{-4.5(35)}%
\setval{J1916+3828}{m1}{0.76\pm0.01}%
\setval{J1916+3828}{m1@paren}{0.765(11)}%
\setval{J1916+3828}{m2}{0.72\pm0.01}%
\setval{J1916+3828}{m2@paren}{0.724(11)}%
\setval{J1916+3828}{omega}{-1.610\pm0.003}%
\setval{J1916+3828}{omega@paren}{-1.6098(30)}%
\setval{J1916+3828}{omega_deg}{-92.2\pm0.2}%
\setval{J1916+3828}{omega_deg@paren}{-92.23(17)}%
\setval{J1916+3828}{pblum_primary}{7.8^{+0.2}_{-0.1}}%
\setval{J1916+3828}{pblum_primary@paren}{7.83(16)}%
\setval{J1916+3828}{period}{7.68258\pm0.00006}%
\setval{J1916+3828}{period@paren}{7.682579(58)}%
\setval{J1916+3828}{phi0}{0.1815\pm0.0005}%
\setval{J1916+3828}{phi0@paren}{0.18154(49)}%
\setval{J1916+3828}{q}{0.948\pm0.009}%
\setval{J1916+3828}{q@paren}{0.9477(92)}%
\setval{J1916+3828}{r1}{1.76\pm0.01}%
\setval{J1916+3828}{r1@paren}{1.760(11)}%
\setval{J1916+3828}{r2}{1.098\pm0.008}%
\setval{J1916+3828}{r2@paren}{1.0978(76)}%
\setval{J1916+3828}{requivsum}{2.86\pm0.02}%
\setval{J1916+3828}{requivsum@paren}{2.858(16)}%
\setval{J1916+3828}{rv_offset_0}{0.3\pm0.5}%
\setval{J1916+3828}{rv_offset_0@paren}{0.28(48)}%
\setval{J1916+3828}{sma}{18.70^{+0.08}_{-0.09}}%
\setval{J1916+3828}{sma@paren}{18.703(85)}%
\setval{J1916+3828}{teff1}{6700\pm200}%
\setval{J1916+3828}{teff1@paren}{6680(190)}%
\setval{J1916+3828}{teff2}{7200\pm200}%
\setval{J1916+3828}{teff2@paren}{7170(210)}%
\setval{J1916+3828}{teffratio}{1.074^{+0.001}_{-0.002}}%
\setval{J1916+3828}{teffratio@paren}{1.0739(14)}%
\setval{J1916+3828}{vgamma}{-185.1\pm0.2}%
\setval{J1916+3828}{vgamma@paren}{-185.14(20)}%
\setval{J1916+3828}{x}{-0.0049\pm0.0003}%
\setval{J1916+3828}{x@paren}{-0.00494(26)}%
\setval{J1916+3828}{y}{-0.127\pm0.004}%
\setval{J1916+3828}{y@paren}{-0.1266(39)}%
\setval{J0350+0805}{age.basti.coeval}{6.4\pm0.3}%
\setval{J0350+0805}{age.basti.coeval@paren}{6.40(29)}%
\setval{J0350+0805}{age.basti.dchi2}{0.51}%
\setval{J0350+0805}{age.basti.primary}{6.5^{+0.4}_{-0.3}}%
\setval{J0350+0805}{age.basti.primary@paren}{6.46(35)}%
\setval{J0350+0805}{age.basti.pval}{0.412}%
\setval{J0350+0805}{age.basti.secondary}{6.2\pm0.6}%
\setval{J0350+0805}{age.basti.secondary@paren}{6.22(57)}%
\setval{J0350+0805}{age.mist.coeval}{7.5\pm0.4}%
\setval{J0350+0805}{age.mist.coeval@paren}{7.45(38)}%
\setval{J0350+0805}{age.mist.dchi2}{0.48}%
\setval{J0350+0805}{age.mist.primary}{7.5^{+0.5}_{-0.4}}%
\setval{J0350+0805}{age.mist.primary@paren}{7.46(45)}%
\setval{J0350+0805}{age.mist.pval}{0.429}%
\setval{J0350+0805}{age.mist.secondary}{7.4\pm0.7}%
\setval{J0350+0805}{age.mist.secondary@paren}{7.43(70)}%
\setval{J0350+0805}{age.padova.coeval}{6.4\pm0.3}%
\setval{J0350+0805}{age.padova.coeval@paren}{6.42(29)}%
\setval{J0350+0805}{age.padova.dchi2}{0.46}%
\setval{J0350+0805}{age.padova.primary}{6.4\pm0.4}%
\setval{J0350+0805}{age.padova.primary@paren}{6.42(36)}%
\setval{J0350+0805}{age.padova.pval}{0.448}%
\setval{J0350+0805}{age.padova.secondary}{6.4^{+0.6}_{-0.5}}%
\setval{J0350+0805}{age.padova.secondary@paren}{6.38(57)}%
\setval{J0350+0805}{b_primary}{0.03^{+0.03}_{-0.02}}%
\setval{J0350+0805}{b_primary@paren}{0.029(23)}%
\setval{J0350+0805}{b_primary_frac}{0.02\pm0.01}%
\setval{J0350+0805}{b_primary_frac@paren}{0.017(13)}%
\setval{J0350+0805}{b_secondary}{0.03^{+0.03}_{-0.02}}%
\setval{J0350+0805}{b_secondary@paren}{0.029(23)}%
\setval{J0350+0805}{b_secondary_frac}{0.02\pm0.01}%
\setval{J0350+0805}{b_secondary_frac@paren}{0.017(13)}%
\setval{J0350+0805}{cosi}{0.002\pm0.002}%
\setval{J0350+0805}{cosi@paren}{0.0023(18)}%
\setval{J0350+0805}{dis.alpha}{-0.34}%
\setval{J0350+0805}{dis.alpha_fe}{0.44}%
\setval{J0350+0805}{dis.l2}{0.31}%
\setval{J0350+0805}{dis.logg1}{4.2}%
\setval{J0350+0805}{dis.logg2}{4.4}%
\setval{J0350+0805}{dis.met}{-0.78}%
\setval{J0350+0805}{dis.rv1}{0}%
\setval{J0350+0805}{dis.rv2}{0}%
\setval{J0350+0805}{dis.teff1}{5800}%
\setval{J0350+0805}{dis.teff2}{5600}%
\setval{J0350+0805}{dis.vmic1}{1.1}%
\setval{J0350+0805}{dis.vmic2}{1.0}%
\setval{J0350+0805}{dis.vsini1}{13.0}%
\setval{J0350+0805}{dis.vsini2}{9.1}%
\setval{J0350+0805}{ecc}{0.0018\pm0.0007}%
\setval{J0350+0805}{ecc@paren}{0.00176(68)}%
\setval{J0350+0805}{incl}{1.569\pm0.002}%
\setval{J0350+0805}{incl@paren}{1.5685(18)}%
\setval{J0350+0805}{incl_deg}{89.87^{+0.09}_{-0.1}}%
\setval{J0350+0805}{incl_deg@paren}{89.87(10)}%
\setval{J0350+0805}{k}{0.758\pm0.002}%
\setval{J0350+0805}{k@paren}{0.7576(18)}%
\setval{J0350+0805}{l3_frac}{0.081^{+0.002}_{-0.003}}%
\setval{J0350+0805}{l3_frac@paren}{0.0810(29)}%
\setval{J0350+0805}{logage.basti.coeval}{9.81\pm0.02}%
\setval{J0350+0805}{logage.basti.coeval@paren}{9.806(20)}%
\setval{J0350+0805}{logage.basti.primary}{9.81\pm0.02}%
\setval{J0350+0805}{logage.basti.primary@paren}{9.811(24)}%
\setval{J0350+0805}{logage.basti.secondary}{9.79\pm0.04}%
\setval{J0350+0805}{logage.basti.secondary@paren}{9.794(40)}%
\setval{J0350+0805}{logage.mist.coeval}{9.87\pm0.02}%
\setval{J0350+0805}{logage.mist.coeval@paren}{9.872(22)}%
\setval{J0350+0805}{logage.mist.primary}{9.87\pm0.03}%
\setval{J0350+0805}{logage.mist.primary@paren}{9.873(26)}%
\setval{J0350+0805}{logage.mist.secondary}{9.87\pm0.04}%
\setval{J0350+0805}{logage.mist.secondary@paren}{9.871(41)}%
\setval{J0350+0805}{logage.padova.coeval}{9.81\pm0.02}%
\setval{J0350+0805}{logage.padova.coeval@paren}{9.807(20)}%
\setval{J0350+0805}{logage.padova.primary}{9.81\pm0.02}%
\setval{J0350+0805}{logage.padova.primary@paren}{9.808(24)}%
\setval{J0350+0805}{logage.padova.secondary}{9.81\pm0.04}%
\setval{J0350+0805}{logage.padova.secondary@paren}{9.805(39)}%
\setval{J0350+0805}{logs1}{-5.0\pm3.0}%
\setval{J0350+0805}{logs1@paren}{-5.0(33)}%
\setval{J0350+0805}{logs2}{-6.0\pm3.0}%
\setval{J0350+0805}{logs2@paren}{-5.7(30)}%
\setval{J0350+0805}{m1}{0.933\pm0.004}%
\setval{J0350+0805}{m1@paren}{0.9334(42)}%
\setval{J0350+0805}{m2}{0.853\pm0.003}%
\setval{J0350+0805}{m2@paren}{0.8528(32)}%
\setval{J0350+0805}{omega}{-1.45^{+0.08}_{-0.03}}%
\setval{J0350+0805}{omega@paren}{-1.451(55)}%
\setval{J0350+0805}{omega_deg}{-83^{+4}_{-2}}%
\setval{J0350+0805}{omega_deg@paren}{-83.1(32)}%
\setval{J0350+0805}{pblum_primary}{7.77^{+0.04}_{-0.03}}%
\setval{J0350+0805}{pblum_primary@paren}{7.775(36)}%
\setval{J0350+0805}{period}{5.329566\pm0.000008}%
\setval{J0350+0805}{period@paren}{5.3295658(79)}%
\setval{J0350+0805}{phi0}{0.83338\pm0.00006}%
\setval{J0350+0805}{phi0@paren}{0.833378(62)}%
\setval{J0350+0805}{q}{0.914\pm0.002}%
\setval{J0350+0805}{q@paren}{0.9137(23)}%
\setval{J0350+0805}{r1}{1.212\pm0.002}%
\setval{J0350+0805}{r1@paren}{1.2118(20)}%
\setval{J0350+0805}{r2}{0.918\pm0.002}%
\setval{J0350+0805}{r2@paren}{0.9179(23)}%
\setval{J0350+0805}{requivsum}{2.130\pm0.004}%
\setval{J0350+0805}{requivsum@paren}{2.1297(35)}%
\setval{J0350+0805}{sma}{15.57\pm0.02}%
\setval{J0350+0805}{sma@paren}{15.572(21)}%
\setval{J0350+0805}{teff1}{6100^{+100}_{-200}}%
\setval{J0350+0805}{teff1@paren}{6150(160)}%
\setval{J0350+0805}{teff2}{5800^{+100}_{-200}}%
\setval{J0350+0805}{teff2@paren}{5840(140)}%
\setval{J0350+0805}{teffratio}{0.950^{+0.002}_{-0.001}}%
\setval{J0350+0805}{teffratio@paren}{0.9502(15)}%
\setval{J0350+0805}{vgamma}{89.54\pm0.07}%
\setval{J0350+0805}{vgamma@paren}{89.543(71)}%
\setval{J0350+0805}{x}{0.0050^{+0.001}_{-0.0009}}%
\setval{J0350+0805}{x@paren}{0.0050(12)}%
\setval{J0350+0805}{y}{-0.042^{+0.009}_{-0.008}}%
\setval{J0350+0805}{y@paren}{-0.0417(84)}%
\setval{J2352-0707}{age.basti.coeval}{9.1^{+0.5}_{-0.4}}%
\setval{J2352-0707}{age.basti.coeval@paren}{9.07(46)}%
\setval{J2352-0707}{age.basti.dchi2}{0.58}%
\setval{J2352-0707}{age.basti.primary}{9.3^{+0.7}_{-0.6}}%
\setval{J2352-0707}{age.basti.primary@paren}{9.28(63)}%
\setval{J2352-0707}{age.basti.pval}{0.376}%
\setval{J2352-0707}{age.basti.secondary}{8.8\pm0.7}%
\setval{J2352-0707}{age.basti.secondary@paren}{8.80(68)}%
\setval{J2352-0707}{age.mist.coeval}{9.4\pm0.5}%
\setval{J2352-0707}{age.mist.coeval@paren}{9.41(53)}%
\setval{J2352-0707}{age.mist.dchi2}{0.55}%
\setval{J2352-0707}{age.mist.primary}{9.7^{+0.7}_{-0.8}}%
\setval{J2352-0707}{age.mist.primary@paren}{9.66(72)}%
\setval{J2352-0707}{age.mist.pval}{0.377}%
\setval{J2352-0707}{age.mist.secondary}{9.1\pm0.8}%
\setval{J2352-0707}{age.mist.secondary@paren}{9.14(80)}%
\setval{J2352-0707}{age.padova.coeval}{9.0\pm0.5}%
\setval{J2352-0707}{age.padova.coeval@paren}{8.96(48)}%
\setval{J2352-0707}{age.padova.dchi2}{0.56}%
\setval{J2352-0707}{age.padova.primary}{9.2\pm0.6}%
\setval{J2352-0707}{age.padova.primary@paren}{9.17(60)}%
\setval{J2352-0707}{age.padova.pval}{0.391}%
\setval{J2352-0707}{age.padova.secondary}{8.7\pm0.7}%
\setval{J2352-0707}{age.padova.secondary@paren}{8.70(73)}%
\setval{J2352-0707}{b_primary}{0.72\pm0.03}%
\setval{J2352-0707}{b_primary@paren}{0.723(30)}%
\setval{J2352-0707}{b_primary_frac}{0.368^{+0.003}_{-0.006}}%
\setval{J2352-0707}{b_primary_frac@paren}{0.3680(44)}%
\setval{J2352-0707}{b_secondary}{0.72\pm0.03}%
\setval{J2352-0707}{b_secondary@paren}{0.723(31)}%
\setval{J2352-0707}{b_secondary_frac}{0.368^{+0.003}_{-0.006}}%
\setval{J2352-0707}{b_secondary_frac@paren}{0.3680(45)}%
\setval{J2352-0707}{cosi}{0.0385^{+0.0004}_{-0.0006}}%
\setval{J2352-0707}{cosi@paren}{0.03851(52)}%
\setval{J2352-0707}{dis.alpha}{-1.00}%
\setval{J2352-0707}{dis.alpha_fe}{0.38}%
\setval{J2352-0707}{dis.l2}{0.48}%
\setval{J2352-0707}{dis.logg1}{4.3}%
\setval{J2352-0707}{dis.logg2}{4.3}%
\setval{J2352-0707}{dis.met}{-1.38}%
\setval{J2352-0707}{dis.rv1}{0}%
\setval{J2352-0707}{dis.rv2}{0}%
\setval{J2352-0707}{dis.teff1}{6100}%
\setval{J2352-0707}{dis.teff2}{6100}%
\setval{J2352-0707}{dis.vmic1}{1.1}%
\setval{J2352-0707}{dis.vmic2}{1.1}%
\setval{J2352-0707}{dis.vsini1}{9.2}%
\setval{J2352-0707}{dis.vsini2}{7.9}%
\setval{J2352-0707}{ecc}{0.0002^{+0.0008}_{-0.0001}}%
\setval{J2352-0707}{ecc@paren}{0.00015(44)}%
\setval{J2352-0707}{incl}{1.5323^{+0.0006}_{-0.0004}}%
\setval{J2352-0707}{incl@paren}{1.53228(52)}%
\setval{J2352-0707}{incl_deg}{87.79^{+0.04}_{-0.02}}%
\setval{J2352-0707}{incl_deg@paren}{87.793(30)}%
\setval{J2352-0707}{k}{0.97^{+0.08}_{-0.06}}%
\setval{J2352-0707}{k@paren}{0.965(72)}%
\setval{J2352-0707}{l3_frac}{0.008^{+0.01}_{-0.006}}%
\setval{J2352-0707}{l3_frac@paren}{0.0076(85)}%
\setval{J2352-0707}{logage.basti.coeval}{9.96\pm0.02}%
\setval{J2352-0707}{logage.basti.coeval@paren}{9.958(22)}%
\setval{J2352-0707}{logage.basti.primary}{9.97\pm0.03}%
\setval{J2352-0707}{logage.basti.primary@paren}{9.968(30)}%
\setval{J2352-0707}{logage.basti.secondary}{9.94^{+0.03}_{-0.04}}%
\setval{J2352-0707}{logage.basti.secondary@paren}{9.945(34)}%
\setval{J2352-0707}{logage.mist.coeval}{9.97^{+0.02}_{-0.03}}%
\setval{J2352-0707}{logage.mist.coeval@paren}{9.974(25)}%
\setval{J2352-0707}{logage.mist.primary}{9.98^{+0.03}_{-0.04}}%
\setval{J2352-0707}{logage.mist.primary@paren}{9.985(33)}%
\setval{J2352-0707}{logage.mist.secondary}{9.96\pm0.04}%
\setval{J2352-0707}{logage.mist.secondary@paren}{9.961(38)}%
\setval{J2352-0707}{logage.padova.coeval}{9.95\pm0.02}%
\setval{J2352-0707}{logage.padova.coeval@paren}{9.952(23)}%
\setval{J2352-0707}{logage.padova.primary}{9.96\pm0.03}%
\setval{J2352-0707}{logage.padova.primary@paren}{9.963(28)}%
\setval{J2352-0707}{logage.padova.secondary}{9.94^{+0.03}_{-0.04}}%
\setval{J2352-0707}{logage.padova.secondary@paren}{9.940(36)}%
\setval{J2352-0707}{logs1}{-0.8\pm0.3}%
\setval{J2352-0707}{logs1@paren}{-0.82(29)}%
\setval{J2352-0707}{logs2}{-1.3^{+0.4}_{-0.7}}%
\setval{J2352-0707}{logs2@paren}{-1.31(56)}%
\setval{J2352-0707}{m1}{0.811^{+0.006}_{-0.005}}%
\setval{J2352-0707}{m1@paren}{0.8115(54)}%
\setval{J2352-0707}{m2}{0.813\pm0.007}%
\setval{J2352-0707}{m2@paren}{0.8135(66)}%
\setval{J2352-0707}{omega}{0^{+1}_{-2}}%
\setval{J2352-0707}{omega@paren}{0.4(16)}%
\setval{J2352-0707}{omega_deg}{20^{+70}_{-110}}%
\setval{J2352-0707}{omega_deg@paren}{23(89)}%
\setval{J2352-0707}{pblum_primary}{6.4^{+0.4}_{-0.5}}%
\setval{J2352-0707}{pblum_primary@paren}{6.42(47)}%
\setval{J2352-0707}{period}{7.89070\pm0.00003}%
\setval{J2352-0707}{period@paren}{7.890705(28)}%
\setval{J2352-0707}{phi0}{0.2574\pm0.0001}%
\setval{J2352-0707}{phi0@paren}{0.25740(10)}%
\setval{J2352-0707}{q}{1.002\pm0.005}%
\setval{J2352-0707}{q@paren}{1.0023(48)}%
\setval{J2352-0707}{r1}{1.04^{+0.03}_{-0.04}}%
\setval{J2352-0707}{r1@paren}{1.044(37)}%
\setval{J2352-0707}{r2}{1.01\pm0.04}%
\setval{J2352-0707}{r2@paren}{1.008(38)}%
\setval{J2352-0707}{requivsum}{2.051^{+0.006}_{-0.007}}%
\setval{J2352-0707}{requivsum@paren}{2.0513(67)}%
\setval{J2352-0707}{sma}{19.60\pm0.05}%
\setval{J2352-0707}{sma@paren}{19.601(46)}%
\setval{J2352-0707}{teff1}{5900\pm200}%
\setval{J2352-0707}{teff1@paren}{5930(190)}%
\setval{J2352-0707}{teff2}{5900\pm200}%
\setval{J2352-0707}{teff2@paren}{5940(200)}%
\setval{J2352-0707}{teffratio}{1.001\pm0.001}%
\setval{J2352-0707}{teffratio@paren}{1.0014(11)}%
\setval{J2352-0707}{vgamma}{-32.5\pm0.1}%
\setval{J2352-0707}{vgamma@paren}{-32.498(96)}%
\setval{J2352-0707}{x}{0.001^{+0.003}_{-0.002}}%
\setval{J2352-0707}{x@paren}{0.0013(30)}%
\setval{J2352-0707}{y}{0.00\pm0.02}%
\setval{J2352-0707}{y@paren}{0.001(19)}%
\setval{J0414-1023}{age.basti.coeval}{4.1\pm0.2}%
\setval{J0414-1023}{age.basti.coeval@paren}{4.07(17)}%
\setval{J0414-1023}{age.basti.dchi2}{0.44}%
\setval{J0414-1023}{age.basti.primary}{4.1\pm0.2}%
\setval{J0414-1023}{age.basti.primary@paren}{4.05(23)}%
\setval{J0414-1023}{age.basti.pval}{0.453}%
\setval{J0414-1023}{age.basti.secondary}{4.1\pm0.2}%
\setval{J0414-1023}{age.basti.secondary@paren}{4.09(25)}%
\setval{J0414-1023}{age.mist.coeval}{4.3\pm0.2}%
\setval{J0414-1023}{age.mist.coeval@paren}{4.29(20)}%
\setval{J0414-1023}{age.mist.dchi2}{0.43}%
\setval{J0414-1023}{age.mist.primary}{4.3\pm0.3}%
\setval{J0414-1023}{age.mist.primary@paren}{4.27(28)}%
\setval{J0414-1023}{age.mist.pval}{0.454}%
\setval{J0414-1023}{age.mist.secondary}{4.3\pm0.3}%
\setval{J0414-1023}{age.mist.secondary@paren}{4.31(29)}%
\setval{J0414-1023}{age.padova.coeval}{4.0\pm0.2}%
\setval{J0414-1023}{age.padova.coeval@paren}{4.00(17)}%
\setval{J0414-1023}{age.padova.dchi2}{0.43}%
\setval{J0414-1023}{age.padova.primary}{4.0\pm0.2}%
\setval{J0414-1023}{age.padova.primary@paren}{3.98(24)}%
\setval{J0414-1023}{age.padova.pval}{0.444}%
\setval{J0414-1023}{age.padova.secondary}{4.0^{+0.3}_{-0.2}}%
\setval{J0414-1023}{age.padova.secondary@paren}{4.01(25)}%
\setval{J0414-1023}{b_primary}{1.11\pm0.02}%
\setval{J0414-1023}{b_primary@paren}{1.108(21)}%
\setval{J0414-1023}{b_primary_frac}{0.610\pm0.002}%
\setval{J0414-1023}{b_primary_frac@paren}{0.6095(20)}%
\setval{J0414-1023}{b_secondary}{1.10\pm0.02}%
\setval{J0414-1023}{b_secondary@paren}{1.102(21)}%
\setval{J0414-1023}{b_secondary_frac}{0.606^{+0.002}_{-0.003}}%
\setval{J0414-1023}{b_secondary_frac@paren}{0.6064(24)}%
\setval{J0414-1023}{cosi}{0.1070^{+0.0005}_{-0.0006}}%
\setval{J0414-1023}{cosi@paren}{0.10702(57)}%
\setval{J0414-1023}{dis.alpha}{-0.38}%
\setval{J0414-1023}{dis.alpha_fe}{0.20}%
\setval{J0414-1023}{dis.l2}{0.49}%
\setval{J0414-1023}{dis.logg1}{4.1}%
\setval{J0414-1023}{dis.logg2}{4.1}%
\setval{J0414-1023}{dis.met}{-0.58}%
\setval{J0414-1023}{dis.rv1}{0}%
\setval{J0414-1023}{dis.rv2}{0}%
\setval{J0414-1023}{dis.teff1}{6300}%
\setval{J0414-1023}{dis.teff2}{6200}%
\setval{J0414-1023}{dis.vmic1}{1.3}%
\setval{J0414-1023}{dis.vmic2}{1.3}%
\setval{J0414-1023}{dis.vsini1}{14.2}%
\setval{J0414-1023}{dis.vsini2}{13.8}%
\setval{J0414-1023}{ecc}{0.003\pm0.002}%
\setval{J0414-1023}{ecc@paren}{0.0027(19)}%
\setval{J0414-1023}{incl}{1.4636\pm0.0006}%
\setval{J0414-1023}{incl@paren}{1.46357(57)}%
\setval{J0414-1023}{incl_deg}{83.86\pm0.03}%
\setval{J0414-1023}{incl_deg@paren}{83.856(33)}%
\setval{J0414-1023}{k}{0.82\pm0.03}%
\setval{J0414-1023}{k@paren}{0.818(29)}%
\setval{J0414-1023}{logage.basti.coeval}{9.61\pm0.02}%
\setval{J0414-1023}{logage.basti.coeval@paren}{9.610(18)}%
\setval{J0414-1023}{logage.basti.primary}{9.61\pm0.03}%
\setval{J0414-1023}{logage.basti.primary@paren}{9.608(25)}%
\setval{J0414-1023}{logage.basti.secondary}{9.61\pm0.03}%
\setval{J0414-1023}{logage.basti.secondary@paren}{9.612(26)}%
\setval{J0414-1023}{logage.mist.coeval}{9.63\pm0.02}%
\setval{J0414-1023}{logage.mist.coeval@paren}{9.632(21)}%
\setval{J0414-1023}{logage.mist.primary}{9.63\pm0.03}%
\setval{J0414-1023}{logage.mist.primary@paren}{9.630(28)}%
\setval{J0414-1023}{logage.mist.secondary}{9.63\pm0.03}%
\setval{J0414-1023}{logage.mist.secondary@paren}{9.634(29)}%
\setval{J0414-1023}{logage.padova.coeval}{9.60\pm0.02}%
\setval{J0414-1023}{logage.padova.coeval@paren}{9.602(19)}%
\setval{J0414-1023}{logage.padova.primary}{9.60\pm0.03}%
\setval{J0414-1023}{logage.padova.primary@paren}{9.600(26)}%
\setval{J0414-1023}{logage.padova.secondary}{9.60\pm0.03}%
\setval{J0414-1023}{logage.padova.secondary@paren}{9.604(27)}%
\setval{J0414-1023}{logs1}{-5.0\pm3.0}%
\setval{J0414-1023}{logs1@paren}{-4.6(33)}%
\setval{J0414-1023}{logs2}{-4^{+3}_{-4}}%
\setval{J0414-1023}{logs2@paren}{-4.3(35)}%
\setval{J0414-1023}{m1}{1.12\pm0.01}%
\setval{J0414-1023}{m1@paren}{1.117(11)}%
\setval{J0414-1023}{m2}{1.11\pm0.01}%
\setval{J0414-1023}{m2@paren}{1.114(11)}%
\setval{J0414-1023}{omega}{-1.68^{+0.05}_{-0.2}}%
\setval{J0414-1023}{omega@paren}{-1.68(11)}%
\setval{J0414-1023}{omega_deg}{-96^{+3}_{-10}}%
\setval{J0414-1023}{omega_deg@paren}{-96.3(62)}%
\setval{J0414-1023}{pblum_primary}{7.6\pm0.2}%
\setval{J0414-1023}{pblum_primary@paren}{7.62(22)}%
\setval{J0414-1023}{period}{6.050367^{+0.000009}_{-0.000010}}%
\setval{J0414-1023}{period@paren}{6.0503672(95)}%
\setval{J0414-1023}{phi0}{0.8899\pm0.0002}%
\setval{J0414-1023}{phi0@paren}{0.88992(21)}%
\setval{J0414-1023}{q}{0.998\pm0.006}%
\setval{J0414-1023}{q@paren}{0.9979(60)}%
\setval{J0414-1023}{r1}{1.77^{+0.02}_{-0.03}}%
\setval{J0414-1023}{r1@paren}{1.767(25)}%
\setval{J0414-1023}{r2}{1.45\pm0.03}%
\setval{J0414-1023}{r2@paren}{1.445(31)}%
\setval{J0414-1023}{requivsum}{3.21\pm0.01}%
\setval{J0414-1023}{requivsum@paren}{3.212(13)}%
\setval{J0414-1023}{requivsum_over_2}{1.606^{+0.007}_{-0.006}}%
\setval{J0414-1023}{requivsum_over_2@paren}{1.6060(64)}%
\setval{J0414-1023}{rv_offset_0}{-0.2\pm0.4}%
\setval{J0414-1023}{rv_offset_0@paren}{-0.22(42)}%
\setval{J0414-1023}{sma}{18.25\pm0.06}%
\setval{J0414-1023}{sma@paren}{18.249(55)}%
\setval{J0414-1023}{teff1}{6200\pm200}%
\setval{J0414-1023}{teff1@paren}{6200(170)}%
\setval{J0414-1023}{teff2}{6100\pm200}%
\setval{J0414-1023}{teff2@paren}{6140(170)}%
\setval{J0414-1023}{teffratio}{0.990\pm0.003}%
\setval{J0414-1023}{teffratio@paren}{0.9903(27)}%
\setval{J0414-1023}{vgamma}{4.5\pm0.2}%
\setval{J0414-1023}{vgamma@paren}{4.50(17)}%
\setval{J0414-1023}{x}{-0.007^{+0.002}_{-0.006}}%
\setval{J0414-1023}{x@paren}{-0.0067(40)}%
\setval{J0414-1023}{y}{-0.05^{+0.03}_{-0.02}}%
\setval{J0414-1023}{y@paren}{-0.051(24)}%
\setval{J1626-1140}{age.basti.coeval}{6.0\pm0.3}%
\setval{J1626-1140}{age.basti.coeval@paren}{6.03(28)}%
\setval{J1626-1140}{age.basti.dchi2}{0.45}%
\setval{J1626-1140}{age.basti.primary}{6.1\pm0.4}%
\setval{J1626-1140}{age.basti.primary@paren}{6.09(37)}%
\setval{J1626-1140}{age.basti.pval}{0.437}%
\setval{J1626-1140}{age.basti.secondary}{6.0\pm0.4}%
\setval{J1626-1140}{age.basti.secondary@paren}{5.97(41)}%
\setval{J1626-1140}{age.mist.coeval}{6.2\pm0.3}%
\setval{J1626-1140}{age.mist.coeval@paren}{6.24(32)}%
\setval{J1626-1140}{age.mist.dchi2}{0.44}%
\setval{J1626-1140}{age.mist.primary}{6.3^{+0.5}_{-0.4}}%
\setval{J1626-1140}{age.mist.primary@paren}{6.29(43)}%
\setval{J1626-1140}{age.mist.pval}{0.446}%
\setval{J1626-1140}{age.mist.secondary}{6.2^{+0.4}_{-0.5}}%
\setval{J1626-1140}{age.mist.secondary@paren}{6.19(45)}%
\setval{J1626-1140}{age.padova.coeval}{6.0\pm0.3}%
\setval{J1626-1140}{age.padova.coeval@paren}{5.99(28)}%
\setval{J1626-1140}{age.padova.dchi2}{0.49}%
\setval{J1626-1140}{age.padova.primary}{6.0\pm0.4}%
\setval{J1626-1140}{age.padova.primary@paren}{6.02(38)}%
\setval{J1626-1140}{age.padova.pval}{0.400}%
\setval{J1626-1140}{age.padova.secondary}{5.9\pm0.4}%
\setval{J1626-1140}{age.padova.secondary@paren}{5.95(41)}%
\setval{J1626-1140}{b_primary}{0.40\pm0.01}%
\setval{J1626-1140}{b_primary@paren}{0.403(11)}%
\setval{J1626-1140}{b_primary_frac}{0.195\pm0.007}%
\setval{J1626-1140}{b_primary_frac@paren}{0.1954(73)}%
\setval{J1626-1140}{b_secondary}{0.40\pm0.01}%
\setval{J1626-1140}{b_secondary@paren}{0.405(11)}%
\setval{J1626-1140}{b_secondary_frac}{0.196\pm0.007}%
\setval{J1626-1140}{b_secondary_frac@paren}{0.1961(68)}%
\setval{J1626-1140}{cosi}{0.0269^{+0.0009}_{-0.001}}%
\setval{J1626-1140}{cosi@paren}{0.02693(97)}%
\setval{J1626-1140}{dis.alpha}{-0.67}%
\setval{J1626-1140}{dis.alpha_fe}{0.19}%
\setval{J1626-1140}{dis.l2}{0.53}%
\setval{J1626-1140}{dis.logg1}{4.2}%
\setval{J1626-1140}{dis.logg2}{4.1}%
\setval{J1626-1140}{dis.met}{-0.86}%
\setval{J1626-1140}{dis.rv1}{0}%
\setval{J1626-1140}{dis.rv2}{0}%
\setval{J1626-1140}{dis.teff1}{6100}%
\setval{J1626-1140}{dis.teff2}{6100}%
\setval{J1626-1140}{dis.vmic1}{1.2}%
\setval{J1626-1140}{dis.vmic2}{1.2}%
\setval{J1626-1140}{dis.vsini1}{9.9}%
\setval{J1626-1140}{dis.vsini2}{9.9}%
\setval{J1626-1140}{ecc}{0.02168^{+0.0002}_{-0.00009}}%
\setval{J1626-1140}{ecc@paren}{0.02168(14)}%
\setval{J1626-1140}{incl}{1.5439^{+0.001}_{-0.0009}}%
\setval{J1626-1140}{incl@paren}{1.54387(97)}%
\setval{J1626-1140}{incl_deg}{88.46^{+0.06}_{-0.05}}%
\setval{J1626-1140}{incl_deg@paren}{88.457(55)}%
\setval{J1626-1140}{k}{1.07^{+0.04}_{-0.05}}%
\setval{J1626-1140}{k@paren}{1.071(42)}%
\setval{J1626-1140}{l3_frac}{0.035\pm0.008}%
\setval{J1626-1140}{l3_frac@paren}{0.0355(81)}%
\setval{J1626-1140}{logage.basti.coeval}{9.78\pm0.02}%
\setval{J1626-1140}{logage.basti.coeval@paren}{9.780(20)}%
\setval{J1626-1140}{logage.basti.primary}{9.78\pm0.03}%
\setval{J1626-1140}{logage.basti.primary@paren}{9.784(27)}%
\setval{J1626-1140}{logage.basti.secondary}{9.78\pm0.03}%
\setval{J1626-1140}{logage.basti.secondary@paren}{9.776(30)}%
\setval{J1626-1140}{logage.mist.coeval}{9.80\pm0.02}%
\setval{J1626-1140}{logage.mist.coeval@paren}{9.795(22)}%
\setval{J1626-1140}{logage.mist.primary}{9.80\pm0.03}%
\setval{J1626-1140}{logage.mist.primary@paren}{9.799(30)}%
\setval{J1626-1140}{logage.mist.secondary}{9.79\pm0.03}%
\setval{J1626-1140}{logage.mist.secondary@paren}{9.792(31)}%
\setval{J1626-1140}{logage.padova.coeval}{9.78\pm0.02}%
\setval{J1626-1140}{logage.padova.coeval@paren}{9.777(20)}%
\setval{J1626-1140}{logage.padova.primary}{9.78\pm0.03}%
\setval{J1626-1140}{logage.padova.primary@paren}{9.779(27)}%
\setval{J1626-1140}{logage.padova.secondary}{9.77\pm0.03}%
\setval{J1626-1140}{logage.padova.secondary@paren}{9.774(30)}%
\setval{J1626-1140}{logs1}{-5.0\pm3.0}%
\setval{J1626-1140}{logs1@paren}{-5.4(32)}%
\setval{J1626-1140}{logs2}{-0.7\pm0.3}%
\setval{J1626-1140}{logs2@paren}{-0.70(26)}%
\setval{J1626-1140}{m1}{0.960\pm0.009}%
\setval{J1626-1140}{m1@paren}{0.9602(88)}%
\setval{J1626-1140}{m2}{0.98\pm0.01}%
\setval{J1626-1140}{m2@paren}{0.976(12)}%
\setval{J1626-1140}{omega}{3.03^{+0.06}_{-0.1}}%
\setval{J1626-1140}{omega@paren}{3.026(89)}%
\setval{J1626-1140}{omega_deg}{173^{+4}_{-6}}%
\setval{J1626-1140}{omega_deg@paren}{173.4(51)}%
\setval{J1626-1140}{pblum_primary}{5.7^{+0.3}_{-0.2}}%
\setval{J1626-1140}{pblum_primary@paren}{5.67(25)}%
\setval{J1626-1140}{period}{7.48687\pm0.00005}%
\setval{J1626-1140}{period@paren}{7.486869(53)}%
\setval{J1626-1140}{phi0}{0.8263^{+0.0007}_{-0.0008}}%
\setval{J1626-1140}{phi0@paren}{0.82626(75)}%
\setval{J1626-1140}{q}{1.016^{+0.007}_{-0.008}}%
\setval{J1626-1140}{q@paren}{1.0164(77)}%
\setval{J1626-1140}{r1}{1.33\pm0.03}%
\setval{J1626-1140}{r1@paren}{1.332(29)}%
\setval{J1626-1140}{r2}{1.43^{+0.02}_{-0.03}}%
\setval{J1626-1140}{r2@paren}{1.427(27)}%
\setval{J1626-1140}{requivsum}{2.76\pm0.01}%
\setval{J1626-1140}{requivsum@paren}{2.758(10)}%
\setval{J1626-1140}{sma}{20.07\pm0.07}%
\setval{J1626-1140}{sma@paren}{20.067(69)}%
\setval{J1626-1140}{teff1}{5900\pm200}%
\setval{J1626-1140}{teff1@paren}{5910(180)}%
\setval{J1626-1140}{teff2}{5900\pm200}%
\setval{J1626-1140}{teff2@paren}{5890(180)}%
\setval{J1626-1140}{teffratio}{0.9965\pm0.0010}%
\setval{J1626-1140}{teffratio@paren}{0.99653(97)}%
\setval{J1626-1140}{vgamma}{-53.2^{+0.2}_{-0.1}}%
\setval{J1626-1140}{vgamma@paren}{-53.16(15)}%
\setval{J1626-1140}{x}{-0.1465^{+0.0007}_{-0.0003}}%
\setval{J1626-1140}{x@paren}{-0.14652(49)}%
\setval{J1626-1140}{y}{0.01\pm0.01}%
\setval{J1626-1140}{y@paren}{0.013(11)}%

\begin{document}

\title{Precise and Accurate Mass and Radius Measurements of Thirty Galactic Metal-Poor Stars in Detached Eclipsing Binaries}

\author[0000-0003-2431-981X]{D. M. Rowan}\thanks{NHFP Hubble Fellow}
\email{dmrowan@berkeley.edu}
\affiliation{Department of Astronomy, University of California, Berkeley, CA 94720, USA}

\author[0009-0001-1470-8400]{K. Z. Stanek}
\email{stanek.32@osu.edu}
\affiliation{Department of Astronomy, The Ohio State University, 140 West 18th Avenue,
Columbus, OH 43210, USA}
\affiliation{Center for Cosmology and Astroparticle Physics, The Ohio State University, 191 W. Woodruff Avenue, Columbus, OH, 43210, USA}

\author[0000-0001-6017-2961]{C. S. Kochanek}
\email{kochanek.1@osu.edu}
\affiliation{Department of Astronomy, The Ohio State University, 140 West 18th Avenue,
Columbus, OH 43210, USA}
\affiliation{Center for Cosmology and Astroparticle Physics, The Ohio State University, 191 W. Woodruff Avenue, Columbus, OH, 43210, USA}

\author[0000-0003-3504-5316]{Benjamin J. Fulton}
\email{bjfulton@ipac.caltech.edu}
\affiliation{NASA Exoplanet Science Institute/Caltech-IPAC, California Institute of Technology, Pasadena, CA 91125, USA}

\author[0000-0002-0551-046X]{Ilya Ilyin}
\email{ilyin@aip.de}
\affiliation{Leibniz-Institut for Astrophysics Potsdam (AIP), An der Sternwarte 16, D14482 Potsdam, Germany}
\author[0000-0002-0531-1073]{Howard Isaacson}
\email{hisaacson@berkeley.edu}
\affiliation{Department of Astronomy, University of California, Berkeley, CA 94720, USA}

\author[0000-0002-7595-6360]{David V. Martin}
\email{david.martin@tufts.edu}
\affiliation{Department of Physics and Astronomy, Tufts University, Medford, MA 02155, USA}

%% Use the \collaboration command to identify collaborations. This command
%% takes an optional argument that is either a number or the word "all"
%% which tells the compiler how many of the authors above the command to
%% show. For example "\collaboration[all]{(DELVE Collaboration)}" wil include
%% all the authors above this command.
%%
%% Mark off the abstract in the ``abstract'' environment. 
\begin{abstract}

Detached eclipsing binaries have long served as fundamental benchmarks for making direct comparisons with stellar models. By combining the light curve and radial velocity orbit of a detached eclipsing binary, precise masses and radii can be measured for both components. However, catalogs of eclipsing binaries are dominated by solar-metallicity stars. Here we characterize fifteen eclipsing binaries selected to have $\mh{}\lesssim -0.5$ using \TESS{} photometry, \Gaia{} XP spectra, large spectroscopic surveys, and Galactic kinematics. We use three spectrographs, PEPSI on the Large Binocular Telescope, the Levy spectrograph on the Automated Planet Finder telescope, and CHIRON on the SMARTS 1.5m telescope. We jointly model the light curves and radial velocities and measure the stellar masses and radii with median fractional uncertainties of $< 1\%$. We use spectral disentangling to extract the spectra of both stars and model them using the spectral synthesis code {\tt Korg}. We find three systems with $\mh{} < -2.0$, two of which are kinematically consistent with the Milky Way's halo. J0001$+$2156 has $\mh{}=\specmet{J0001+2156}$, making it the most metal-poor detached eclipsing binary characterized to date. Eleven of the twelve other systems have $\mh{} \lesssim -0.5$ and are split between being highly ($\alpham{} > 0.25$) and moderately ($\alpham{} < 0.25$) $\alpha$-enhanced. We compare our measured masses and radii to predictions from theoretical isochrones and estimate the ages of the binaries. We estimate ages with typical uncertainties of $\lesssim 0.5$~Gyr, and three of the binaries have estimated ages greater than $10$~Gyr, up to $13.8^{+0.8}_{-0.7}$~Gyr. Finally, we update the empirical relations for stellar masses and radii on the main sequence using a sample of 621 stars.

\end{abstract}

%% Keywords should appear after the \end{abstract} command. 
%% PASP uses Unified Astronomy Thesaurus (UAT) concepts:
%% https://astrothesaurus.org
%% You will be asked to selected these concepts during the submission process
%% but this old "keyword" functionality is maintained in case authors want
%% to include these concepts in their preprints.
%%
%% You can use the \uat command to link your UAT concepts back its source.
%\keywords{\uat{Galaxies}{573} --- \uat{Cosmology}{343} --- \uat{High Energy astrophysics}{739} --- \uat{Interstellar medium}{847} --- \uat{Stellar astronomy}{1583} --- \uat{Solar physics}{1476}}

%\keywords{\uat{Classical Novae}{251} --- \uat{Ultraviolet astronomy}{1736} --- \uat{History of astronomy}{1868} --- \uat{Interdisciplinary astronomy}{804}}

%% From the front matter, we move on to the body of the paper.
%% Sections are demarcated by \section and \subsection, respectively.
%% Observe the use of the LaTeX \label
%% command after the \subsection to give a symbolic KEY to the
%% subsection for cross-referencing in a \ref command.
%% You can use LaTeX's \ref and \label commands to keep track of
%% cross-references to sections, equations, tables, and figures.
%% That way, if you change the order of any elements, LaTeX will
%% automatically renumber them.

\section{Introduction} \label{sec:intro}

Detached eclipsing binaries (EBs) offer the most effective means to measure accurate and precise model-independent masses and radii of stars. From the EB light curve, we measure the orbital period, eccentricity, inclination, and the fractional radii of the binary components. From multi-epoch spectroscopy, we can measure the radial velocity semi-amplitudes of both stars. By combining the light curve and RV orbits, we can measure the physical masses and radii of both stars with fractional uncertainties of $\sim 1$--3\% \citep[e.g.,][]{Andersen91, Torres10}.

Since the stars in detached EBs have evolved effectively in isolation, the mass and radius measurements from EBs have historically been used as important benchmarks for testing predictions from stellar theory \citep[e.g.,][]{Paxton13,Swayne22}. Mass and radius measurements from EBs also serve as an anchor for other methods of mass estimation, such as asteroseismology \citep{Hekker10, Gaulme16}. The empirical relations between stellar parameters such as mass and luminosity are calibrated with EBs \citep[e.g.,][]{Fernandes21}, and these relations can be used to estimate stellar properties \citep[e.g.,][]{Serenelli21}. 

Despite the importance of EBs for making benchmark mass and radius measurements, only a few hundred detached eclipsing binaries have precise masses and radii \citep{Southworth15, Maxted23}. The distribution of the existing sample is also far from uniform across the Hertzsprung--Russell diagram, and is dominated by main sequence stars at solar metallicity. In particular, there are few well-characterized low-metallicity ($[\rm{M/H}] \leq -0.5$) EBs.

Most of the well-characterized metal-poor eclipsing binaries are found in globular clusters \citep[e.g.,][]{Thompson20, Rozyczka22}. \citet{Kaluzny13} measured masses and radii for three main sequence eclipsing binaries in M4, which has $[\rm{Fe/H}] \simeq -1.2$. An even more metal-poor system was characterized in M55, which has $[\rm{Fe/H}] \simeq -1.9$ \citep{Kaluzny14}. The Magellanic Clouds offer a promising source of metal-poor eclipsing binaries, but the large distance has limited detailed studies to bright evolved and massive stars \citep[e.g.,][]{Graczyk12, Pietrzynski11, Graczyk14, Taormina25}.

There have been some metal-poor eclipsing binaries identified in the Galactic field, such as the evolved thick-disk binary KIC 4054905 \citep{Brogaard22} and V432~Aur \citep{Siviero04}, both of which have $[\rm{Fe/H}] \approx -0.6$. In total, there are only six Galactic field binaries with $[\rm{M/H}] \leq -0.5$ in DEBCat \citep{Southworth15}, the most up-to-date catalog of detached eclipsing binaries with robust mass and radius measurements. 

The lack of metal-poor EBs in these samples is in part a consequence of target selection strategies. Eclipsing binaries are typically selected photometrically, often as part of large catalogs of variable stars \citep[e.g.,][]{Jayasinghe19, Rowan22}. With large all-sky spectroscopic and photometric surveys, we have the opportunity to invert the selection, starting from metal-poor stars and then identifying eclipsing binaries. Here we present masses and radii for 15 metal-poor eclipsing binaries using \TESS{} photometry and ground-based spectroscopy. Section \ref{sec:target_selection} describes our target selection strategy and the \TESS{} photometry. We measure RVs with three spectrographs (Section \ref{sec:spec_obs}) and jointly model the light curves and radial velocities in Section \ref{sec:models}. We use spectral disentangling and spectral synthesis to measure the metallicity and $\alpha$-abundance of binaries in Section \ref{sec:disentangling}. We use the \Gaia{} astrometry to characterize the Galactic kinematics of the sample in Section \ref{sec:kinematics}, and estimate the age of the binaries using theoretical isochrones in Section \ref{sec:ages}. We discuss each target individually in Section \ref{sec:individual_targets}. Finally, in Sections \ref{sec:discussion} and \ref{sec:conclusions} we examine the dependence of stellar masses and radii on metallicity, compare our measured metallicities to the predictions from the \Gaia{} XP spectra, discuss tidal synchronization and circularization, and look ahead to \Gaia{} DR4.

\section{Target Selection and TESS Light Curves} \label{sec:target_selection}

\newcommand{\tablecaptiontext}{Summary of the metal-poor eclipsing binaries. The orbital period is from the joint light curve and radial velocity fit in Section \ref{sec:models}. The probability that a binary is a member of the thin disk is given by $p_{\rm{thin}}$ and is based on the \Gaia{} parallaxes, proper motions, and the center-of-mass velocity. \label{tab:summarytable}}
\begin{deluxetable*}{llrrrrrrrrrr}[t]
\tablecaption{\tablecaptiontext}
\tablehead{
\colhead{Source} & \colhead{Name} & \colhead{TIC} & \colhead{RA} & \colhead{DEC} & \colhead{$G$} & \colhead{Sector} & \colhead{$P$} & \colhead{$p_{\rm{thin}}$} & \colhead{$N_{\rm{APF}}$} & \colhead{$N_{\rm{PEPSI}}$} & \colhead{$N_{\rm{CHIRON}}$} \\
\colhead{} & \colhead{} & \colhead{} & \colhead{[deg]} & \colhead{[deg]} & \colhead{[mag]} & \colhead{} & \colhead{[d]} & \colhead{} & \colhead{} & \colhead{} & \colhead{}
}
\startdata
2847197557833077504 & J0001$+$2156 & 238282623 & 0.3064 & 21.9365 & 12.7 & 57 & 9.594 & 0.00 & 14 & 8 & 0 \\
362498302094814720 & J0043$+$3505 & 267806749 & 10.9807 & 35.0947 & 12.9 & 57 & 3.203 & 0.00 & 10 & 5 & 0 \\
300726119643476992 & J0157$+$2928 & 28331914 & 29.4482 & 29.4793 & 12.1 & 58 & 8.197 & 0.00 & 9 & 4 & 0 \\
3249407677302959872 & J0345$-$0407 & 333642126 & 56.3637 & -4.1238 & 11.6 & 31 & 9.660 & 0.99 & 8 & 0 & 1 \\
3277480820501314816 & J0350$+$0805 & 345412359 & 57.5155 & 8.0976 & 13.0 & 70 & 5.331 & 0.03 & 0 & 1 & 13 \\
3191963314353757952 & J0414$-$1023 & 67745301 & 63.6537 & -10.3873 & 11.7 & 32 & 6.051 & 1.00 & 0 & 2 & 11 \\
864979548396586880 & J0727$+$2018 & 60841563 & 111.7628 & 20.3075 & 12.6 & 46 & 4.216 & 1.00 & 10 & 3 & 0 \\
1098324086498919552 & J0807$+$6945 & 71520019 & 121.9004 & 69.7648 & 11.7 & 40 & 15.193 & 0.01 & 6 & 6 & 0 \\
832384629268579584 & J1047$+$4815 & 53478754 & 161.9541 & 48.2545 & 12.1 & 21+75 & 12.889 & 0.08 & 9 & 5 & 0 \\
1191216433749408128 & J1556$+$1218 & 377430449 & 239.1201 & 12.3041 & 11.2 & 51 & 6.262 & 0.00 & 11 & 7 & 0 \\
4331604848802095360 & J1626$-$1140 & 163851609 & 246.7386 & -11.6786 & 12.8 & 91 & 7.487 & 1.00 & 10 & 0 & 0 \\
2252263532714971136 & J1905$+$6233 & 230392332 & 286.3735 & 62.5543 & 11.8 & 79 & 2.504 & 1.00 & 22 & 3 & 0 \\
2100618270538909056 & J1909$+$4025 & 121273032 & 287.4587 & 40.4285 & 11.9 & 81 & 3.475 & 0.00 & 13 & 4 & 0 \\
2099410530033714560 & J1916$+$3828 & 121938117 & 289.0928 & 38.4786 & 12.3 & 80 & 7.683 & 0.00 & 11 & 2 & 0 \\
2442415483099264384 & J2352$-$0707 & 9709345 & 358.1456 & -7.1264 & 13.0 & 70 & 7.890 & 0.02 & 0 & 0 & 9 \\
\enddata
\end{deluxetable*}

\begin{figure}
    \centering
    \includegraphics[width=\linewidth]{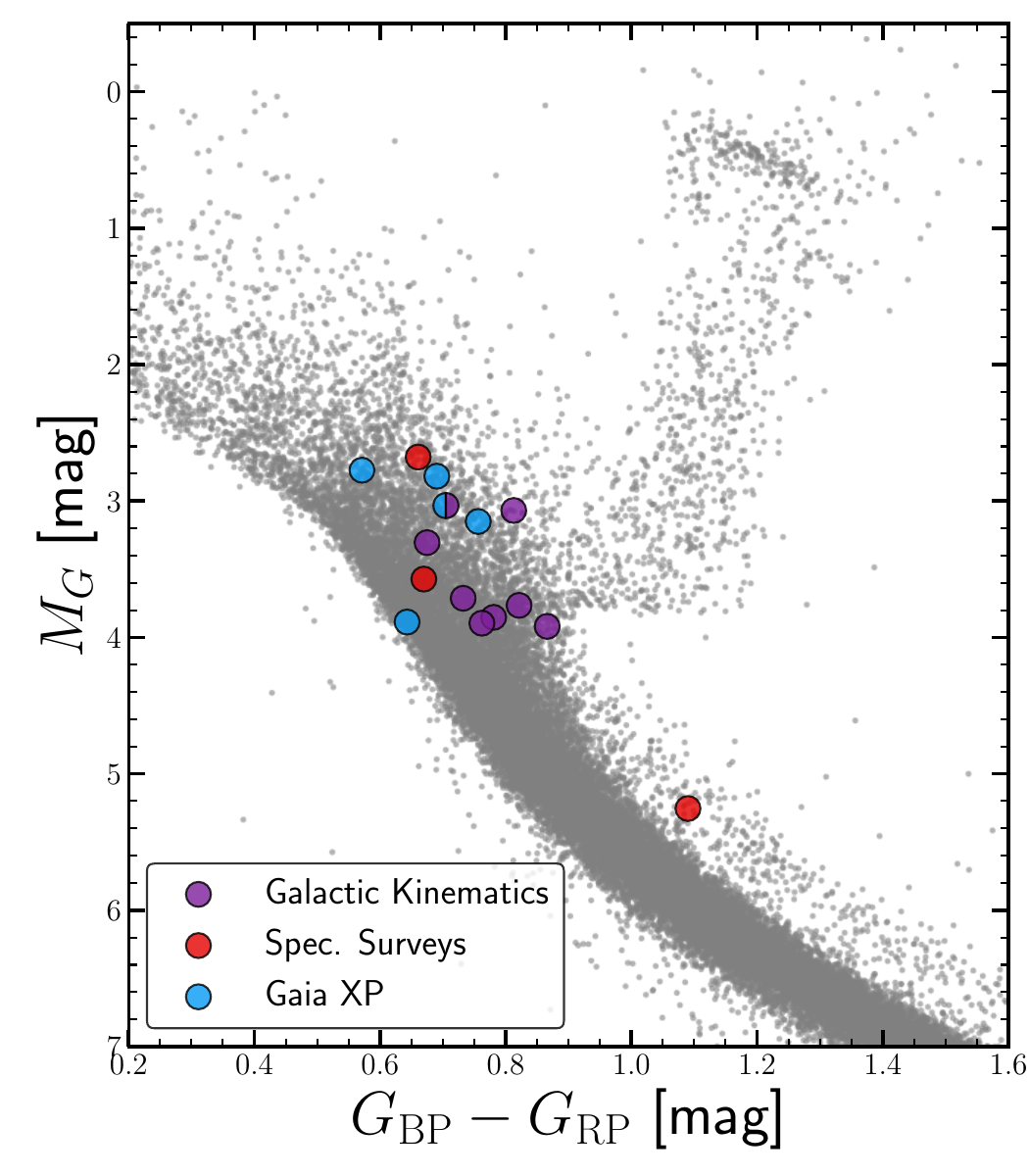}
    \caption{\Gaia{} color-magnitude diagram of the metal-poor EBs colored by their selection method. The background shows a random sample of stars in \Gaia{} DR3.}
    \label{fig:cmd}
\end{figure}

We select metal-poor eclipsing binaries by visually inspecting \TESS{} light curves of metal-poor stars. We use three methods to select metal-poor stars. First, we select stars with $[\rm{M/H}] < -1.0$ from four large spectroscopic surveys: APOGEE~DR17 \citep{Abdurrouf22}, LAMOST~DR8 \citep{Cui12}, RAVE~DR6 \citep{Steinmetz20}, and GALAH~DR3 \citep{Buder21}. Even though these abundances are derived using single-star models, the uncertainties from the presence of a binary companion are estimated to be $\sigma_{\mh} \lesssim 0.2$~dex \citep{ElBadry18}, which is sufficient to identify binaries as low-metallicity candidates. We select stars with $\log g > 3.5$ and $T_{\rm{eff}} < 6500$~K to exclude evolved and early-type stars. As detached eclipsing binaries, evolved and early-type stars would have periods too long to be characterized in a 27.4~day \TESS{} sector. Based on the limiting magnitudes of our spectrographs, we also select stars with \Gaia{} $G < 13$~mag. This results in a sample of 8{,}679 unique stars. We also select stars with metallicity estimates of $\mh{} < -1.0$ from the low resolution \Gaia{} XP spectra \citep{Andrae23} and apply the same $\log g$, $T_{\rm{eff}}$, and $G$ cuts. This results in 33{,}911 low-metallicity stars. 

We supplement our chemical selection with kinematic selection. We use the \Gaia{} DR3 parallaxes, proper motions, and mean RVs to compute the three-dimensional space velocities, $U$, $V$, and $W$. We require the \Gaia{} parallax $\varpi$ to have $\varpi / \sigma_\varpi > 20$, and $G < 13$~mag. We only select stars with $\log g > 3.5$ and $T_{\rm{eff}} < 6500$~K estimated from the \Gaia{} GSP-Phot models \citep{Andrae23_gsp} to be consistent with the spectroscopic selection. To reduce the size of the visual inspection sample, we only select stars identified as photometric variables in \Gaia{} DR3. We then use the relations from \citet{Ramirez07} to calculate thin disk, thick disk, and halo membership probabilities from the $UVW$ velocities. We select 1{,}165 stars that have thin disk membership probabilities $p_{\rm{thin}} < 0.1$. 

There is some overlap between the selection methods, and we identify 41{,}994 unique metal-poor candidates. For each source, we download the \TESS{} light curves from the Quick-Look Pipeline \citep[QLP,][]{Huang20}. For each target, we run a Lomb-Scargle periodogram \citep{Lomb76, Scargle82} on each available sector of \TESS{} data and phase fold the light curve at twice the period corresponding to the maximum power. We only consider targets where the period has false alarm probability $<1\times 10^{-5}$. We visually inspect the phase-folded light curves of these candidates to select detached eclipsing binaries for spectroscopic follow-up. 

During visual inspection, we identified $\sim 120$ EB candidates. We then preferentially selected EBs for spectroscopic follow-up without significant variability from star spots or pulsations that can introduce additional challenges in the light curve modeling. While these features can be modeled, for example with Gaussian processes \citep[e.g.,][]{Wang22}, we identified a large enough sample of targets without these features to avoid introducing additional complexity into the model. We also avoid EBs with very large differences in eclipse depths \citep[e.g., Figure A5 of][]{Rowan22} which are less likely to be detected as double-lined spectroscopic binaries. In total, we selected a final sample of 15 detached metal-poor eclipsing binary candidates for spectroscopic follow-up. Table \ref{tab:summarytable} reports the properties of these targets.

We cross-match these 15 targets with the International Variable Stars Index \citep[VSX,][]{Watson06}. We find that 11 systems are known EBs, discovered in the Northern Sky Variability Survey \citep[NSVS,][]{Wozniak04}, the Czech Variable Star Catalog \citep[CzeV,][]{Skarka17}, the All-Sky Automated Survey for Supernovae \citep[ASAS-SN,][]{Shappee14, Kochanek17, Jayasinghe19}, and \TESS{} \citep{Prsa22}. We indicate which targets are known EBs in Section \ref{sec:individual_targets}.

Figure \ref{fig:cmd} shows the targets on a \Gaia{} color-magnitude diagram. We use distances from \citet{BailerJones21} and correct for extinction using the three-dimensional ``Combined19'' dust map in {\tt mwdust} \citep{Bovy16, Drimmel03, Marshall06, Green15}. The binaries are above the main sequence, as expected, and have colors consistent with FGK stars. Eight of the targets are selected based on kinematics, five are selected with \Gaia{} XP spectra, and three with the large spectroscopic surveys. One target, J0001$+$2156, was selected based on both Galactic kinematics and the metallicity inferred from the \Gaia{} XP spectra.

\renewcommand{\tablecaptiontext}{Properties of synthetic spectral templates used to derive RVs. \label{tab:templates}}
\begin{deluxetable*}{llrrrrrrrr}[t]
\tablecaption{\tablecaptiontext}
\tablehead{
\colhead{Source} & \colhead{Name} & \colhead{$T_{\rm eff,1}$} & \colhead{$T_{\rm eff,2}$} & \colhead{$\log g_1$} & \colhead{$\log g_2$} & \colhead{$v\sin i_1$} & \colhead{$v\sin i_2$} & \colhead{$[\rm M/H]$} & \colhead{$[\alpha/\rm M]$} \\
\colhead{} & \colhead{} & \colhead{[K]} & \colhead{[K]} & \colhead{} & \colhead{} & \colhead{[km\,s$^{-1}$]} & \colhead{[km\,s$^{-1}$]} & \colhead{} & \colhead{}
}
\startdata
2847197557833077504 & J0001$+$2156 & 5800 & 5200 & 3.6 & 3.6 & 5.0 & 5.0 & $-2.2$ & 0.4 \\
362498302094814720 & J0043$+$3505 & 5500 & 5500 & 3.9 & 4.1 & 29.4 & 23.0 & $-0.9$ & 0.1 \\
300726119643476992 & J0157$+$2928 & 5600 & 5500 & 4.1 & 4.1 & 2.5 & 2.7 & $-1.1$ & 0.4 \\
3249407677302959872 & J0345$-$0407 & 4600 & 4700 & 3.5 & 3.6 & 5.3 & 5.6 & $-0.6$ & 0.2 \\
3277480820501314816 & J0350$+$0805 & 5700 & 5300 & 4.3 & 4.1 & 11.6 & 8.1 & $-0.7$ & 0.3 \\
3191963314353757952 & J0414$-$1023 & 6100 & 6100 & 3.9 & 4.1 & 13.7 & 13.4 & $-0.6$ & 0.1 \\
864979548396586880 & J0727$+$2018 & 6200 & 5400 & 4.3 & 4.0 & 14.9 & 12.0 & $-0.8$ & 0.2 \\
1098324086498919552 & J0807$+$6945 & 5700 & 5600 & 4.4 & 4.3 & 3.0 & 3.1 & $-0.7$ & 0.4 \\
832384629268579584 & J1047$+$4815 & 5700 & 5500 & 4.4 & 4.4 & 10.6 & 6.0 & $-0.6$ & 0.4 \\
1191216433749408128 & J1556$+$1218 & 6200 & 5700 & 4.1 & 4.2 & 9.1 & 6.1 & $-1.8$ & 0.4 \\
4331604848802095360 & J1626$-$1140 & 5800 & 5700 & 3.7 & 3.5 & 11.9 & 11.4 & $-0.7$ & 0.1 \\
2252263532714971136 & J1905$+$6233 & 6400 & 6200 & 4.1 & 3.9 & 30.0 & 25.9 & $-1.0$ & 0.2 \\
2100618270538909056 & J1909$+$4025 & 5800 & 5800 & 4.0 & 4.2 & 18.6 & 18.1 & $-1.1$ & 0.1 \\
2099410530033714560 & J1916$+$3828 & 6500 & 6000 & 4.1 & 3.6 & 5.0 & 5.0 & $-1.6$ & 0.4 \\
2442415483099264384 & J2352$-$0707 & 6100 & 5900 & 4.1 & 3.5 & 3.0 & 3.0 & $-0.6$ & 0.2 \\
\enddata
\end{deluxetable*}

For each selected target, we then choose one \TESS{} sector to be used for light curve modeling. Even if multiple \TESS{} sectors are available, we only use one sector to limit systematics from variations in the amount of blended light. The exception is J1047$+$4815 (\Gaia{} DR3 832384629268579584), where we use two sectors. We discuss this target in more detail in Section \ref{sec:individual_targets}. Since our binaries are short period ($P\sim 2$--15~days), multiple orbits are still covered in a single \TESS{} sector, and the joint fit with the radial velocities still enables precise orbital period measurements. 

During the \TESS{} prime mission, full-frame images (FFIs) were taken with a 30~min cadence. This cadence was shortened in the first extended mission to 10~min, and was later shortened to 200s. We generally aim to select sectors with higher FFI cadence but avoid sectors where eclipses are cut off by sector and spacecraft orbit gaps or where the binary was near the edge of the detector. Table \ref{tab:summarytable} reports the \TESS{} sector used for each target. 

Finally, to remove long-term signals from \TESS{} systematics, stellar activity, or variations in blended light from nearby stars, we use {\tt wotan} \citep{Hippke19} to detrend the light curves. We mask eclipses by first fitting a two-Gaussian model to the phase-folded light curve using \PHOEBE{} \citep{Prsa16, Conroy20}, and then use the {\tt wotan} cosine detrending model. While it is possible for detrending to remove astrophysical signals from ellipsoidal variability or reflection, we expect the amplitude of these effects to be small for the detached main sequence FGK binaries selected here. We discuss this in more detail for the individual targets in Section \ref{sec:individual_targets}.

\section{Spectroscopic Observations} \label{sec:spec_obs}

\renewcommand{\tablecaptiontext}{Radial velocity measurements for the observed eclipsing binaries derived with TODCOR. The full table is available as part of the supplemental material.\label{tab:rvs}}
\begin{deluxetable*}{lrrrrrl}[t]
\tablecaption{\tablecaptiontext}
\tablehead{
\colhead{Name} & \colhead{JD} & \colhead{$RV_1$} & \colhead{$\sigma_{RV_1}$} & \colhead{$RV_2$} & \colhead{$\sigma_{RV_2}$} & \colhead{Spectrograph} \\
\colhead{} & \colhead{$-2460000\ [\rm{d}]$} & \colhead{[km\,s$^{-1}$]} & \colhead{[km\,s$^{-1}$]} & \colhead{[km\,s$^{-1}$]} & \colhead{[km\,s$^{-1}$]} & \colhead{}
}
\startdata
J1556$+$1218 & 354.02977 & $48.04$ & $0.20$ & $-82.47$ & $0.20$ & PEPSI \\
J1556$+$1218 & 381.99198 & $-74.51$ & $0.20$ & $50.96$ & $0.20$ & PEPSI \\
J1556$+$1218 & 396.99589 & $20.65$ & $0.20$ & $-53.55$ & $0.20$ & APF \\
J1556$+$1218 & 402.93478 & $1.74$ & $0.20$ & $-33.17$ & $0.20$ & APF \\
J1556$+$1218 & 409.89996 & $36.77$ & $0.20$ & $-71.69$ & $0.20$ & APF \\
J1556$+$1218 & 410.91481 & $41.80$ & $0.20$ & $-77.38$ & $0.20$ & APF \\
J1556$+$1218 & 417.92186 & $10.19$ & $0.20$ & $-43.32$ & $0.20$ & APF \\
J1556$+$1218 & 418.89324 & $-47.97$ & $0.20$ & $22.08$ & $0.20$ & PEPSI \\
J1556$+$1218 & 418.9256 & $-50.32$ & $0.20$ & $23.78$ & $0.20$ & APF \\
J1556$+$1218 & 419.95372 & $-76.91$ & $0.20$ & $52.93$ & $0.20$ & APF \\
J1556$+$1218 & 446.73499 & $-2.14$ & $0.20$ & $-27.88$ & $0.20$ & PEPSI \\
J1556$+$1218 & 449.96873 & $-32.80$ & $0.20$ & $5.57$ & $0.20$ & PEPSI \\
J1556$+$1218 & 450.84138 & $-72.29$ & $0.20$ & $48.73$ & $0.20$ & PEPSI \\
J1556$+$1218 & 1056.02125 & $47.73$ & $0.20$ & $-82.31$ & $0.20$ & PEPSI \\
J1556$+$1218 & 1057.06989 & $21.44$ & $0.20$ & $-55.24$ & $0.20$ & APF \\
J1556$+$1218 & 1059.06543 & $-76.96$ & $0.20$ & $51.60$ & $0.20$ & APF \\
J1556$+$1218 & 1062.0707 & $43.13$ & $0.20$ & $-78.88$ & $0.20$ & APF \\
J1556$+$1218 & 1074.03926 & $23.85$ & $0.20$ & $-58.40$ & $0.20$ & APF \\
\enddata
\end{deluxetable*}

To measure the radial velocities and spectroscopic parameters of our target stars, we collected spectra using three different spectrographs. We obtained at least nine epochs for each target. Here we briefly describe the three instruments. 

We obtained high-resolution ($R\approx 43{,}000$) spectra for 12 targets with the Potsdam Echelle Polarimetric and Spectroscopic Instrument \citep[PEPSI,][]{Strassmeier15} on the Large Binocular Telescope. We used the 300$\mu$m fiber and two cross-dispersers (CDs) covering 4758--5416~\AA{} (CD3) and 6244--7427~\AA{} (CD5). We only use the CD3 data for RV determination because there is little contamination from telluric features. The typical exposure times were 300--900s to reach a signal-to-noise ratio of $S/N \approx 100$ at 500nm. The 2D echelle spectra are reduced following the procedure outlined in \citet{Strassmeier18}. 

We obtained high-resolution ($R\approx 80{,}000$) spectra for 12 targets with the Automated Planet Finder (APF) Levy spectrograph on the Lick Observatory 2.4m \citep{Vogt14}. The observations used the $2\arcsec\times3\arcsec$ Decker-T slit, covered a wavelength range of 3730--10206~\AA{}, and did not use the iodine cell. The raw 2D echelle spectra are reduced to 1D spectra through the California Planet Survey \citep[CPS,][]{Howard10} pipeline. The 1D spectra are then de-spiked to remove signals from cosmic rays and are blaze-corrected by fitting polynomials to the continuum in each order. The APF observations had a typical exposure time of 1700s. We exclude orders affected by telluric lines when measuring the RVs. 

Finally, we obtained high-resolution spectra ($R\approx 28{,}000$) for four targets using CHIRON on the SMARTS 1.5m telescope \citep{Tokovinin13, Schwab12}. The spectra were taken in the fiber mode using $4\times4$~pixel binning and a Th-Ar comparison lamp. The CHIRON observations used a typical exposure time of 1500s. As with the APF observations, we extract 1D spectra and correct for the blaze function by fitting polynomials to the continuum in each order. We use 36 orders spanning 4700--7792~\AA{} for the RV analysis, avoiding regions containing telluric lines. 

\begin{figure}
    \centering
    \includegraphics[width=\linewidth]{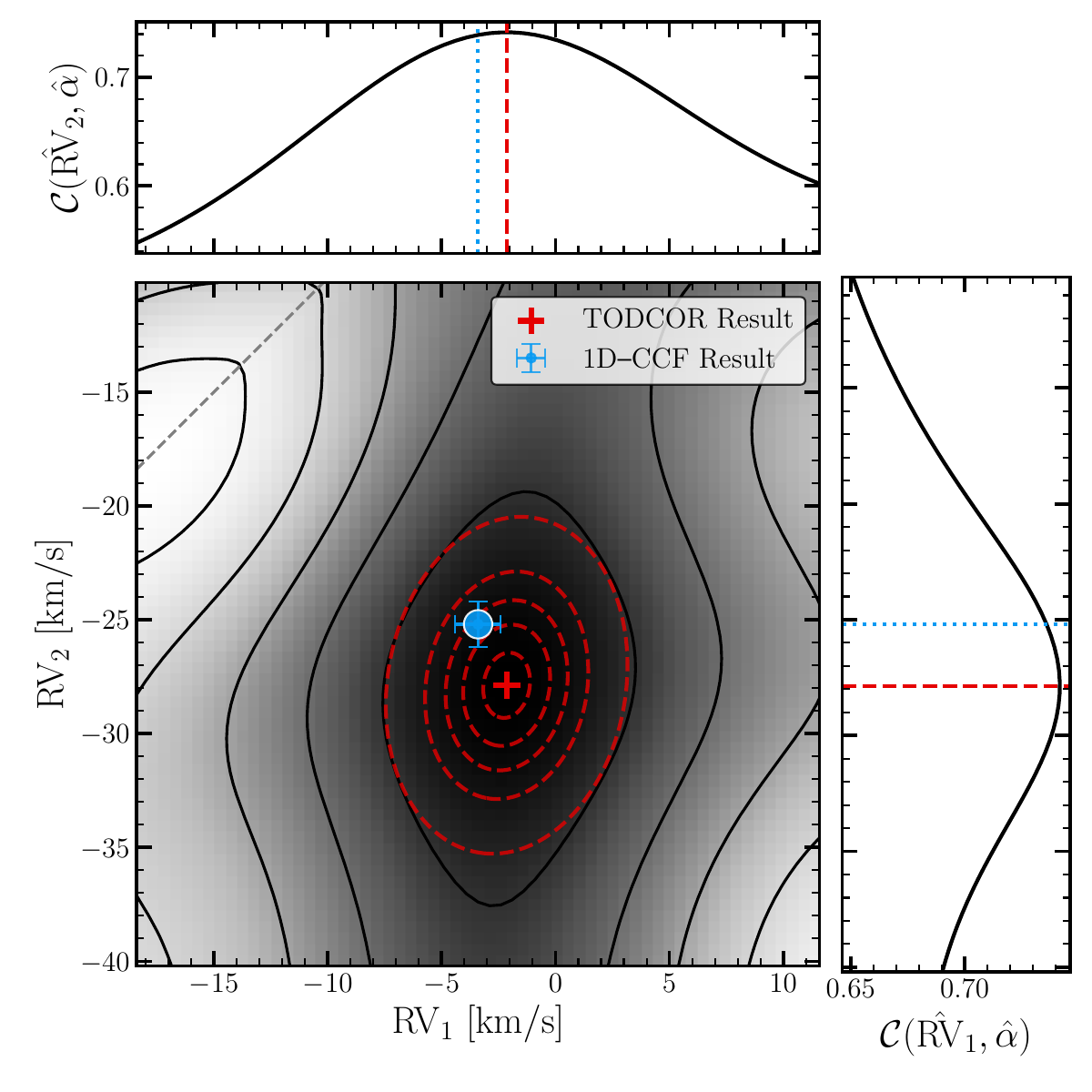}
    \caption{Center: TODCOR profile for a PEPSI spectrum of J1556$+$1218. We model the surface with a 2D Gaussian to measure the velocities of both components (red contours). The top and right panels show slices through this maximum, which are used to measure the uncertainty in the radial velocity (Equation \ref{eqn:rv_err}). For comparison, the blue point and blue lines show the measured velocities from the 1D-CCF, which uses a single template corresponding to the primary star.}
    \label{fig:todcor_example}
\end{figure}

We measure radial velocities using a two-dimensional cross-correlation function \citep[TODCOR,][]{Zucker94}. The standard one-dimensional cross-correlation function (1D-CCF) uses a single template typically chosen to match the spectral type of the photometric primary. While double-lined spectroscopic binaries (SB2s) can produce two peaks in the 1D-CCF, modeling the CCF profile of SB2s with low flux ratios or small velocity differences between the components can bias the measured RVs (e.g., Figure \ref{fig:todcor_example}). 

TODCOR requires two template spectra for the cross-correlation. For the best results, the template spectra should match the spectral types of the binary. We select templates by fitting a two-star spectrum model with \Korg{} \citep{Wheeler23} to the spectrum taken closest to RV quadrature. We minimize the $\chi^2$ and solve for the effective temperatures, $T_{\rm{eff}}$, surface gravities, $\log g$, and projected rotational velocities $v \sin i$ of both components, and a shared metallicity and $\alpha$-abundance. We set the microturbulent velocities using an analytic relation implemented in \iSpec{} \citep{BlancoCuaresma14, Jofre14}. The properties of the templates used with TODCOR are reported in Table \ref{tab:templates}. We derive more robust spectroscopic parameters using spectral disentangling in Section \ref{sec:disentangling}.

For one target, J0001$+$2156, we use a modified template selection strategy. Since the flux ratio of this system is small ($F_2 / F_1 \lesssim 10\%$), we find that we cannot reliably fit a two-star model to the spectrum taken near quadrature. Instead, we fit a single-star model to construct a template of the photometric primary. We use a synthetic template for the secondary with the same surface gravity and projected rotational velocity. We set the effective temperature of the secondary template for J0001$+$2156 based on the effective temperature ratio $T_{\rm{eff},2}/T_{\rm{eff},1}$ reported in \citet{Rowan22}. 

We use the synthetic spectral templates to derive RVs with TODCOR. The templates are broadened with a Gaussian kernel to match the resolution of the instruments. We calculate the two-dimensional cross-correlation function, $\mathcal{R}$, over a velocity grid with a step size of 0.5~km/s. For the APF and CHIRON spectra, we apply TODCOR to each echelle order independently and combine the TODCOR profiles of the individual orders following \citet{Zucker03}. We then fit a two-dimensional Gaussian to the peak of the TODCOR profile to measure the RVs of both components. 

We take slices through the maximum of the TODCOR profile and measure the RV uncertainties as 
\begin{equation} \label{eqn:rv_err}
    \sigma_{\rm{RV}}^2 = - \left( N \frac{C^{\prime\prime}(\hat{s})}{C(\hat{s})}\frac{C^2(\hat{s})}{1-C^2(\hat{s})}\right)^{-1},
\end{equation}
where $C$ is the slice through $\mathcal{R}$, $\hat{s}$ are the velocities where $\mathcal{R}$ is maximized, and $N$ is the number of points in the spectra \citep{Zucker03}. For the APF and CHIRON spectra where $\mathcal{R}$ is combined from all the echelle orders, Equation \ref{eqn:rv_err} is modified to replace the factor $N$ with $NM$, where $M$ is the number of orders. Figure \ref{fig:todcor_example} shows an example of the TODCOR profile for one of the J1556$+$1218 PEPSI spectra.

We measure RVs with typical uncertainties of $\sigma_{\rm{RV}}\sim 100$~m/s. To account for systematic uncertainties in template selection \citep[see Figure 7 of][]{Rowan25}, we set an uncertainty floor of $\sigma_{\rm{RV}} = 200$~m/s. Table \ref{tab:rvs} reports the measured radial velocities for each target. 

\section{Light Curve and RV Orbit Models} \label{sec:models}

\begin{figure*}
    \centering
    \includegraphics[trim={2cm 0 0 0},clip,width=\linewidth]{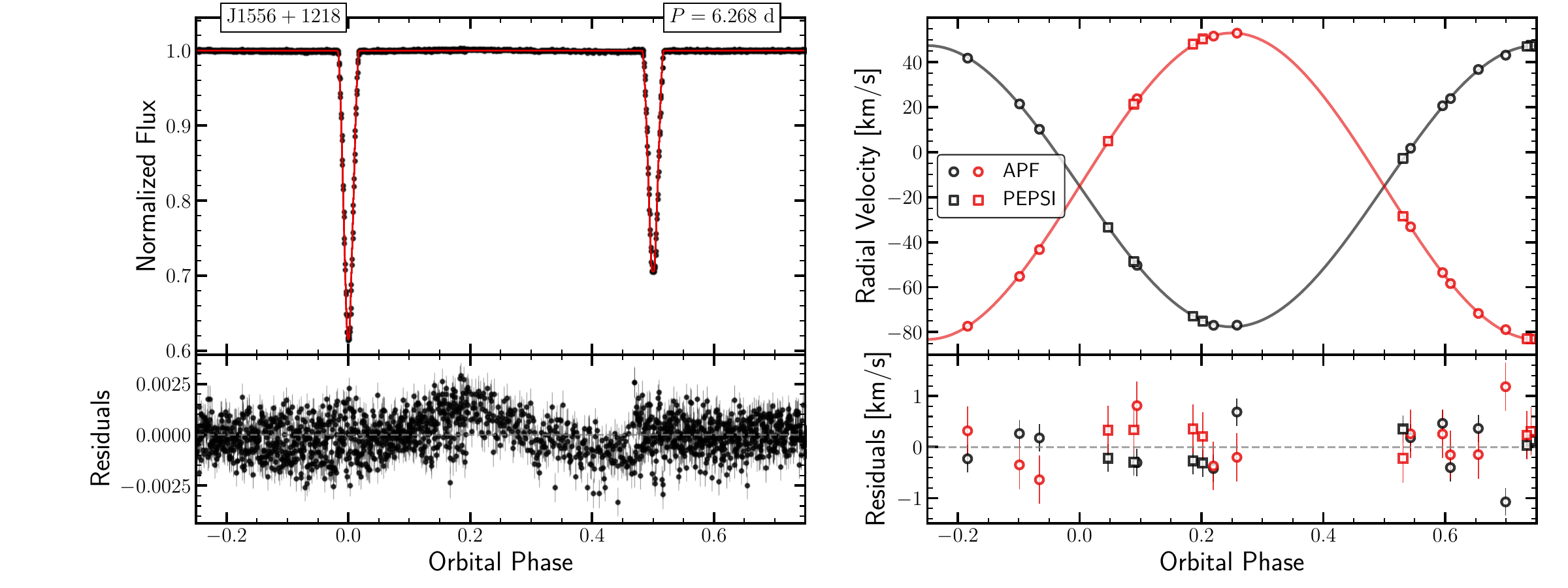}
    \includegraphics[trim={1cm 0 0 0},clip,width=0.8\linewidth]{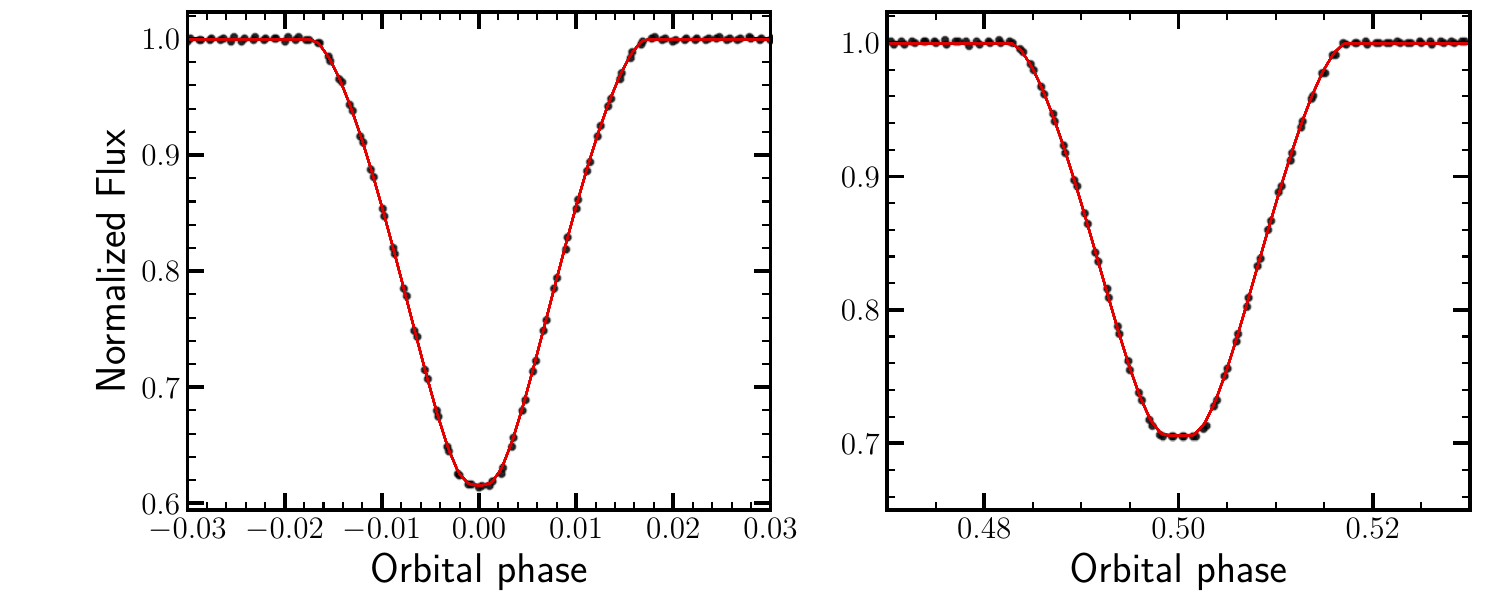}
    \caption{Light curve (top left) and RV orbit (top right) for J1556$+$1218. The red and black lines show random samples from the MCMC posteriors. The lower panels show the light curve fit zoomed in on the primary (left) and secondary (right) eclipses.}
    \label{fig:lc_rv}
\end{figure*}

\setlength{\tabcolsep}{8pt}
\renewcommand{\arraystretch}{1.3}
\renewcommand{\tablecaptiontext}{Median and 1$\sigma$ values from MCMC posteriors of joint light curve and RV orbit fit with \ellc{} and \PHOEBE{}. A machine-readable table is available as part of the supplemental material. The table is continued in Table \ref{tab:mcmc_posteriors_2}. \label{tab:mcmc_posteriors}} 
\begin{deluxetable*}{lcccccc}
\tablecaption{\tablecaptiontext}
\tablehead{\colhead{Parameter} & \colhead{J0001$+$2156} & \colhead{J0043$+$3505} & \colhead{J0157$+$2928} & \colhead{J0345$-$0407} & \colhead{J0350$+$0805} & \colhead{J0414$-$1023}}
\startdata
$M_1$\,[$M_\odot$] & $0.821(10)$ & $0.9381(95)$ & $0.8190(33)$ & $0.819(16)$ & $0.9334(42)$ & $1.117(11)$ \\
$M_2$\,[$M_\odot$] & $0.6583(55)$ & $0.9262(83)$ & $0.8194(27)$ & $0.7831(89)$ & $0.8528(32)$ & $1.114(11)$ \\
$R_1$\,[$R_\odot$] & $2.0311(81)$ & $1.575(13)$ & $0.9103(74)$ & $\cdots$ & $1.2118(20)$ & $1.6060(64)$ \\
$R_2$\,[$R_\odot$] & $0.6243(26)$ & $1.527(13)$ & $0.9103(74)$ & $\cdots$ & $0.9179(23)$ & $1.6060(64)$ \\
$P$\,[d] & $9.58581(15)$ & $3.2027554(73)$ & $8.199489(16)$ & $9.653224(14)$ & $5.3295658(79)$ & $6.0503672(95)$ \\
$\phi_0$ & $0.70114(94)$ & $0.754934(97)$ & $0.128710(26)$ & $0.60911(13)$ & $0.833378(62)$ & $0.88992(21)$ \\
$e$ & $0.0005(11)$ & $0.00060(92)$ & $0.00011(27)$ & $0.0149(17)$ & $0.00176(68)$ & $0.0027(19)$ \\
$\omega$\,[deg] & $-90(100)$ & $-89(71)$ & $75(95)$ & $-107.4(20)$ & $-83.1(32)$ & $-96.3(62)$ \\
$i$\,[deg] & $89.85(14)$ & $84.494(13)$ & $87.503(98)$ & $87.5140(72)$ & $89.87(10)$ & $83.856(33)$ \\
$T_{\rm eff,2}/T_{\rm{eff,1}}$ & $0.9633(15)$ & $1.02843(66)$ & $1.0019(10)$ & $0.9685(35)$ & $0.9502(15)$ & $0.9903(27)$ \\
$\gamma$\,[km\,s$^{-1}$] & $-228.03(14)$ & $-157.93(25)$ & $-134.616(66)$ & $-6.063(86)$ & $89.543(71)$ & $4.50(17)$ \\
$\delta_{\rm{RV}_0}$\,[km\,s$^{-1}$] & $0.63(22)$ & $0.99(42)$ & $0.70(12)$ & \nodata & \nodata & $-0.22(42)$ \\
$\ell_3$ & \nodata & \nodata & $0.087(33)$ & \nodata & $0.0810(29)$ & \nodata \\
\hline
\colhead{Parameter} & \colhead{J0727$+$2018} & \colhead{J0807$+$6945} & \colhead{J1047$+$4815} & \colhead{J1556$+$1218} & \colhead{J1626$-$1140} & \colhead{J1905$+$6233} \\
\hline
$M_1$\,[$M_\odot$] & $0.9759(75)$ & $0.9036(72)$ & $0.9610(46)$ & $0.7593(39)$ & $0.9602(88)$ & $0.9992(67)$ \\
$M_2$\,[$M_\odot$] & $0.8675(51)$ & $0.9058(75)$ & $0.8897(53)$ & $0.6956(34)$ & $0.976(12)$ & $0.9560(64)$ \\
$R_1$\,[$R_\odot$] & $\cdots$ & $1.0530(47)$ & $1.2761(23)$ & $1.0325(17)$ & $1.332(29)$ & $1.3870(81)$ \\
$R_2$\,[$R_\odot$] & $\cdots$ & $1.0439(46)$ & $0.9705(18)$ & $0.7380(13)$ & $1.427(27)$ & $1.3324(81)$ \\
$P$\,[d] & $4.2161435(71)$ & $15.196523(57)$ & $12.88945117(78)$ & $6.268171(10)$ & $7.486869(53)$ & $2.5036783(62)$ \\
$\phi_0$ & $0.90341(18)$ & $0.61627(15)$ & $0.4893067(34)$ & $0.739618(76)$ & $0.82626(75)$ & $0.98332(46)$ \\
$e$ & $0.0080(28)$ & $0.00007(24)$ & $0.42806(19)$ & $0.00004(13)$ & $0.02168(14)$ & $0.00042(54)$ \\
$\omega$\,[deg] & $90.5(10)$ & $90(100)$ & $-154.145(65)$ & $-63(90)$ & $173.4(51)$ & $-19(76)$ \\
$i$\,[deg] & $83.95(11)$ & $89.6452(41)$ & $89.721(13)$ & $89.0595(49)$ & $88.457(55)$ & $78.738(13)$ \\
$T_{\rm eff,2}/T_{\rm{eff,1}}$ & $0.9526(69)$ & $0.99990(35)$ & $0.96926(55)$ & $0.93856(42)$ & $0.99653(97)$ & $0.9966(18)$ \\
$\gamma$\,[km\,s$^{-1}$] & $22.74(18)$ & $-23.05(14)$ & $34.871(83)$ & $-15.145(98)$ & $-53.16(15)$ & $29.26(15)$ \\
$\delta_{\rm{RV}_0}$\,[km\,s$^{-1}$] & $1.37(27)$ & $0.70(21)$ & $0.73(14)$ & $0.68(16)$ & \nodata & $0.75(43)$ \\
$\ell_3$ & $0.126(27)$ & \nodata & \nodata & \nodata & $0.0355(81)$ & \nodata \\
\enddata
\end{deluxetable*}

\renewcommand{\tablecaptiontext}{Same as Table \ref{tab:mcmc_posteriors} for the remaining targets.\label{tab:mcmc_posteriors_2}}
\begin{deluxetable*}{lccc}[t]
\tablecaption{\tablecaptiontext}
\tablehead{
\colhead{Parameter} & \colhead{J1909$+$4025} & \colhead{J1916$+$3828} & \colhead{J2352$-$0707}
}
\startdata
$M_1$\,[$M_\odot$] & $0.8951(56)$ & $0.765(11)$ & $0.8115(54)$ \\
$M_2$\,[$M_\odot$] & $0.8941(49)$ & $0.724(11)$ & $0.8135(66)$ \\
$R_1$\,[$R_\odot$] & $1.194(12)$ & $1.760(11)$ & $1.044(37)$ \\
$R_2$\,[$R_\odot$] & $1.196(12)$ & $1.0978(76)$ & $1.008(38)$ \\
$P$\,[d] & $3.4755451(61)$ & $7.682579(58)$ & $7.890705(28)$ \\
$\phi_0$ & $0.65788(26)$ & $0.18154(49)$ & $0.25740(10)$ \\
$e$ & $0.00095(60)$ & $0.0161(10)$ & $0.00015(44)$ \\
$\omega$\,[deg] & $82(13)$ & $-92.23(17)$ & $23(89)$ \\
$i$\,[deg] & $86.677(23)$ & $87.40(10)$ & $87.793(30)$ \\
$T_{\rm eff,2}/T_{\rm{eff,1}}$ & $0.99928(16)$ & $1.0739(14)$ & $1.0014(11)$ \\
$\gamma$\,[km\,s$^{-1}$] & $-92.85(14)$ & $-185.14(20)$ & $-32.498(96)$ \\
$\delta_{\rm{RV}_0}$\,[km\,s$^{-1}$] & $0.90(27)$ & $0.28(48)$ & \nodata \\
$\ell_3$ & $0.0377(33)$ & $0.075(16)$ & $0.0076(85)$ \\
\enddata
\end{deluxetable*}

We use \ellc{} \citep{Maxted16} implemented within Physics Of Eclipsing BinariES \citep[\PHOEBE{},][]{Prsa05, Conroy20} to simultaneously model the \TESS{} light curves and RVs to measure masses and radii. \ellc{} models the stellar surfaces as triaxial ellipsoids, which is less robust than the full triangulated mesh of the equipotential surfaces. Since our targets are detached eclipsing binaries with little or no evidence of ellipsoidal modulations, \ellc{} offers a less computationally expensive alternative to the default \PHOEBE{} backend. 

We sample over the binary parameters with Markov Chain Monte Carlo (MCMC) using {\tt emcee} \citep{ForemanMackey13}. The free parameters include the orbital period, $P$, the phase of the superior conjunction, $\phi_0$, and the cosine of the inclination angle, $\cos i$. The orbital eccentricity, $e$, and argument of periastron, $\omega$, are parameterized as $\sqrt{e}\cos\omega$ and $\sqrt{e}\sin\omega$. We also sample the primary mass, $M_1$, the mass ratio, $q=M_2/M_1$, the sum of the radii, $R_1+R_2$, the ratio of the radii, $k=R_2/R_1$, and the effective temperatures, $T_{\rm{eff},1}$ and $T_{\rm{eff},2}$. We include both temperatures, rather than their ratio, to allow explorations of limb darkening coefficients. The center-of-mass velocity, $\gamma$, and the passband luminosity of the photometric primary, $L_{\rm{pb}}$, are included as additional free parameters. For targets with RVs from multiple instruments, we additionally fit for an RV offset to account for differences in RV zero-points, and we fit for the stellar velocity jitter of both components, $s_1$ and $s_2$, which are added in quadrature to the measured $\sigma_{\rm{RV}}$.

Nearby stars, whether bound companions or chance alignments, can contaminate the EB light curve as ``third-light''. This can bias the measured inclination and stellar parameters. Since the \TESS{} pixels are large ($21\arcsec$), we use \Gaia{} DR3 to search for nearby stars that have a magnitude difference of $\Delta G < 5.0$ within $21\arcsec$. We also look for evidence of wide tertiary companions to the binaries using the \Gaia{} {\tt ruwe}, which is the renormalized unit weight error in the \Gaia{} astrometric solution and find that J0157$+$2928 has ${\tt ruwe} = 3.1$. Finally, the \Gaia{} {\tt ipd\_frac\_multi\_peak} parameter reports the fraction of \Gaia{} transits with multiple peaks detected in the \Gaia{} 1D line-spread function. All 15 EBs have ${\tt ipd\_frac\_multi\_peak} = 0$. For targets with possible contaminating sources, we include a fractional third-light parameter, $\ell_3=F_3/(F_1+F_2+F_3)$, in the \ellc{} model.

Since the majority of the photometric measurements are taken outside of eclipse and offer little to no constraint on the system geometry, we bin the out-of-eclipse light curves to reduce the computational cost. We include all points in the eclipses and bin the out-of-eclipse data in time by a factor of 20--30, depending on the FFI cadence of the selected sector. This corresponds to out-of-eclipse time bins of 3--18~hours, depending on the orbital period.

We use model atmospheres from \citet{Castelli03} for passband luminosity normalization. The limb darkening coefficients are interpolated from the \citet{Claret22} grids for the ``power-2'' limb-darkening law \citep{Hestroffer97}. We include a Gaussian prior on the effective temperature of the primary centered on the RV template temperature with $\sigma = 200$~K, since only the ratio of temperatures is constrained with a single-band light curve. We also include a Gaussian prior on the RV offsets centered on zero with a 1.5~km/s dispersion. Finally, we place a half-normal prior with a scale $0.5$~km/s on each of the per-component RV jitter terms to prevent the jitter from inflating to allow an otherwise poor orbital fit.

\begin{figure*}
    \centering
    \includegraphics[width=\linewidth]{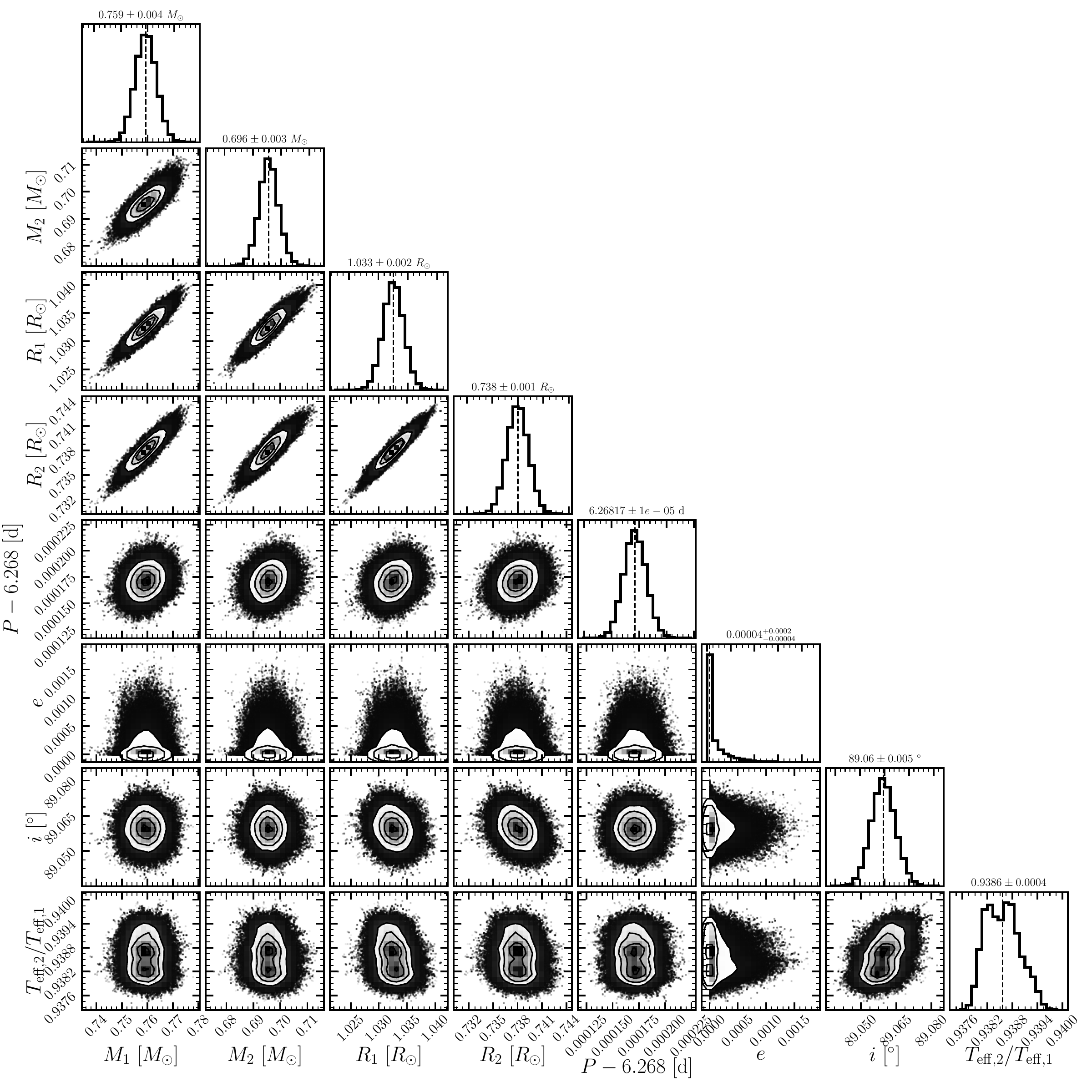}
    \caption{MCMC posteriors for J1556$+$1218.}
    \label{fig:corner_example}
\end{figure*}

We use 48 walkers and run the MCMC chains for 50{,}000 iterations. We visually inspect the walker probabilities to select a burn-in for each target, typically 10{,}000 iterations. Table \ref{tab:mcmc_posteriors} reports the median posteriors and 1$\sigma$ uncertainties. Figure \ref{fig:lc_rv} shows an example light curve and RV orbit fit and Figure \ref{fig:corner_example} shows an example of the MCMC posteriors for the same target. Appendix \ref{sec:appendix_lc_rvs} includes the light curves and orbit fits for the remaining targets. 

We measure fractional uncertainties on the masses of $< 2\%$ for all binaries, and $< 1\%$ for 10 binaries. We measure fractional uncertainties on the radii of $\leq 1\%$ for 10 systems and $< 4\%$ for the remaining systems. We discuss the properties of each target individually in Section \ref{sec:individual_targets}.

For a circular orbit with perfectly measured $P$ and $i$, the fractional uncertainty on the masses depends only on the uncertainty of the velocity semi-amplitudes, $K_1$ and $K_2$, 
\begin{equation}
\begin{split}
\left( \frac{\sigma_{M_i}}{M_i} \right)^2 = &\left[ 3 \frac{\sqrt{\sigma_{K_1}^2 + \sigma_{K_2}^2}}{K_1 + K_2} \right]^2 \\
&+ \left[ \frac{K_i}{K_1 + K_2} \right]^2 \left[ \left( \frac{\sigma_{K_1}}{K_1} \right)^2 + \left( \frac{\sigma_{K_2}}{K_2} \right)^2 \right].
\end{split}
\end{equation}
\noindent For a totally-eclipsing binary where the ingress and egress times are precisely measured, the uncertainties on the radii depend on the uncertainty on the semi-major axis. This also depends on $K_1$ and $K_2$ as
\begin{equation}
    \dfrac{\sigma_{R_1}}{R_1} = \dfrac{\sigma_{R_2}}{R_2} = \dfrac{\sqrt{\sigma_{K_1}^2+\sigma_{K_2}^2}}{K_1+K_2}
\end{equation}

For $K_1 = 62.5$~km/s and $K_2=68$~km/s and $\sigma_K = 0.1$~km/s, which is what we measure from the RVs of J1556$+$1218, this gives fractional uncertainties $\sigma_M / M = 0.34\%$ and $\sigma_R / R = 0.10\%$. For comparison, we measure $\sigma_{M_1} / M_1 = 0.52\%$ and $\sigma_{R_1} / R_1 = 0.17\%$. We are approaching the limit of what is achievable with the available RVs, and an improved constraint on $K_1$ and $K_2$ through additional RVs would lower the mass and radius uncertainties even further. 

\section{Spectral Disentangling} \label{sec:disentangling}

In order to measure the metallicities and $\alpha$-abundances of the stars, we disentangle the multi-epoch spectra and solve for the spectra of the binary components. We use the ``Shift and Add'' grid disentangling method \citep{Gonzalez06} as implemented in \citet{Shenar20, Shenar22}\footnote{\url{https://github.com/TomerShenar/Disentangling_Shift_And_Add}}. The advantage of this approach, as opposed to simply fitting an observed spectrum with a two-star model, is that we can effectively increase the signal-to-noise ratio by $\sqrt{N}$ for spectroscopic parameter estimation, where $N$ is the number of observations. This method does not rely on any underlying line list, spectral synthesis code, or model atmospheres. It iteratively solves for the two component spectra that, when shifted according to a Keplerian orbit, reproduce the observed multi-epoch spectra. We fit the output spectra as single stars with spectral synthesis models, reducing correlations between the spectroscopic parameters of both stars. 

\begin{figure*}
    \centering
    \includegraphics[trim={2cm 0 0 0},clip,width=\linewidth]{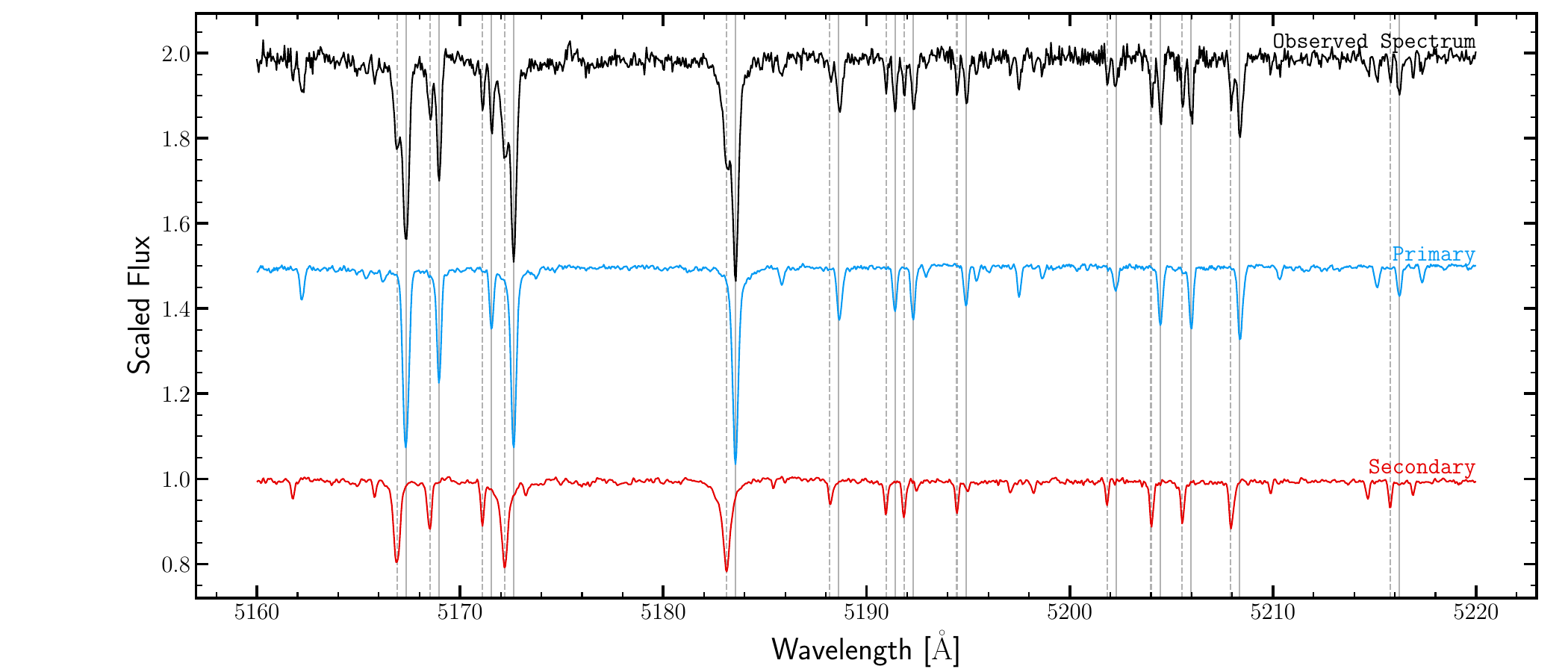}
    \caption{Output of spectral disentangling for J1556$+$1218. The black line shows one of the PEPSI spectra. The blue and red lines show the disentangled component spectra, which are based on the full set of PEPSI spectra. The component spectra have been shifted to match the velocities of the stars at the epoch of the example PEPSI spectrum. The vertical gray lines show a set of absorption features with solid lines corresponding to the primary and dashed lines corresponding to the secondary. We label these and other absorption lines in Figure \ref{fig:disentangle_fit}.}
    \label{fig:disentangling_example}
\end{figure*}

Since we have already solved for the binary orbit, we fix the orbital parameters and solve for the component spectra with 10{,}000 iterations of grid disentangling. We use a wavelength range of $\lambda\lambda$5100--5300~\AA, which covers the \ion{Mg}{1}~b triplet and a number of Fe lines. For targets that were observed with multiple spectrographs, we analyze the data from each instrument separately to reduce systematics from differences in the wavelength coverage and resolution. Figure \ref{fig:disentangling_example} shows the disentangled component spectra of J1556$+$1218 and one of the representative PEPSI spectra. 

Next we model the disentangled component spectra as single stars with \Korg{} \citep{Wheeler23} and the line list from \citet{Heiter21} to measure the stellar atmospheric parameters. The two components are coupled through a shared metallicity and $\alpha$-abundance, since we expect the two stars to be born from gas with the same metallicity. The disentangled component spectra are also coupled through a flux ratio parameter. The component spectra derived from spectral disentangling can be arbitrarily scaled to match any flux ratio $\ell_2 = F_2 / (F_1 + F_2)$, since we do not know \textit{a priori} if a star has intrinsically weak lines or if it is being diluted by the flux of its companion.

Rather than fitting $\ell_2$ directly, we compute it at every optimization step from the LC+RV solution and model atmospheres. From the joint light curve and RV orbit model, we measure the ratio of the stellar radii and the ratio of the effective temperatures. We compute the luminosity ratio and generate models of the spectral energy distribution using BaSeL model atmospheres \citep{Lejeune98} implemented with {\tt pystellibs}\footnote{\url{https://github.com/mfouesneau/pystellibs}} and extinctions from {\tt mwdust} \citep{Bovy16} to compute the wavelength-specific flux ratio.

\begin{figure*}
    \centering
    \includegraphics[trim={2cm 0 0 0},clip,width=\linewidth]{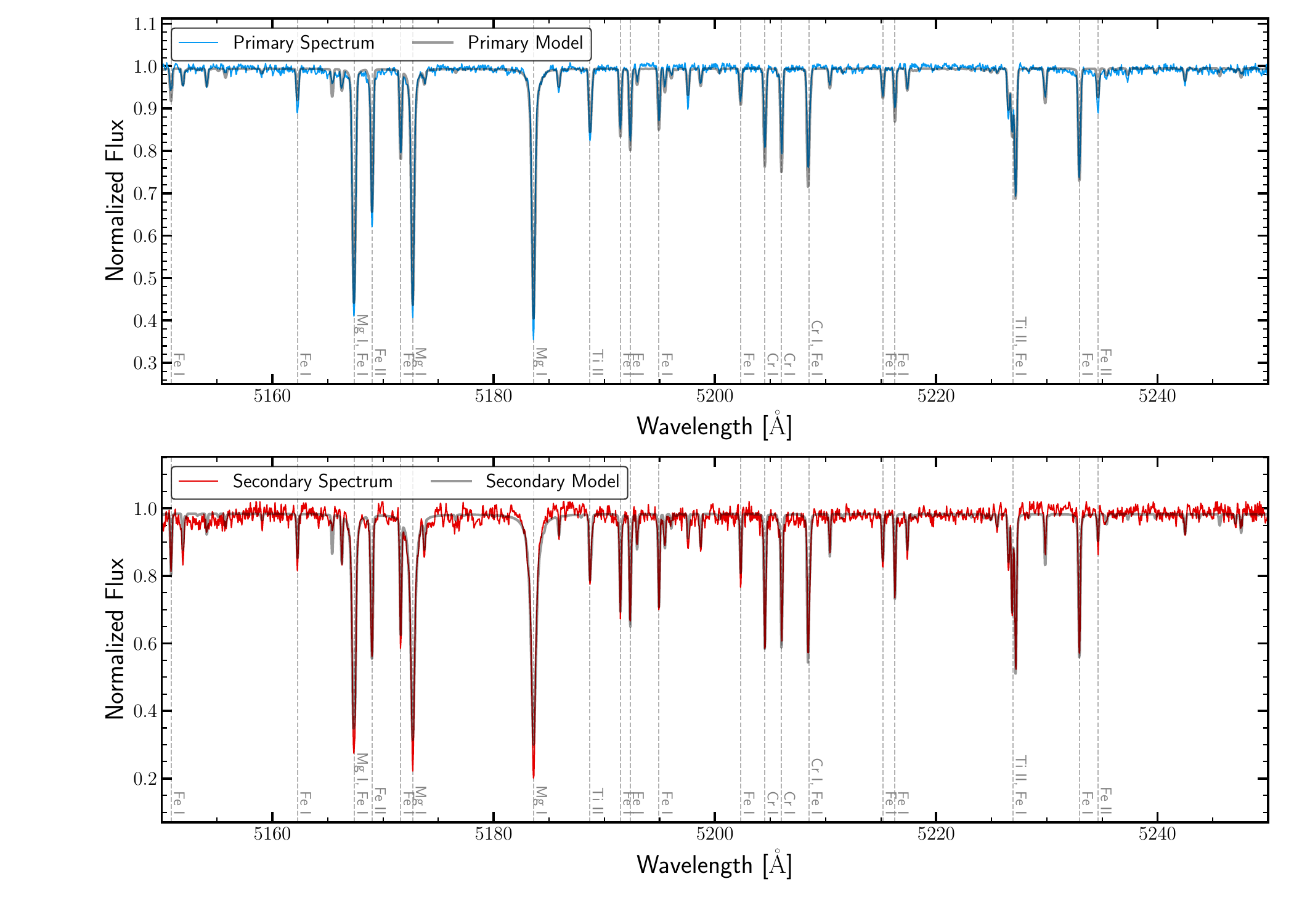}
    \caption{Fits to the disentangled spectra of J1556$+$1218 using \Korg{}. The primary and secondary spectra (blue and red) correspond to the components extracted from disentangling shown in Figure \ref{fig:disentangling_example}. Key absorption lines are marked with vertical lines.}
    \label{fig:disentangle_fit}
\end{figure*}

Similarly, since the ratio $T_{\rm{eff},2} / T_{\rm{eff},1}$ is constrained from the light curve, we only optimize for $T_{\rm{eff},1}$. The surface gravities of both components are fixed using the measured mass and radius. We use empirical relations from \citet{Jofre14} to determine the microturbulent velocities based on the effective temperature, surface gravity, and metallicity. The free parameters in the optimization are therefore the effective temperature of the primary, $T_{\rm{eff},1}$, the projected rotational velocities of both stars, $v \sin i_{1,2}$, the metallicity, and the $\alpha$-abundance. We minimize the sum of the $\chi^2$ fit statistic of the components with the limited-memory bounded Broyden--Fletcher--Goldfarb--Shanno (L-BFGS-B) algorithm. Figure \ref{fig:disentangle_fit} shows an example of the fit to the disentangled component spectra of J1556$+$1218. Table \ref{tab:spec_table} reports the derived spectroscopic parameters for all targets.

The component spectra that we fit with {\tt Korg} are not the observed spectra, but rather outputs from the iterative spectral disentangling procedure, which depends on the light curve and RV orbit model. The systematic uncertainties on the fluxes in the component spectra are therefore poorly calibrated. We generate representative synthetic binary systems with similar spectral signal-to-noise and orbit model uncertainties and attempt to recover the injected stellar parameters. We use these tests to set conservative uncertainties of $\sigma_{T_{\rm{eff}}} = 200$~K, $\sigma_{v \sin i} = 3$~km/s, and $\sigma_{[\rm{X/H}]}=0.1$~dex. For targets with sufficient spectra from both APF and PEPSI (no system with CHIRON spectra had enough APF or PEPSI observations), we perform the disentangling and component fitting separately with each spectrograph, yielding two independent solutions that serve as a consistency check on the derived spectroscopic parameters. We find that the solutions agree at or below these adopted conservative uncertainties. We discuss the results for each target in Section \ref{sec:individual_targets}.

\begin{figure*}
    \centering
    \begin{tabular}{cc}
        \includegraphics[width=0.48\linewidth]{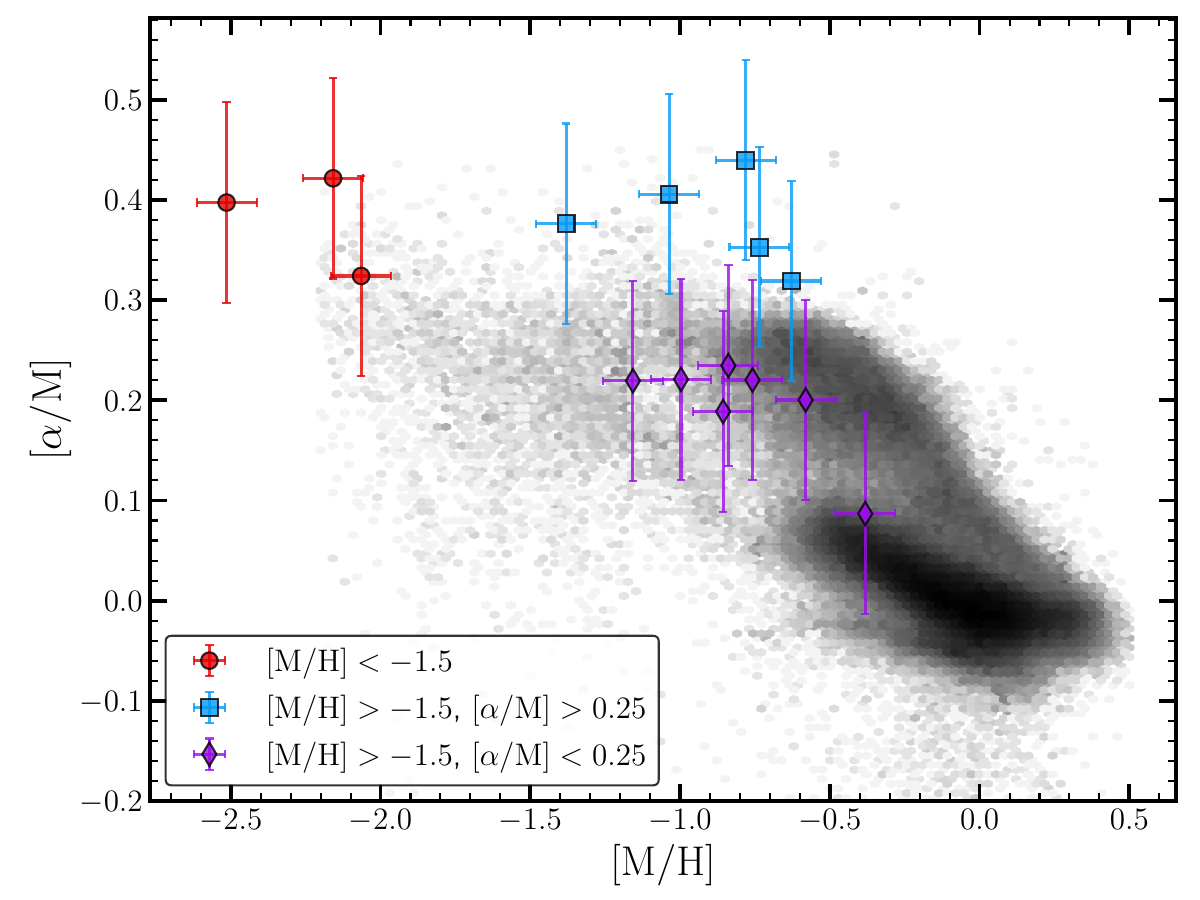} &
        \includegraphics[width=0.48\linewidth]{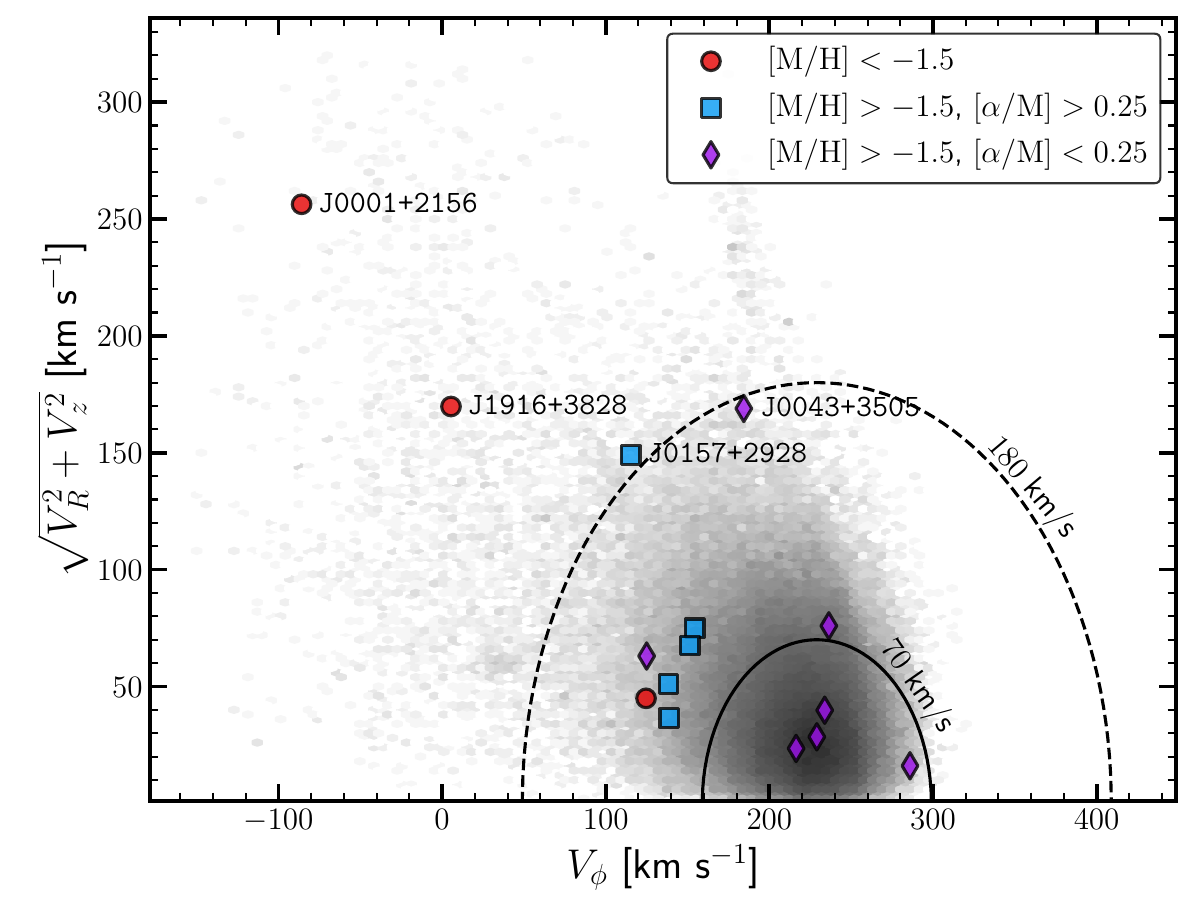} \\
    \end{tabular}
    \caption{Left: The 15 eclipsing binaries in the $[\alpha/\mathrm{M}]$--$[\mathrm{M/H}]$ plane. Right: Toomre diagram showing the Galactic kinematics of the binaries. The solid and dashed lines show boundaries of the thick disk and halo populations \citep{Bensby14}. In both panels, stars from SDSS-V are shown in gray for comparison.}
    \label{fig:alpha_iron_toomre}
\end{figure*}

Figure \ref{fig:alpha_iron_toomre} shows the measured metallicity and $\alpha$-abundance compared to stars in SDSS-V Milky Way Mapper \citep{Meszaros25}. We find three groups of targets. There are three systems at very low metallicity, $\mh{}<-1.5$, and two populations with $\mh{}>-1.5$ split by $\alpha$-abundance. Five systems are highly $\alpha$-enhanced with $\alpham{} > 0.25$, and seven are only moderately $\alpha$-enhanced with $\alpham{} < 0.25$. The most metal-poor system, J0001$+$2156, has $\mh{} = \specmet{J0001+2156}$, making it the most metal-poor detached eclipsing binary characterized to date. 

\setlength{\tabcolsep}{4pt}
\renewcommand{\tablecaptiontext}{Spectroscopic parameters determined from disentangled component spectra fit with spectral synthesis using \Korg{}. \label{tab:spec_table}}
\begin{deluxetable*}{lrrrrrrrrrrr}[t]
\tablecaption{\tablecaptiontext}
\tablehead{
\colhead{Name} & \colhead{$T_{\rm eff,1}$} & \colhead{$T_{\rm eff,2}$} & \colhead{$\log g_1$} & \colhead{$\log g_2$} & \colhead{$v\sin i_1$} & \colhead{$v\sin i_2$} & \colhead{$\xi_1$} & \colhead{$\xi_2$} & \colhead{$[\rm M/H]$} & \colhead{$[\alpha/\rm M]$} & \colhead{$\ell_2$} \\
\colhead{} & \colhead{[K]} & \colhead{[K]} & \colhead{} & \colhead{} & \colhead{[km\,s$^{-1}$]} & \colhead{[km\,s$^{-1}$]} & \colhead{[km\,s$^{-1}$]} & \colhead{[km\,s$^{-1}$]} & \colhead{} & \colhead{} & \colhead{}
}
\startdata
J0001$+$2156 & 5700 & 5500 & 3.7 & 4.7 & 10.4 & 5.0 & 1.1 & 0.9 & $-2.51$ & $0.40$ & $0.08$ \\
J0043$+$3505 & 5600 & 5800 & 4.0 & 4.0 & 26.9 & 26.6 & 1.1 & 1.1 & $-0.76$ & $0.22$ & $0.52$ \\
J0157$+$2928 & 5700 & 5700 & 4.4 & 4.4 & 3.6 & 3.8 & 1.0 & 1.0 & $-1.11$ & $0.38$ & $0.50$ \\
J0345$-$0407 & 5000 & 4900 & 4.6 & 4.6 & 4.1 & 4.9 & 0.9 & 0.9 & $-0.48$ & $0.23$ & $0.45$ \\
J0350$+$0805 & 5800 & 5600 & 4.2 & 4.4 & 13.0 & 9.1 & 1.1 & 1.0 & $-0.78$ & $0.44$ & $0.31$ \\
J0414$-$1023 & 6300 & 6200 & 4.1 & 4.1 & 14.2 & 13.8 & 1.3 & 1.3 & $-0.58$ & $0.20$ & $0.49$ \\
J0727$+$2018 & 6300 & 6000 & 4.4 & 4.4 & 13.0 & 13.6 & 1.2 & 1.1 & $-0.85$ & $0.23$ & $0.39$ \\
J0807$+$6945 & 5700 & 5700 & 4.3 & 4.4 & 5.1 & 5.2 & 1.0 & 1.0 & $-0.73$ & $0.35$ & $0.50$ \\
J1047$+$4815 & 5700 & 5500 & 4.2 & 4.4 & 11.7 & 7.0 & 1.0 & 1.0 & $-0.63$ & $0.32$ & $0.33$ \\
J1556$+$1218 & 6000 & 5600 & 4.3 & 4.5 & 8.5 & 5.0 & 1.1 & 0.9 & $-2.06$ & $0.32$ & $0.27$ \\
J1626$-$1140 & 6100 & 6100 & 4.2 & 4.1 & 9.9 & 9.9 & 1.2 & 1.2 & $-0.86$ & $0.19$ & $0.53$ \\
J1905$+$6233 & 6400 & 6400 & 4.2 & 4.2 & 26.5 & 27.9 & 1.4 & 1.3 & $-1.00$ & $0.22$ & $0.48$ \\
J1909$+$4025 & 6000 & 6000 & 4.2 & 4.2 & 17.8 & 17.7 & 1.1 & 1.1 & $-1.16$ & $0.22$ & $0.50$ \\
J1916$+$3828 & 6000 & 6400 & 3.8 & 4.2 & 11.9 & 8.0 & 1.2 & 1.3 & $-2.16$ & $0.42$ & $0.36$ \\
J2352$-$0707 & 6100 & 6100 & 4.3 & 4.3 & 9.2 & 7.9 & 1.1 & 1.1 & $-1.38$ & $0.38$ & $0.48$ \\
\enddata
\end{deluxetable*}

\section{Galactic Kinematics} \label{sec:kinematics}

Galactic kinematics offer an independent check on the metallicity and $\alpha$-abundance of our targets. We use the \Gaia{} DR3 five-dimensional astrometric solution and the center-of-mass velocity $\gamma$ measured from the radial velocity orbit to infer the Galactic kinematics of the binaries and assess whether they are kinematically consistent with forming in the thick disk or halo. Figure \ref{fig:alpha_iron_toomre} shows the Toomre diagram for the characterized EBs. We use rotational velocities of 70~km/s and 180~km/s relative to the Solar tangential velocity to separate the thin disk, thick disk, and halo populations \citep{Bensby14}. We find two targets on extreme halo orbits, both of which have $\mh{} < -1.5$ from the spectral disentangling analysis. The four stars with Galactic orbits most similar to the Sun have higher metallicities $\mh{} > -1.2$ and lower $\alpha$-abundance $\alpham{} < 0.25$, as expected for a thin disk population. All of the binaries with $\alpham{} > 0.25$ are kinematically consistent with the thick disk or halo. 

\begin{figure*}
    \centering
    \includegraphics[width=\linewidth]{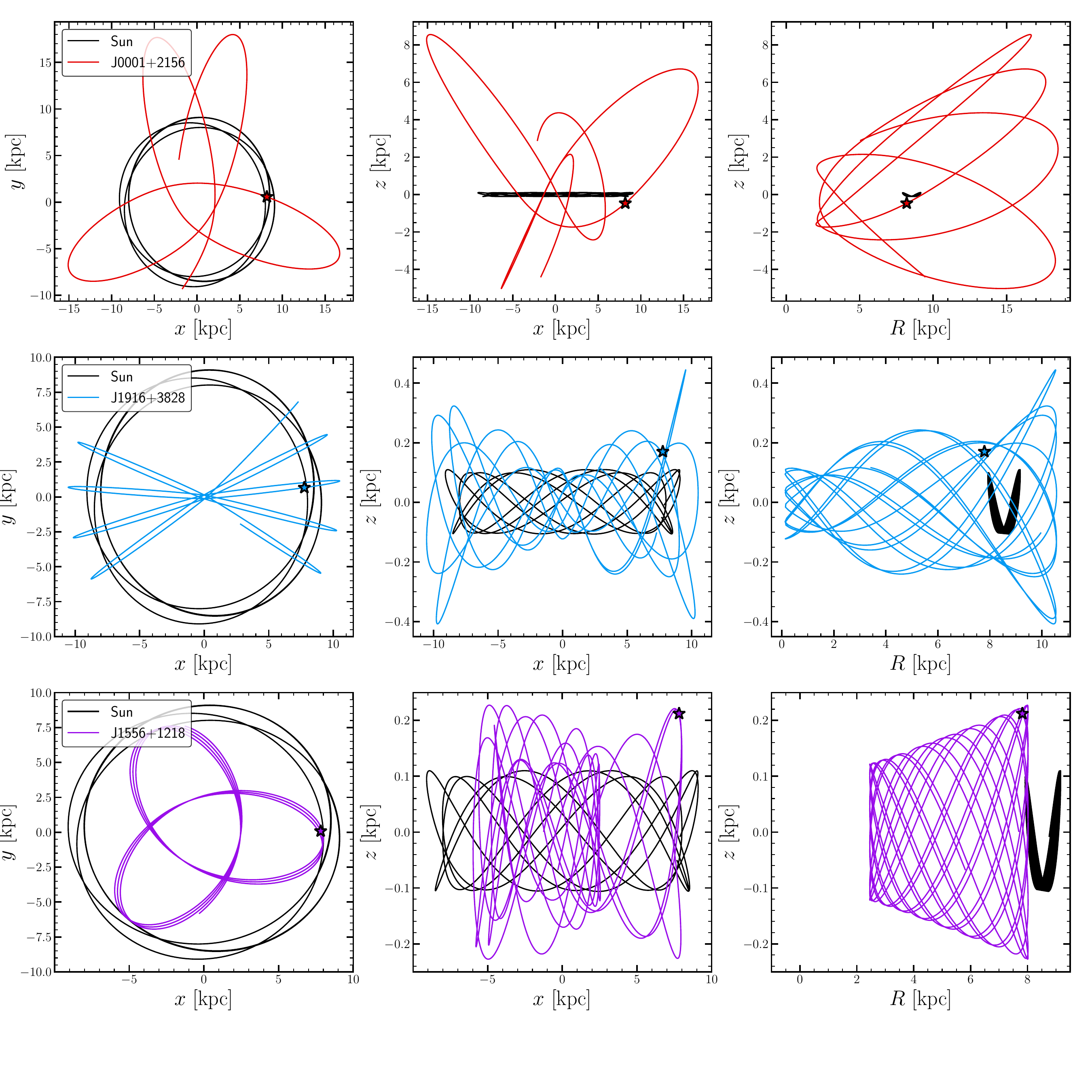}
    \caption{Kinematic orbits for the three most metal-poor systems in the $x$-$y$ (left), $x$-$z$ (middle), and $R$-$z$ (right) plane, where $x$, $y$, $z$ are Galactic Cartesian coordinates and $R$ is the distance from the Galactic center. The orbits are integrated over a 1~Gyr ($\pm 500$~Myr) window centered on the present epoch. The current location of the binary is marked with the star. In all panels, the orbit of the Sun over the same time period is shown in black.}
    \label{fig:kinematic_orbits}
\end{figure*}

Figure \ref{fig:kinematic_orbits} plots the Galactic orbits for the three most metal-poor systems integrated with {\tt galpy} using the {\tt MWPotential2014} potential. J0001$+$2156 and J1916$+$3828 both have orbits consistent with the Galactic halo (Figure \ref{fig:alpha_iron_toomre}), but are kinematically distinct. While J0001$+$2156 reaches as high as 8~kpc above the Galactic plane, J1916$+$3828 stays within $\sim 400$~pc and instead has a highly eccentric orbit ($e_{\rm{gal}} = 0.98$) that extends to within $\sim 1$~kpc of the Galactic center. J1556$+$1218 is kinematically consistent with the thick disk population, and has a moderately eccentric Galactic orbit ($e_{\rm{gal}} = 0.53$) with a maximum vertical departure from the Galactic midplane $z_{\rm{max}} = 230$~pc.

\section{Age Estimation and Model Comparison} \label{sec:ages}

We expect the stars in these close binaries to have been born from gas with the same metallicity at the same time. Theoretical isochrones should therefore predict the same age for both binary components given their measured masses and radii. We use isochrones from three stellar evolution codes: PARSEC v2.0 \citep{Bressan12}, which we access using {\tt ezpadova}\footnote{\url{https://github.com/mfouesneau/ezpadova}}, MESA Isochrones and Stellar Tracks (MIST) v2.5 \citep{Choi16, Dotter16, Dotter26}, and ``a Bag of Stellar Tracks and Isochrones'' \citep[BaSTI,][]{Hidalgo18, Pietrinferni21}, to test for coevality. 

For each binary, we construct a grid of isochrones spanning $\log(\mathrm{age/yr}) = 8.0\text{--}10.2$ at the measured metallicity with $\Delta \log(\mathrm{age/yr}) = 0.01 $. The MIST and BaSTI isochrones can also vary $\alpha$-abundance on a fixed grid of $[\alpha/\rm{Fe}]=\{-0.2,\ 0.0,\ +0.2,\ +0.4,\ +0.6\}$ for MIST and $[\alpha/\rm{Fe}]=\{-0.2,\ 0.0,\ +0.4\}$ for BaSTI. We use the value closest to the measured $\alpha$-abundance. We interpolate each isochrone in the grid in the mass-radius plane and find the minimum $\chi^2$ for both binary components relative to the isochrone model
\begin{equation}
    \chi^2_i = \min_{j}\left[\left(\frac{M_i - M_j}{\sigma_{M_i}}\right)^2 + \left(\frac{R_i - R_j}{\sigma_{R_i}}\right)^2\right],
\end{equation}
where $i=1$ or 2 is the index of the component (primary or secondary) and $j$ is the index along the isochrone. We then minimize $\chi^2_{\rm{coeval}} = \chi^2_1 + \chi^2_2$ over the age grid.

In order to see if the binary is consistent with being coeval, we compare this $\chi^2_{\rm{coeval}}$ to a $\chi^2_{\rm{free}}$, which instead allows for the two stars to have independent ages. We recompute $\chi^2_{\rm{coeval}}$ and $\chi^2_{\rm{free}}$, sampling over the masses $M_i$ and $R_i$ from our MCMC posteriors, to compute the distribution of predicted stellar ages and $\Delta \chi^2 = \chi^2_{\rm{coeval}} - \chi^2_{\rm{free}}$. 

To calibrate the significance of the observed $\Delta \chi^2$ distribution, we construct a null distribution by Monte Carlo sampling of synthetic coeval binaries. We draw masses from the observed mass posteriors and inject realistic measurement noise. We use an age prior restricting ages to those where both components are no longer on the pre-main sequence but have radii $R_{\rm{synth},i} < 2 R_i$, to exclude binaries where the primary is a red giant. The fraction of synthetic coeval systems with $\Delta \chi^2$ greater than the median of the measured $\Delta \chi^2$ distribution defines a $p$-value, where $p > 0.05$ indicates that the observed binary is consistent with being coeval. 

\begin{figure*}
    \centering
    \includegraphics[width=\linewidth]{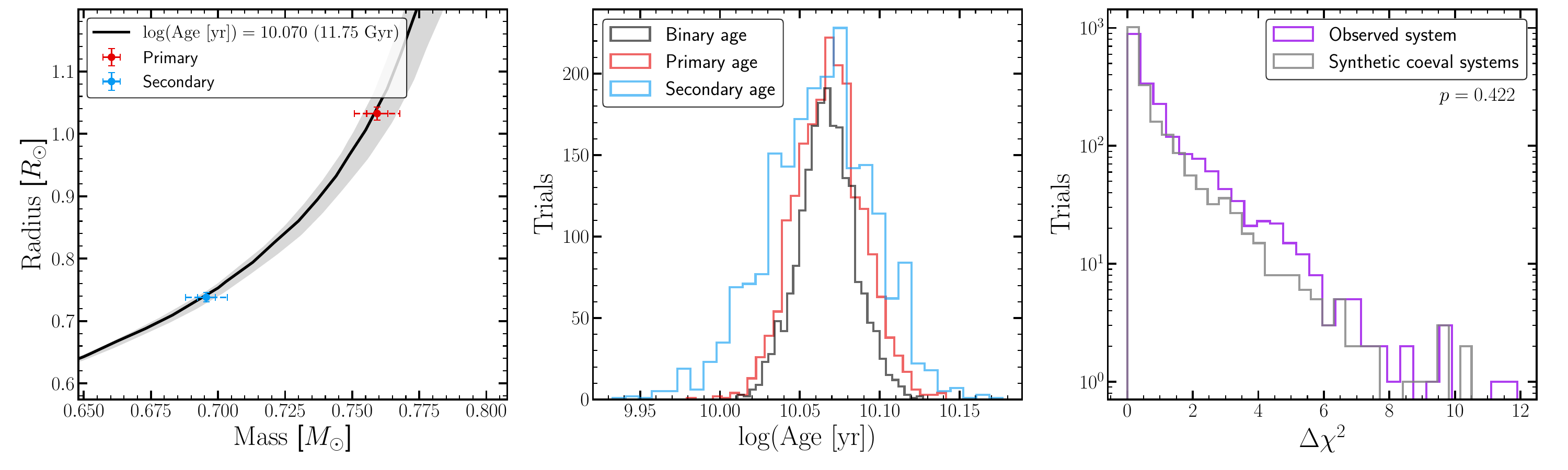}
    \caption{Example of the age determination and model comparison for J1556$+$1218. The left panel shows the masses and radii measured from the light curves and RV orbits. The best-fitting binary isochrone is shown in black and the $1\sigma$ age range is shown in gray. The middle panels show the age posteriors, sampling over the mass and radius posteriors, for both the binary age and the age estimated for each component individually. The right panel shows the $\Delta \chi^2$ distribution between this coeval binary isochrone fit and the fit that assigns an independent age to each star. The gray line shows random coeval synthetic systems sampled using a mass and age prior representative of the target at the same metallicity. We use the median of the observed $\Delta \chi^2$ distribution to assess the significance of the difference between these distributions.}
    \label{fig:age_example}
\end{figure*}

Figure \ref{fig:age_example} shows the best-fitting coeval isochrone in the mass-radius plane, the age posteriors for the coeval and free-age fits, and the $\Delta \chi^2$ distribution for J1556$+$1218 using the PARSEC isochrone grid. We find that the measured masses and radii for the two components agree well with an isochrone of age $\age{J1556+1218}{padova}$~Gyr. The independent age constraint from the secondary is broader than that of the primary, which is expected since it is less evolved, but the primary, secondary, and binary ages all agree well within $1\sigma$. The observed $\Delta \chi^2$ distribution is very similar to the predicted distribution from synthetic binaries, and we find a $p$-value of $0.438$.

Table \ref{tab:age_table} reports the median and $1\sigma$ age posteriors and $p$-values from the $\Delta \chi^2$ test for all targets with the three isochrone grids. We find that 13 out of 15 systems are consistent with being coeval with $p > 0.05$. We discuss these systems in detail in Section \ref{sec:individual_targets}. For the three most metal-poor systems, we measure $\Delta \chi^2 = 0.48$, $1.13$, and $1.40$ (using PARSEC isochrones), which supports the accuracy of stellar models at low metallicity. 

\setlength{\tabcolsep}{4pt}
\renewcommand{\tablecaptiontext}{Age estimates from comparison to theoretical isochrone grids. For each target, we estimate the age by minimizing the $\chi^2_{\rm{coeval}}$ (Section \ref{sec:ages}). We compute a $\Delta \chi^2$ distribution between a model that uses a coeval age and an independent age for both binary components, and determine a $p$-value based on comparing the distribution of $\Delta \chi^2$ to simulated binaries representative of each target. \label{tab:age_table}}
\begin{deluxetable*}{lrrrrrrrrr}[t]
\tablecaption{\tablecaptiontext}
\tablehead{
\colhead{Name} & \colhead{$t_{\rm Padova}$} & \colhead{$\Delta\chi^2_{\rm Padova}$} & \colhead{$p_{\rm Padova}$} & \colhead{$t_{\rm MIST}$} & \colhead{$\Delta\chi^2_{\rm MIST}$} & \colhead{$p_{\rm MIST}$} & \colhead{$t_{\rm BaSTI}$} & \colhead{$\Delta\chi^2_{\rm BaSTI}$} & \colhead{$p_{\rm BaSTI}$} \\
\colhead{} & \colhead{[Gyr]} & \colhead{} & \colhead{} & \colhead{[Gyr]} & \colhead{} & \colhead{} & \colhead{[Gyr]} & \colhead{} & \colhead{}
}
\startdata
J0001$+$2156 & $10.2^{+0.4}_{-0.5}$ & 1.13 & 0.226 & $10.2^{+0.7}_{-0.5}$ & 1.07 & 0.227 & $10.3^{+0.6}_{-0.5}$ & 1.07 & 0.226 \\
J0043$+$3505 & $7.7\pm0.3$ & 0.58 & 0.372 & $8.2\pm0.3$ & 0.62 & 0.365 & $7.7\pm0.3$ & 0.60 & 0.370 \\
J0157$+$2928 & $7.5\pm0.4$ & 0.48 & 0.444 & $8.5^{+0.5}_{-0.4}$ & 0.48 & 0.435 & $7.4\pm0.4$ & 0.48 & 0.438 \\
J0345$-$0407 & $4.0\pm2.0$ & 0.77 & 0.304 & $4.0\pm2.0$ & 0.72 & 0.319 & $3.0\pm2.0$ & 0.77 & 0.305 \\
J0350$+$0805 & $6.4\pm0.3$ & 0.46 & 0.448 & $7.5\pm0.4$ & 0.48 & 0.429 & $6.4\pm0.3$ & 0.51 & 0.412 \\
J0414$-$1023 & $4.0\pm0.2$ & 0.43 & 0.444 & $4.3\pm0.2$ & 0.43 & 0.454 & $4.1\pm0.2$ & 0.44 & 0.453 \\
J0727$+$2018 & $4.3^{+0.4}_{-0.5}$ & 5.01 & 0.009 & $4.3\pm0.5$ & 4.96 & 0.009 & $4.2\pm0.5$ & 4.10 & 0.018 \\
J0807$+$6945 & $6.1\pm0.4$ & 0.47 & 0.431 & $7.2^{+0.5}_{-0.4}$ & 0.49 & 0.438 & $6.0\pm0.4$ & 0.53 & 0.425 \\
J1047$+$4815 & $6.2\pm0.3$ & 0.54 & 0.408 & $7.5\pm0.4$ & 0.52 & 0.435 & $6.1\pm0.3$ & 0.92 & 0.288 \\
J1556$+$1218 & $11.7^{+0.5}_{-0.4}$ & 0.50 & 0.422 & $11.8\pm0.5$ & 0.48 & 0.424 & $11.8\pm0.5$ & 0.48 & 0.421 \\
J1626$-$1140 & $6.0\pm0.3$ & 0.49 & 0.400 & $6.2\pm0.3$ & 0.44 & 0.446 & $6.0\pm0.3$ & 0.45 & 0.437 \\
J1905$+$6233 & $5.5\pm0.2$ & 4.24 & 0.035 & $5.6\pm0.2$ & 4.27 & 0.026 & $5.6\pm0.2$ & 4.57 & 0.024 \\
J1909$+$4025 & $6.9\pm0.3$ & 0.42 & 0.458 & $7.1\pm0.3$ & 0.45 & 0.428 & $7.1\pm0.3$ & 0.43 & 0.458 \\
J1916$+$3828 & $13.8^{+0.8}_{-0.7}$ & 1.40 & 0.181 & $14.0\pm0.7$ & 1.44 & 0.198 & $13.9^{+0.8}_{-0.6}$ & 1.51 & 0.161 \\
J2352$-$0707 & $9.0\pm0.5$ & 0.56 & 0.391 & $9.4\pm0.5$ & 0.55 & 0.377 & $9.1^{+0.5}_{-0.4}$ & 0.58 & 0.376 \\
\enddata
\end{deluxetable*}

\begin{figure}
    \centering
    \includegraphics[width=\linewidth]{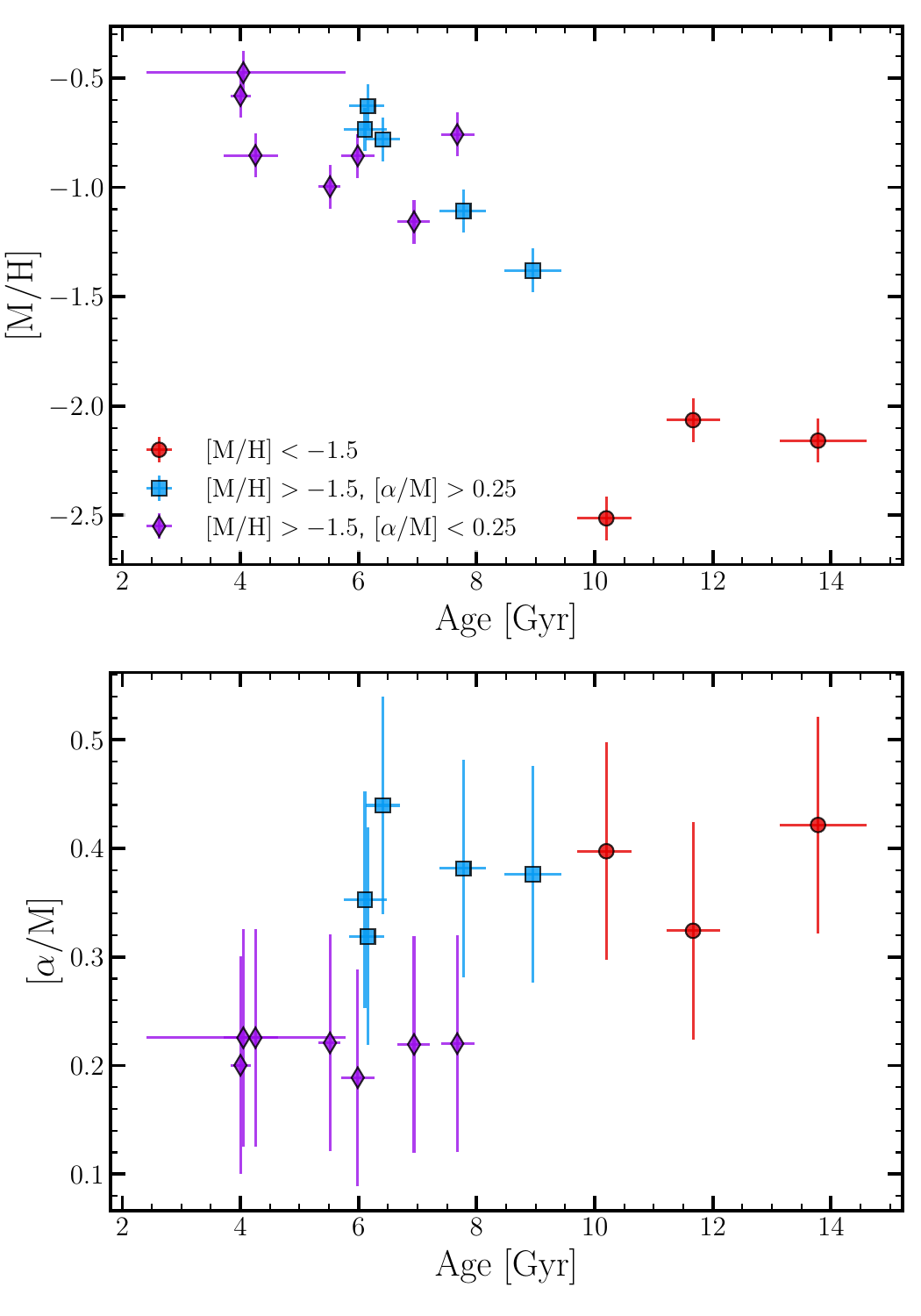}
    \caption{Distribution of measured metallicities (top) and $\alpha$-abundance (bottom) and stellar ages using PARSEC isochrones. We find that the oldest stars in the sample are also the most metal-poor.}
    \label{fig:age_metallicity}
\end{figure}

Figure \ref{fig:age_metallicity} shows the distribution of metallicities and $\alpha$-abundances and measured ages from the binary PARSEC isochrone fits. The three systems with $\mh{} < -1.5$ are the oldest three binaries in the sample. Stars with $\mh{} > -1.5$ and $\alpham{} > 0.25$ have ages $6 \lesssim \tau \lesssim 9$~Gyr, and stars with $\alpham{} < 0.25$ have $\tau \lesssim 8$~Gyr. The measured metallicities, $\alpha$-abundances, kinematics, and ages therefore complete a self-consistent picture for the three populations shown in Figure \ref{fig:alpha_iron_toomre}.

\section{Individual Targets} \label{sec:individual_targets}

Sections \ref{sec:target_selection} through \ref{sec:ages} describe the standard analysis procedure applied to all targets. Here we describe the nuances in the analysis of each target and the measured properties of each system. 

\textbf{J0001$+$2156 (\Gaia{} DR3 2847197557833077504)} is the most metal-poor system in the sample with $\mh{} = \specmet{J0001+2156}$. This is a known eclipsing binary discovered in ASAS-SN photometry \citep{Jayasinghe19} and included in the ASAS-SN eclipsing binaries catalog \citep{Rowan22}. We selected this as a candidate metal-poor eclipsing binary using the \Gaia{} XP metallicity catalog from \citet{Andrae23}, which reports $\mh{} = -2.79$. Using the \Gaia{} proper motions, parallax, and mean RV, we also recovered this target as part of the kinematics-based search with halo membership probability $p_{\rm{halo}}\approx 1.0$.

The \TESS{} light curve shows that this target has eclipses of roughly similar depth (Figure \ref{fig:lc_rv_appendix}). From the light curve model, the effective temperature ratio is $T_{\rm{eff},2}/T_{\rm{eff},1} \sim 0.96$, but the 1D-CCF shows that this system is a single-lined spectroscopic binary, suggesting that the binary has a low flux ratio. For this target, we determine the template spectrum of the binary by fitting an observed PEPSI spectrum with a single-star model. We then set the secondary template using the effective temperature ratio estimated in \citet{Rowan22}, which is based on the ASAS-SN light curve. The combination of a low flux ratio binary and a metal-poor spectrum makes extracting the secondary velocities of this binary more challenging than in the other systems. We found that the secondary velocities could be recovered only with TODCOR. As a result of this low flux ratio, we measure secondary RV residuals that are $\sim 2\times$ larger than those of the primary (Figure \ref{fig:lc_rv_appendix}) and set an uncertainty floor of $500$~m/s on the secondary RVs. 

With spectral disentangling, we are able to extract the subtle signatures of the secondary. We find the binary has a flux ratio of $\ell_2 = 0.08$ at 520nm. This low flux ratio can be explained by the large radius ratio. The photometric primary with $M_1=\mass{J0001+2156}{1}\ M_\odot$ and radius $R_1=\radius{J0001+2156}{1}\ R_\odot$ is near the end of its main sequence lifetime, while the lower-mass secondary with $M_2 = \mass{J0001+2156}{2}\ M_\odot$ is still firmly on the main sequence with $R_2 = \radius{J0001+2156}{2}\ R_\odot$. The primary is the largest star in the sample and the secondary is the least massive (Figure \ref{fig:mass_radius}).

\begin{figure*}
    \centering
    \includegraphics[width=\linewidth]{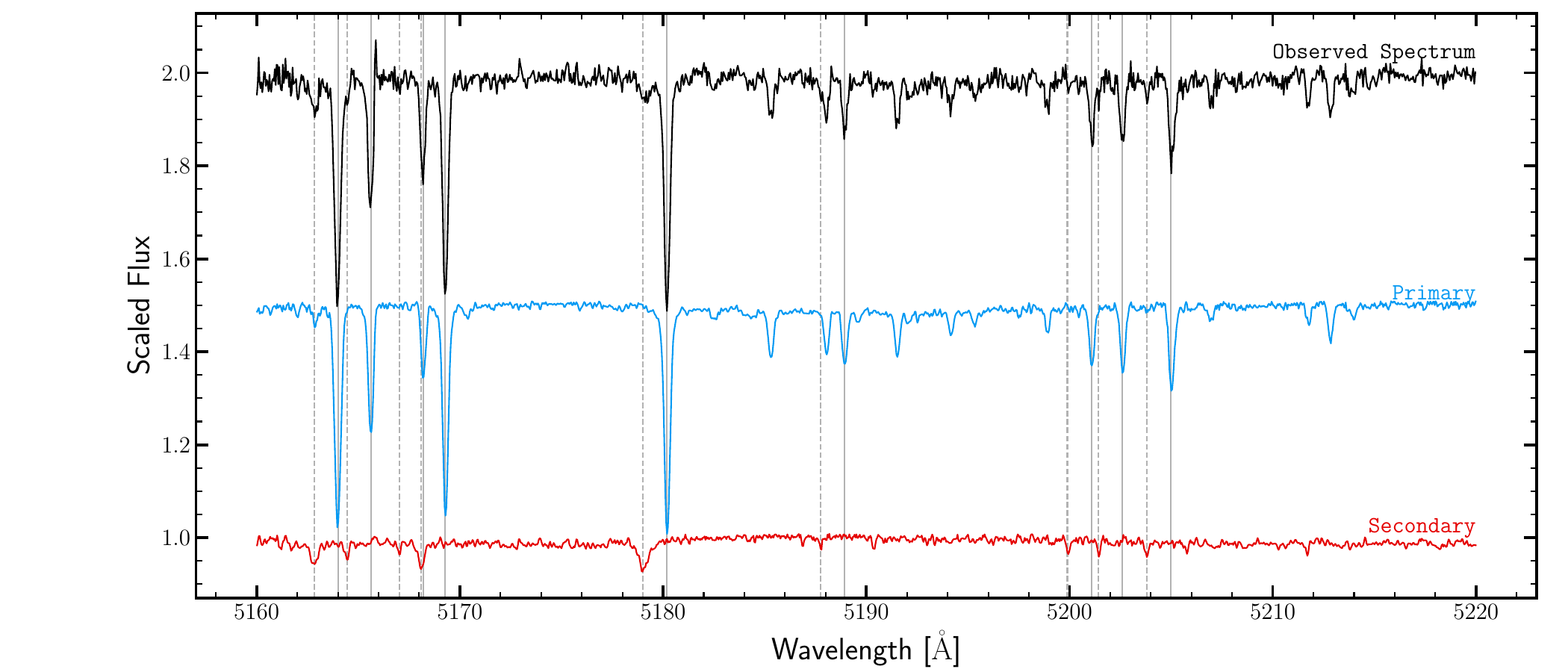}
    \caption{Same as Figure \ref{fig:disentangling_example} except for J0001$+$2156, which is both the most metal-poor target in our sample and the lowest flux ratio binary. The combination of the low flux ratio and extreme metallicity results in a secondary that is challenging to identify. The contributions from the secondary to the observed spectrum can be most clearly seen in the wings of the \ion{Mg}{1}~b triplet. The vertical gray lines show a selection of Mg, Fe, and Cr lines for the primary (solid) and secondary (dashed).}
    \label{fig:J0001_disentangling}
\end{figure*}

\begin{figure*}
    \centering
    \includegraphics[width=\linewidth]{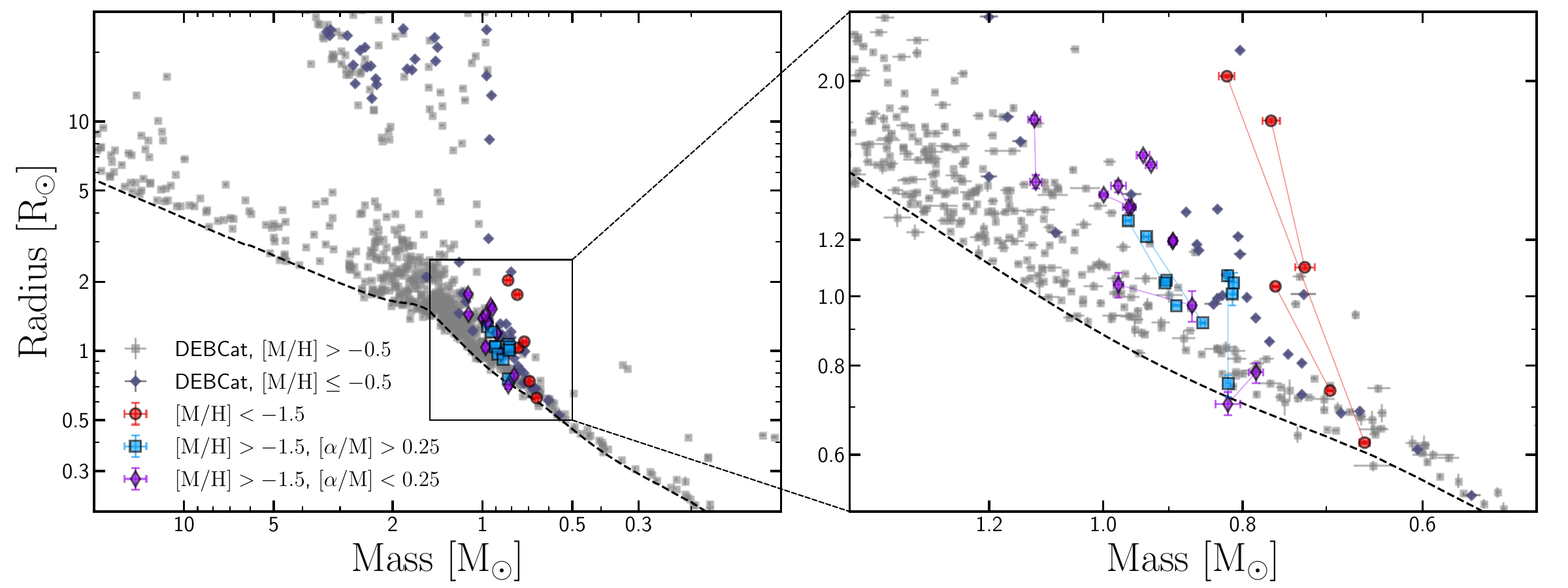}
    \includegraphics[width=\linewidth]{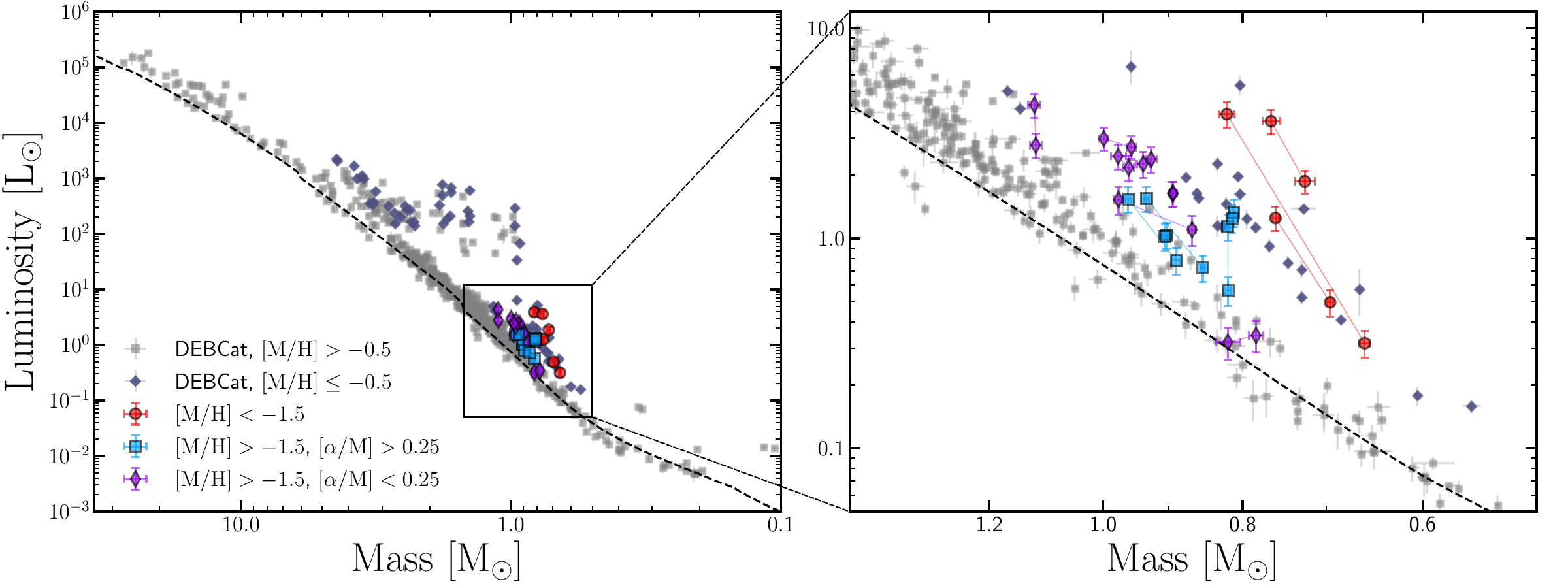}
    \caption{Measured masses and radii (top) and luminosities (bottom) for the sample of EBs characterized here compared to the full DEBCat \citep{Southworth15}. The zero-age main sequence (ZAMS) from MIST \citep{Dotter16, Choi16} is shown in black. The right panel shows a zoom-in to the mass range of our targets. Colored lines connect the individual binary systems. Many of our binaries are more evolved than the bulk of the population at similar masses, suggesting that they are older than the solar-metallicity population.}
    \label{fig:mass_radius}
\end{figure*}

We show the disentangled component spectra in Figure \ref{fig:J0001_disentangling}. The binary is clearly very metal-poor, with only a few observed lines in the composite spectrum. Aside from the \ion{Mg}{1}~b triplet, weak lines of \ion{Fe}{1}, \ion{Ti}{2}, \ion{Cr}{1}, and \ion{Ca}{1} are observed. Even though we can detect the secondary, the low flux ratio means that the secondary component spectrum has very low signal-to-noise $S/N < 10$. Despite this, the contribution from the secondary spectrum to the observed composite spectrum is clear in Figure \ref{fig:J0001_disentangling}. 

For the other targets in this sample, we jointly fit the disentangled component spectra as single stars, coupled through the flux ratio $\ell_2$, which rescales the disentangled components (Section \ref{sec:disentangling}). We take a different approach for this target. Because of the low $S/N$ of the secondary component spectrum, we only minimize the $\chi^2$ of the primary. We emphasize that the disentangled component spectrum is still scaled according to $\ell_2$, which is determined at each optimization step based on the effective temperature ratio, radius ratio, and synthetic model atmospheres. We measure a spectroscopic metallicity $\mh{} = \specmet{J0001+2156}$ and $\alpham{}=\specalpha{J0001+2156}$. This analysis is done using the eight PEPSI spectra. We repeated the disentangling and spectral synthesis fit using the 14 APF spectra to find a metallicity $\mh{}=-2.54$ and $\alpham{}=0.38$ that are consistent within our $\sigma_{[\rm{X/H}]} = 0.1$~dex error estimate.

This target also has the largest $\sqrt{V_R^2 + V_Z^2}=256$~km/s of the binaries in our sample and has a halo orbit consistent with large $\sim 8$~kpc departures from the Galactic midplane. All three theoretical isochrones, PARSEC, MIST, and BaSTI, find that the binary is consistent with being coeval. We measure an age $\tau = \age{J0001+2156}{padova}$~Gyr with the PARSEC isochrones, making this the third oldest binary in the sample.

\textbf{J0043$+$3505 (\Gaia{} DR3 362498302094814720)} is a partially eclipsing, approximately equal-mass binary in a short $P=3.203$~day orbit discovered in ASAS-SN photometry \citep{Jayasinghe19}. We selected this as a thick disk candidate based on Galactic kinematics. We use {\tt galpy} to integrate the orbit and find that it reaches a maximum $z_{\rm{max}} = 6$~kpc from the Galactic midplane. We calculate membership probabilities for the thick disk and halo $p_{\rm{thick}} = 0.61$ and $p_{\rm{halo}} = 0.39$. The light curve fit shows correlated residuals outside of the eclipse (Figure \ref{fig:lc_rv_appendix}), suggesting that the detrending procedure (Section \ref{sec:target_selection}) has overfit and removed part of the binary signal from ellipsoidal variability. The amplitude of this ellipsoidal effect is driven by a combination of the masses, the fractional radii, and the inclination, which are all robustly measured from the RVs and the eclipses, so the effect of detrending the light curve on the measured masses and radii is minimal. As a result of the short orbital period, the stars have broad lines consistent with large projected rotational velocities $v \sin i \approx 27$~km/s. We measure a metallicity $\mh{} = \specmet{J0043+3505}$ and $\alpham{}=\specalpha{J0043+3505}$ using the PEPSI spectra. 

Even though we only have five PEPSI spectra for this target (Table \ref{tab:summarytable}), they sample both quadrature phases, and the orbit is precisely known from the joint light curve and RV orbit fit. We repeat the disentangling and component spectrum fit using the set of 10 APF spectra, measuring $\mh{}=-0.74$ and $\alpham{} = 0.23$, consistent with the PEPSI results. We also find that the measured projected rotational velocities from the PEPSI and APF spectra agree to within $0.5$~km/s for both components. 

Since this binary is nearly a twin, with $M_1 = \mass{J0043+3505}{1}\ M_\odot$ and $M_2 = \mass{J0043+3505}{2}\ M_\odot$, the comparison between the component ages predicted by evolutionary tracks is not as effective as for J0001$+$2156. The PARSEC and BaSTI isochrones estimate an age of $\tau = \age{J0043+3505}{padova}$~Gyr, while the MIST isochrones prefer a slightly older $\tau=\age{J0043+3505}{mist}$~Gyr binary age (Table \ref{tab:age_table}).

\textbf{J0157$+$2928 (\Gaia{} DR3 300726119643476992)} is a $P=8.2$~day twin binary on a thick disk orbit. We selected this system based on a thick disk membership probability of $p_{\rm{thick}} = 0.7$ and a halo membership probability of $p_{\rm{halo}} = 0.3$. We find that the two components have masses that agree at $| M_1  - M_2 | < 0.01\ M_\odot$ and the eclipses are of equal depth. Since this binary is grazing, with impact parameter $b=a \cos i / (R_1+R_2) = 0.5$, and the orbit is consistent with circular $e=\ecc{J0157+2928}$, we find that we are only able to constrain the sum of the radii, and not $R_2 / R_1$. We therefore report $R_1 = R_2 = (R_1 + R_2) / 2$ in Table \ref{tab:mcmc_posteriors}, when computing $\log g$ for spectral disentangling, and in estimating the age of the binary. 

We measure $\mh{} = \specmet{J0157+2928}$ and $\alpham{} = \specalpha{J0157+2928}$, consistent with a star on the boundary of the thick disk and halo in the Toomre diagram (Figure \ref{fig:alpha_iron_toomre}). These results use the APF spectra. We also analyzed the four PEPSI spectra, since they cover a range of velocity separations (Figure \ref{fig:lc_rv_appendix}). We find a similar $\mh{} = -1.05$ and $\alpham{} = 0.41$, in good agreement with the APF results.

\textbf{J0345$-$0407 (\Gaia{} DR3 3249407677302959872)} is a $P=9.66$~day binary selected based on RAVE DR6 spectroscopy \citep{Steinmetz20}, which reports $\mh{}=-1.5$. J0345$-$0407 was included in the \TESS{} EB catalog \citep{Prsa22}. We measure masses $M_1 = \mass{J0345-0407}{1}\ M_\odot$ and $M_2 = \mass{J0345-0407}{2}\ M_\odot$ but find we are only able to constrain the sum of the radii $R_1 + R_2$ from the grazing eclipses. For this system, we instead solve for the ratio of the radii, $k=R_2 / R_1$, spectroscopically. We modify our disentangling approach described in Section \ref{sec:disentangling} to instead solve for $T_{\rm{eff},1},\ v\sin i_{1,2},\ \mh{},\ \alpham{}$, and $k$. At each step of the optimization, we use the $R_1+R_2$ constraint from the light curve to solve for $R_1$ and $R_2$, which then sets $\log g_{1,2}$ and $\ell_2$. The spectral synthesis fit finds $k=0.99$, which corresponds to $R_1 = 0.75\ R_\odot$ and $R_2 = 0.74\ R_\odot$. We measure a metallicity $\mh{} = \specmet{J0345-0407}$, which is significantly higher than the $\mh{}=-1.5$ reported by RAVE. We construct a binary model from our spectral synthesis result with component velocities determined for the epoch of the RAVE observation, and find our fit reproduces the RAVE spectrum with $\chi^2_\nu = 1.03$. This binary has the lowest $T_{\rm{eff}}$ in our sample, with $T_{\rm{eff},1} = \specteff{J0345-0407}{1}$~K and $T_{\rm{eff},2} = \specteff{J0345-0407}{2}$~K, and sits below the other binaries on the \Gaia{} CMD (Figure \ref{fig:cmd}). 

We find that the binary is highly $\alpha$-enhanced, $\alpham{}=\specalpha{J0345-0407}$. Kinematically, this binary sits on the boundary of the thin disk and thick disk in the Toomre diagram. The Galactic orbit has a slightly larger guiding center radius than the Sun, $R_c = 9.0$~kpc, and is on an eccentric $e_{\rm{gal}} = 0.22$ orbit that reaches a maximum $z_{\rm{max}}=840$~pc from the Galactic midplane. 

\textbf{J0350$+$0805 (\Gaia{} DR3 3277480820501314816)} is a $P=5.33$~day totally eclipsing binary first discovered in ASAS-SN photometry \citep{Jayasinghe19}. We measure a small but non-zero eccentricity $e=\ecc{J0350+0805}$. We selected this binary using \Gaia{} kinematics with thick disk probability $p_{\rm{thick}} = 0.96$ and measure a metallicity from the disentangled component spectra of $\mh{} = \specmet{J0350+0805}$. The binary is also consistent with being $\alpha$-enhanced, $\alpham{} = \specalpha{J0350+0805}$, as expected for a thick disk star. Despite a mass difference of only $\sim 0.1\ M_\odot$, the primary is $\sim 0.3\ R_\odot$ larger than the secondary, allowing for a robust age determination. We find a binary age of $\age{J0350+0805}{padova}$~Gyr, $\age{J0350+0805}{mist}$~Gyr, and $\age{J0350+0805}{basti}$~Gyr using PARSEC, MIST, and BaSTI theoretical isochrones, respectively. 

\textbf{J0414$-$1023 (\Gaia{} DR3 3191963314353757952)} is a $P=6.05$~day binary selected using archival RAVE spectra. This binary was originally discovered with ASAS-SN \citep{Jayasinghe19}. RAVE DR6 \citep{Steinmetz20} reports $\mh{}=-1.0$. The masses are consistent with a stellar twin, $M_1 = \mass{J0414-1023}{1}\ M_\odot$ and $M_2 = \mass{J0414-1023}{2}\ M_\odot$. Like J0157$+$2928, we find that the grazing eclipses prevent precise determination of $k$. We report $R_1 = R_2 = (R_1 + R_2) / 2$ in Table \ref{tab:mcmc_posteriors}, when computing $\log g$ for spectral disentangling, and in estimating the age of the binary. 

We disentangle the CHIRON spectra and find $\mh{} = \specmet{J0414-1023}$, higher than the $\mh{}=-1.0$ reported by RAVE, but we note that the RAVE spectrum has a low signal-to-noise $S/N\sim25$. We find that the binary is somewhat $\alpha$-enhanced, with $\alpham{} = \specalpha{J0414-1023}$. Kinematically, this binary is consistent with the thin disk and has membership probability $p_{\rm{thin}} = 1.0$. 

\textbf{J0727$+$2018 (\Gaia{} DR3 864979548396586880)} is a $P=4.22$~day binary selected using the LAMOST LRS catalog, which reports $\feh{}=-1.1$. The eclipses are shallow ($\lesssim 10\%$) and grazing. We measure an impact parameter $b=a \cos i / (R_1+R_2) = 0.6$. We also find a non-zero third light fraction, $\ell_3 = \thirdlight{J0727+2018}$, which is unsurprising given that the binary is $\sim 3\farcm4$ (10 \TESS{} pixels) away from a $V=5.9$~mag star and $\sim 1\farcm1$ from a $V=11.9$~mag star. This third light fraction is consistent with the $\sim10\%$ contamination predicted by the SPOC {\tt CROWDSAP}. 

While the sum of the radii is constrained by the eclipse durations, the ratio of the radii is not, so we do not report individual radii in Table \ref{tab:mcmc_posteriors}. Instead, we determine $k=R_2/R_1$ spectroscopically, as described above for J0345$-$0407. We find $k=0.88$, which corresponds to $R_1 = 1.07\ R_\odot$ and $R_2 = 0.94\ R_\odot$ using the $R_1 + R_2 = \radiussum{J0727+2018}\ R_\odot$ measurement from the light curve.

We measure a metallicity $\mh{}=\specmet{J0727+2018}$ and $\alpham{}=\specalpha{J0727+2018}$. The Galactic kinematics are consistent with a thin disk orbit, and we find a membership probability $p_{\rm{thin}} = 0.99$.

\textbf{J0807$+$6945 (\Gaia{} DR3 1098324086498919552)} is a $P=15.2$~day binary, the longest period system in our sample. We selected this as a low-metallicity candidate based on Galactic kinematics with $p_{\rm{thick}} = 0.98$. Despite only seeing two primary and two secondary eclipses in a single \TESS{} sector, we are able to precisely measure the orbital period with twelve RVs. The binary is nearly totally eclipsing with deep $\sim 0.65$~mag eclipses, and we measure the inclination $i = \inclination{J0807+6945}^{\circ}$. We measure a metallicity $\mh{} = \specmet{J0807+6945}$ and $\alpham{} = \specalpha{J0807+6945}$, chemically consistent with being part of the thick disk. We find an age $\age{J0807+6945}{padova}$~Gyr with PARSEC and BaSTI, and $\age{J0807+6945}{mist}$~Gyr with MIST. 

\begin{figure*}[h]
    \centering
    \includegraphics[width=\linewidth]{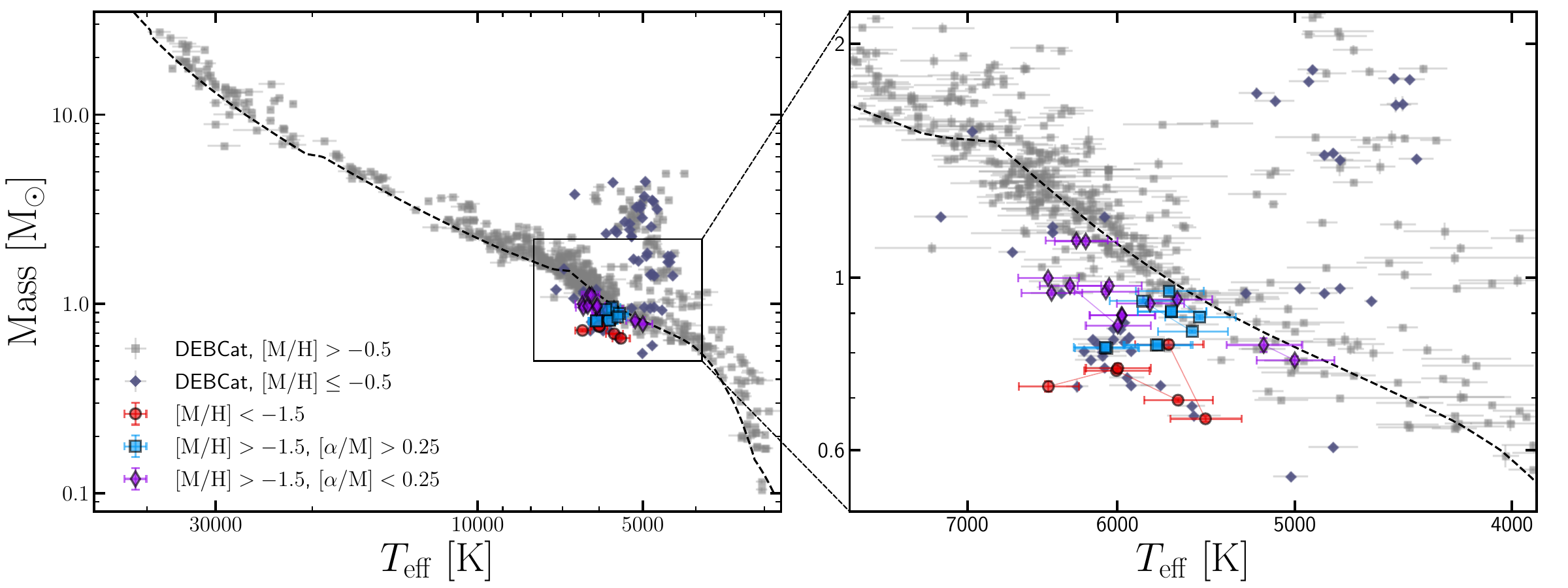}
    \includegraphics[width=\linewidth]{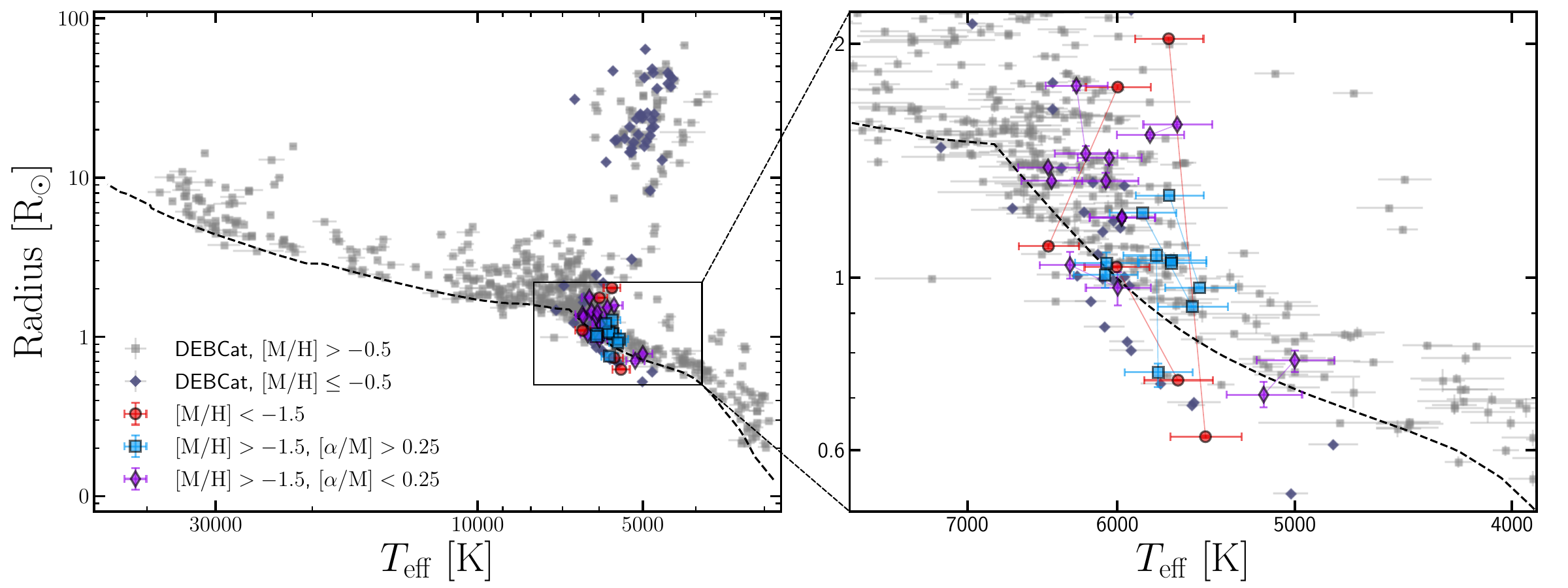}
    \includegraphics[width=\linewidth]{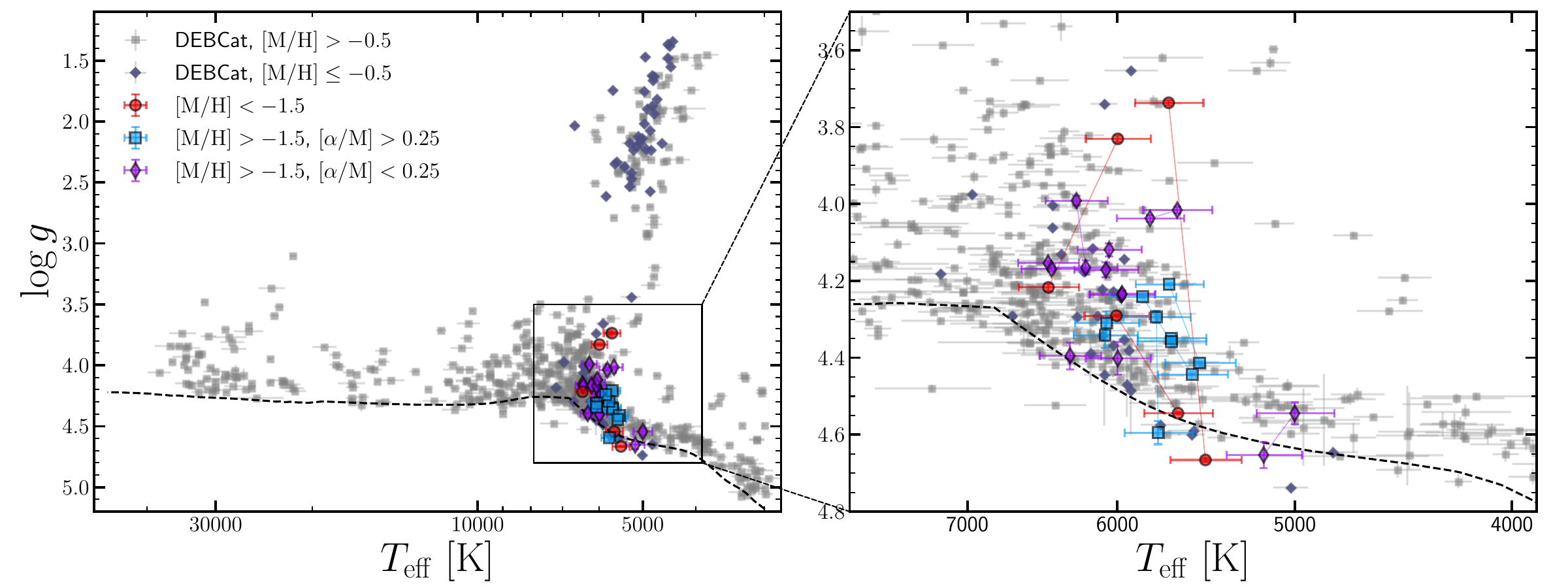}
    \caption{Same as Figure \ref{fig:mass_radius}, but for effective temperature and mass (top), radius (middle), and $\log g$ (bottom). In all cases, the dashed line shows the ZAMS for the Solar metallicity MIST evolutionary tracks.}    \label{fig:debcat}
\end{figure*}

\textbf{J1047$+$4815 (\Gaia{} DR3 832384629268579584)} is a $P=12.9$~day binary. We selected this based on Galactic kinematics, with a predicted $p_{\rm{thick}} = 0.91$ using the \Gaia{} mean RV. This binary has been observed in two \TESS{} sectors, 21 (Feb 2020) and 75 (Feb 2024). In sector 75, J1047$+$4815 was in camera 1, which comes within $25^\circ{}$ of Earth. Approximately 30\% of the \TESS{} data are flagged as poor quality from scattered Earth light, and all of the secondary eclipses are affected. Since sector 21 includes two secondary eclipses but only one primary eclipse, we include both sectors in our analysis. 

This binary is also the only one in our sample with a moderately eccentric orbit, $e=\ecc{J1047+4815}$. The binary is totally eclipsing, with $i=\inclination{J1047+4815}^{\circ}$. We measure masses $M_1=\mass{J1047+4815}{1}\ M_\odot$ and $M_2=\mass{J1047+4815}{2}\ M_\odot$ and radii $R_1=\radius{J1047+4815}{1}\ R_\odot$ and $R_2 = \radius{J1047+4815}{2}\ R_\odot$. The fit of the disentangled component spectra finds a metallicity $\mh{} = \specmet{J1047+4815}$ and $\alpham{}=\specalpha{J1047+4815}$, consistent with the thick disk membership probability $p_{\rm{thick}} = 0.91$. We find $\Delta \chi^2 < 1$ for all theoretical isochrones, with ages of $\age{J1047+4815}{padova}$~Gyr, $\age{J1047+4815}{mist}$~Gyr, and $\age{J1047+4815}{basti}$~Gyr for PARSEC, MIST, and BaSTI, respectively.

\textbf{J1556$+$1218 (\Gaia{} DR3 1191216433749408128)} is a $P=6.3$~day binary discovered in ASAS photometry \citep[ASAS J155629+1218.3,][]{Pojmanski02} and recovered in ASAS-SN \citep{Jayasinghe19, Rowan22}. This is the brightest binary in our sample with $G=11.2$~mag. We selected this target based on the \citet{Andrae23} catalog of metallicities from \Gaia{} XP spectra, which reports $\mh{}=-1.6$. This is a totally eclipsing system (Figure \ref{fig:lc_rv}). Only one sector of data is available, and $\sim 60\%$ of the photometry is contaminated by scattered Earth light. We therefore choose not to bin the \TESS{} data and fit all of the available uncontaminated \TESS{} photometry. The light curve residuals show correlated structure between the eclipses, which could be evidence for rotational variability on one of the stars, contamination from nearby stars, or systematics from scattered light. Given that the scale of these residuals is $<0.5\%$, we do not expect this signal, regardless of origin, to bias the measured masses and radii. We measure masses $M_1 = \mass{J1556+1218}{1}\ M_\odot$ and $M_2 = \mass{J1556+1218}{2}\ M_\odot$ and radii $R_1 = \radius{J1556+1218}{1}\ R_\odot$ and $R_2 = \radius{J1556+1218}{2}\ R_\odot$. The orbital eccentricity, $e=\ecc{J1556+1218}$, is consistent with a circular orbit. 

The disentangled component spectra for this target (Figure \ref{fig:disentangling_example}) are consistent with two stars of nearly the same spectral type and a luminosity ratio $\ell_2=\speclightratio{J1556+1218}$. We measure a metallicity $\mh{}=\specmet{J1556+1218}$ and $\alpham{}=\specalpha{J1556+1218}$. We show the fit to the disentangled component spectra of this target in Figure \ref{fig:disentangle_fit}. The spectrum has only a small number of narrow \ion{Fe}{1}, \ion{Fe}{2}, \ion{Mg}{1}, \ion{Ti}{2}, and \ion{Cr}{1} lines. These results are based on the seven PEPSI spectra. We repeat the disentangling and spectral synthesis analysis using the 11 APF spectra. We measure $\mh{}=-2.04$ and $\alpham{}=0.28$ from the APF observations, which agree with the PEPSI results at $<0.1$~dex (Table \ref{tab:spec_table}). The effective temperatures, projected rotational velocities, and flux ratios are also consistent between the PEPSI and APF fits at $1\sigma$. 

The other two EBs in our sample with $\mh{} < -1.5$ are kinematically consistent with the Milky Way halo. This target instead has kinematics matching the thick disk population (Figure \ref{fig:alpha_iron_toomre}). The Galactic orbit, shown in Figure \ref{fig:kinematic_orbits}, has only small departures from the Galactic midplane with $z_{\rm{max}} \sim 200$~pc, but is on a more eccentric orbit than the Sun. Using the center-of-mass velocity (Table \ref{tab:mcmc_posteriors}), the \Gaia{} parallaxes and proper motions, and the relations from \citet{Ramirez07}, we calculate a thick disk membership probability of $p_{\rm{thick}} = 0.98$. 

This is the second oldest binary in our sample with a PARSEC age $\tau = \age{J1556+1218}{padova}$~Gyr, which is consistent with the ages measured with MIST and BaSTI. Figure \ref{fig:age_example} shows the best-fitting PARSEC isochrone for this target. Even though the primary is low mass, it is near the end of its main sequence lifetime, enabling this precise age constraint. The joint age constraint is dominated by the primary, but the independent age estimate from the secondary agrees within 0.2~dex in $\log\tau$, and the $\Delta\chi^2$ distribution between the independent and joint age fits is strongly peaked at zero. 

\textbf{J1626$-$1140 (\Gaia{} DR3 4331604848802095360)} is a $P=7.5$~day binary selected using \Gaia{} XP spectra and was originally detected by NSVS \citep[NSVS 16380920,][]{Wozniak04} and later recovered in ASAS-SN. We measure a small, but non-zero $e=\ecc{J1626-1140}$, and find $M_1 = \mass{J1626-1140}{1}\ M_\odot$, $M_2 = \mass{J1626-1140}{2}\ M_\odot$, $R_1 = \radius{J1626-1140}{1}\ R_\odot$, and $R_2 = \radius{J1626-1140}{2}\ R_\odot$. From the disentangled component spectra, we measure $\mh{} = \specmet{J1626-1140}$, which is consistent with the $\mh{} = -0.82$ reported in the \citet{Andrae23} \Gaia{} XP metallicity catalog. The binary is somewhat $\alpha$-enhanced, with $\alpham{} = \specalpha{J1626-1140}$, but is kinematically consistent with the thin disk. We estimate coeval ages of $\age{J1626-1140}{padova}$~Gyr, $\age{J1626-1140}{mist}$~Gyr, and $\age{J1626-1140}{basti}$~Gyr, with PARSEC, MIST, and BaSTI isochrones, respectively.

\textbf{J1905$+$6233 (\Gaia{} DR3 2252263532714971136)} is the shortest period binary in the sample with $P=2.5$~days and was originally discovered in ASAS-SN photometry \citep{Jayasinghe19}. This target was selected using the \citet{Andrae23} catalog of \Gaia{} XP metallicities, which reports $\mh{} = -1.03$. We use the APF spectra for disentangling, and find $\mh{} = \specmet{J1905+6233}$ and $\alpham{} = \specalpha{J1905+6233}$. The binary is kinematically consistent with the thin disk, with $p_{\rm{thin}} = 0.99$, but is on a prograde orbit with $V_\phi > V_{\rm{LSR}}$ and guiding center radius $R_{c} = 11$~kpc. 

We find a relatively poor fit to the masses and radii with binary isochrones for PARSEC, MIST, and BaSTI, with median $\Delta \chi^2 \sim 4.2$ between the single-age and independent age fits. This modest disagreement is likely due to rapid rotation. We measure large projected rotational velocities $v \sin i > 25$~km/s for both components. We also find evidence of ellipsoidal variability in the out-of-eclipse light curve (Figure \ref{fig:lc_rv_appendix}). While the {\tt ellc} triaxial ellipsoid model is not as robust as the {\tt PHOEBE} backend in modeling ellipsoidal modulations, we do not expect the tidal distortions to bias our measured masses and radii. Furthermore, the mass ratio and inclination are independently constrained through the RV semi-amplitudes and eclipse depth, respectively. 

\textbf{J1909$+$4025 (\Gaia{} DR3 2100618270538909056)} is an equal-mass binary with $P = 3.48$~days. This is a known EB detected in the Northern Sky Variability Survey \citep[NSVS,][]{Wozniak04} and later recovered in ASAS-SN \citep{Jayasinghe19} and \TESS{} \citep{Prsa22}. We selected this target with Galactic kinematics based on suspected thick disk membership. The Galactic orbit, using the \Gaia{} parallax and proper motion and the measured center-of-mass velocity, has moderate eccentricity $e_{\rm{gal}} = 0.52$ and a maximum $z_{\rm{max}}=1.0$~kpc from the Galactic plane. We measure a metallicity of $\mh{} = \specmet{J1909+4025}$ and $\alpha$-abundance of $\alpham{} = \specalpha{J1909+4025}$, consistent with the thick disk population. The disentangling fit also recovers $\ell_2 = 0.50$, as expected. We measure component masses and radii consistent with each other at the $1\sigma$ level. We estimate an age of $\age{J1909+4025}{padova}$~Gyr using PARSEC, and $\age{J1909+4025}{mist}$~Gyr using MIST and BaSTI. 

\textbf{J1916$+$3828 (\Gaia{} DR3 2099410530033714560)} is in a $P = 7.68$~day orbit and was first reported as an EB in the Czech Variable Star Catalog \citep[CzeV2574,][]{Skarka17} and later recovered by ASAS-SN. We selected this target from the \citet{Andrae23} catalog of data-driven metallicities from \Gaia{} XP spectra, which reports $\mh{} = -2.14$. We measure masses $M_1 = \mass{J1916+3828}{1}\ M_\odot$ and $M_2 = \mass{J1916+3828}{2}\ M_\odot$ and radii $R_1 = \radius{J1916+3828}{1}\ R_\odot$ and $R_2 = \radius{J1916+3828}{2}\ R_\odot$ and measure a non-zero orbital eccentricity $e=\ecc{J1916+3828}$. 

From the disentangled component spectra, we measure a metallicity $\mh{} = \specmet{J1916+3828}$, in good agreement with the \Gaia{} XP metallicity, and $\alpham{} = \specalpha{J1916+3828}$. Unlike the other two systems with $\mh{} < -2.0$, we only disentangle and fit the APF spectra since only two PEPSI epochs are available, neither of which were obtained near RV quadrature. 

Its Galactic orbit is unique among the systems presented here. Figure \ref{fig:kinematic_orbits} shows the integrated orbit using {\tt galpy} with the {\tt MWPotential2014} potential. The binary is on a highly eccentric orbit $e_{\rm{gal}} = 0.98$, passing within $\sim 1$~kpc of the Galactic center. In the Toomre diagram (Figure \ref{fig:alpha_iron_toomre}), this binary is consistent with a halo orbit, and the $\sqrt{V_R^2 + V_Z^2} = 170$~km/s is only surpassed by J0001$+$2156. 

The stars are both $M < 0.8 \ M_\odot$ but we find $R > 1\ R_\odot$, suggesting they come from an old population. The predicted coeval ages are $\age{J1916+3828}{padova}$~Gyr, $\age{J1916+3828}{mist}$~Gyr, and $\age{J1916+3828}{basti}$~Gyr, for PARSEC, MIST, and BaSTI isochrones, respectively, making this the oldest binary in our sample. Out of the three binaries with $\mh{} < -1.5$, we find that this system has the worst agreement between the measured masses and radii and the theoretical isochrones. We compare the $\Delta \chi^2$ distribution between a binary coeval age fit and a model with independent ages for both components. We find $\Delta \chi^2 = \agedeltachisq{J1916+3828}{padova}$, $\Delta \chi^2 = \agedeltachisq{J1916+3828}{mist}$, and $\Delta \chi^2 = \agedeltachisq{J1916+3828}{basti}$ for PARSEC, MIST, and BaSTI, respectively. For all three sets of theoretical isochrones we find $p > 0.1$, which indicates that the measured masses and radii are still consistent with being coeval within the measurement uncertainties. 

This binary is qualitatively similar to HD 140283 \citep[the ``Methuselah'' star,][]{Bond13}, which has an asteroseismic age $\tau = 14.2\pm0.4$~Gyr \citep{Lundkvist25} and is on a Galactic orbit with $e_{\rm{gal}} = 0.92$. That star is also $\alpha$-enhanced and metal-poor with $\feh{} = -2.3$. The old age $\tau = \age{J1916+3828}{padova}$~Gyr, low metallicity $\mh{} = \specmet{J1916+3828}$, $\alpha$-enhancement $\alpham{}=\specalpha{J1916+3828}$, and halo orbit all form a consistent picture of J1916$+$3828 as a binary formed early in the history of the Milky Way.

\textbf{J2352$-$0707 (\Gaia{} DR3 2442415483099264384)} is a $P=7.89$~day binary discovered in ASAS-SN photometry \citep{Jayasinghe19}. We selected this as a metal-poor candidate using Galactic kinematics with thick disk probability $p_{\rm{thick}} = 0.95$. The measured masses and radii are consistent with being a twin equal-mass binary, and we measure a metallicity $\mh{} = \specmet{J2352-0707}$ and $\alpha$-abundance $\alpham{} = \specalpha{J2352-0707}$. The binary is consistent with an age of $\age{J2352-0707}{mist}$~Gyr using MIST isochrones. 

\begin{figure*}
    \centering
    \includegraphics[width=\linewidth]{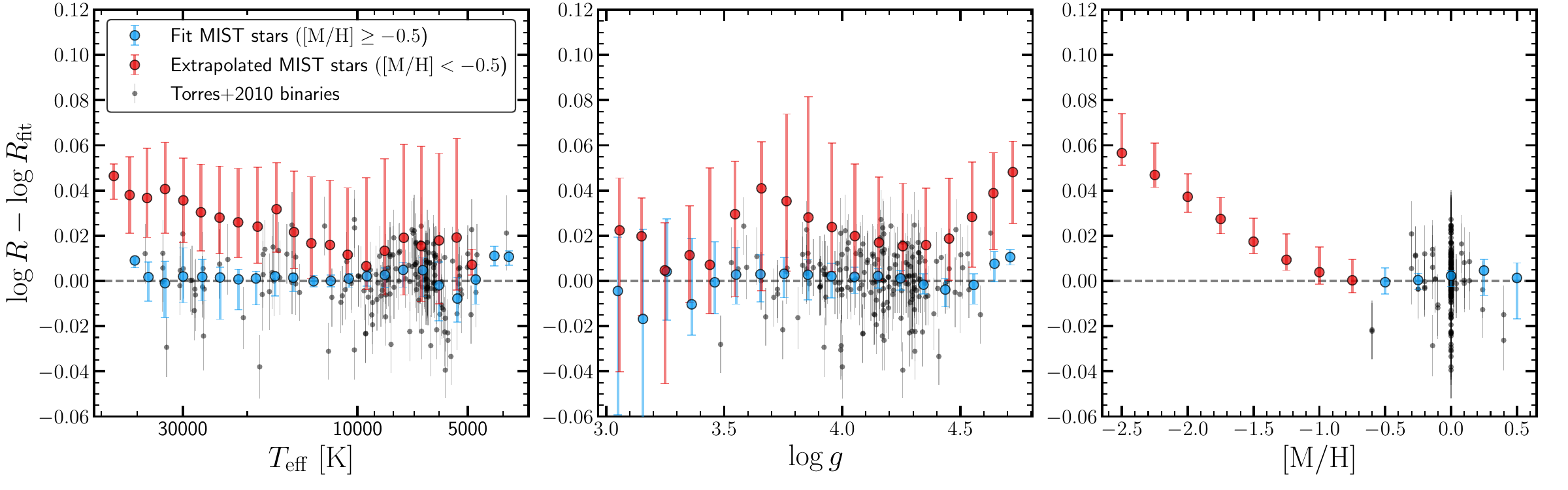}
    \caption{Residuals of the $\log R$ fit to a simulated MIST stellar population. We use the functional form given in Equation \ref{eqn:empirical_torres} from \citetalias{Torres10}. We only fit the stars with metallicities $\mh{} > -0.5$ corresponding to the metallicity range of the EBs used in \citetalias{Torres10}. We find that this model is effective near solar metallicity, but underpredicts the radii of metal-poor stars when extrapolated, motivating the need for benchmark EBs at low metallicity.}
    \label{fig:model_extrapolation}
\end{figure*}

\section{Discussion} \label{sec:discussion}

The 15 binaries characterized here yield masses and radii for 30 stars with $0.66 \leq M/M_\odot \leq 1.12$, $0.62 \leq R/R_\odot \leq 2.03$, and a median fractional uncertainties of 0.78\% and 0.65\% for the masses and radii, respectively. We measure metallicities from the disentangled component spectra and find that 14 have $\mh{} \leq -0.5$, including three systems with $\mh{} < -2.0$, compared to only six Milky Way field binaries with spectroscopic metallicity measurements $\mh{} < -0.5$ that are currently included in DEBCat \citep{Southworth15}\footnote{KIC 4054905 \citep{Brogaard22}; TYC 5227-1023-1 \citep{Traven17}; V432 Aur \citep{Siviero04}; KIC 11285625 \citep{Debosscher13}; CN Lyn \citep{Yucel25}; KIC 10001167 \citep{Thomsen25}}. Of the DEBCat binaries, KIC 10001167 \citep{Thomsen25} is the most metal-poor, with $\mh{} = -0.73$. 

In this section, we compare our measured stellar parameters to other binaries in DEBCat and update the empirical mass and radius relations for main sequence stars from \citet{Torres10}. We also compare our measured metallicities to the estimates from a data-driven model using \Gaia{} XP spectra, interpret the tidal evolution of these systems, and make predictions for the detectability of similar systems in \Gaia{} DR4. 

\subsection{Empirical Relations for Stellar Parameters}

Figures \ref{fig:mass_radius} and \ref{fig:debcat} show the measured masses, radii, effective temperatures, surface gravities, and luminosities of our 15 systems compared to those in DEBCat. At fixed mass, the metal-poor systems we characterize here are hotter and more luminous than the majority of the sample, which is expected for metal-poor stars with lower opacity in their atmospheres.

\citet{Torres10} (hereafter \citetalias{Torres10}) reports fits to stellar mass and radius of the form
\begin{equation} \label{eqn:empirical_torres}
    \begin{split}
        \log M = \Bigl[{}&a_1 + a_2 X + a_3 X^2 + a_4 X^3 + \\
        &a_5 (\log g)^2 + a_6 (\log g)^3 + a_7 \mh{}\Bigr] \\
        \log R = \Bigl[{}&b_1 + b_2 X + b_3 X^2 + b_4 X^3 + \\
        &b_5 (\log g)^2 + b_6 (\log g)^3 + b_7 \mh{}\Bigr],
    \end{split} 
\end{equation}

\noindent where $X \equiv \log T_{\rm{eff}} - 4.1$. They report scatters in these calibrations of $\sigma_{\log M} = 0.027$ and $\sigma_{\log R} = 0.014$. The \citetalias{Torres10} fit excludes stars with $M < 0.6\ M_\odot$ and pre-main-sequence stars, but does include three evolved stars where $\log g < 3.0$. 

The simple polynomials used in \citetalias{Torres10} are effective for main sequence stars with $\mh{} \sim 0$ but become less accurate at low metallicity.  To illustrate this, we generated a stellar population with MIST evolutionary tracks using {\tt BRUTUS} \citep{Speagle25} spanning masses $0.6 < M/M_\odot < 30$, and metallicities $-2.5 < \mh{} < 0.5$ at $\alpham{} = 0$. We use a uniform distribution of ages, which corresponds to a constant star formation rate. We fit the Equation \ref{eqn:empirical_torres} model to only the sample with $\mh{} \geq -0.5$ and extrapolate down to the low-metallicity population. Figure \ref{fig:model_extrapolation} plots the residuals of this model in $\log R$ as a function of effective temperature, surface gravity, and metallicity. For comparison, we also show residuals of the same model fit to the stars included in \citetalias{Torres10} with $\log g > 3$. As expected, the MIST solar-metallicity population and the \citetalias{Torres10} sample are fit well by the relations in Equation \ref{eqn:empirical_torres}. However, when this model is extrapolated to low metallicity, it underpredicts the model radius by as much as $\sim 13\%$, with the errors steadily increasing to lower metallicity.

We update the coefficients of Equation \ref{eqn:empirical_torres} using the stellar parameters reported in DEBCat and those that we measure here. We again exclude stars with $M < 0.6\ M_\odot$ and six pre-main sequence stars\footnote{EK Cep \citep{Popper87}; RS Cha \citep{Clausen80}; V1174 Ori \citep{Stassun04}; NP Per \citep{Lacy16}; V1200 Cen \citep{Marcadon20}; 2MASS J12220147$-$5737565 \citep{Stassun22}}. We also only consider stars with $\log g > 3.0$, since evolved stars do not follow a defined relation between mass and effective temperature, surface gravity, and metallicity. Only half of the binaries in DEBCat have a measured metallicity. Following \citetalias{Torres10}, we assume that those systems without reported metallicities have $\mh{} = 0$, but we find that excluding these systems does not significantly change the fit coefficients. 

Some of the EBs in DEBCat have very small uncertainties on $\log M$ and $\log R$. For example, FM~Leo has reported primary mass and radius uncertainties $\sigma_{\log M} = 0.0002$ and $\sigma_{\log R} = 0.0006$ \citep{Helminiak21}. Since a small number of such binaries would then dominate the weighted least squares fit, we also fit for an intrinsic scatter, $\sigma_{\rm{int}}$, that is added in quadrature to the measured uncertainty, $\sigma_{\rm{meas}}$, to get the uncertainty used in the fit, $\sigma^2 = \sigma_{\rm{meas}}^2 + \sigma_{\rm{int}}^2$. Table \ref{tab:empirical_relation} reports the RMS scatters in $\log R$ and $\log M$ for both the \citetalias{Torres10} and newly optimized coefficients, along with the values of the new coefficients and the estimates of the intrinsic scatter, $\sigma_{\rm{int}}$.

\begin{figure*}
    \centering
    \includegraphics[width=\linewidth]{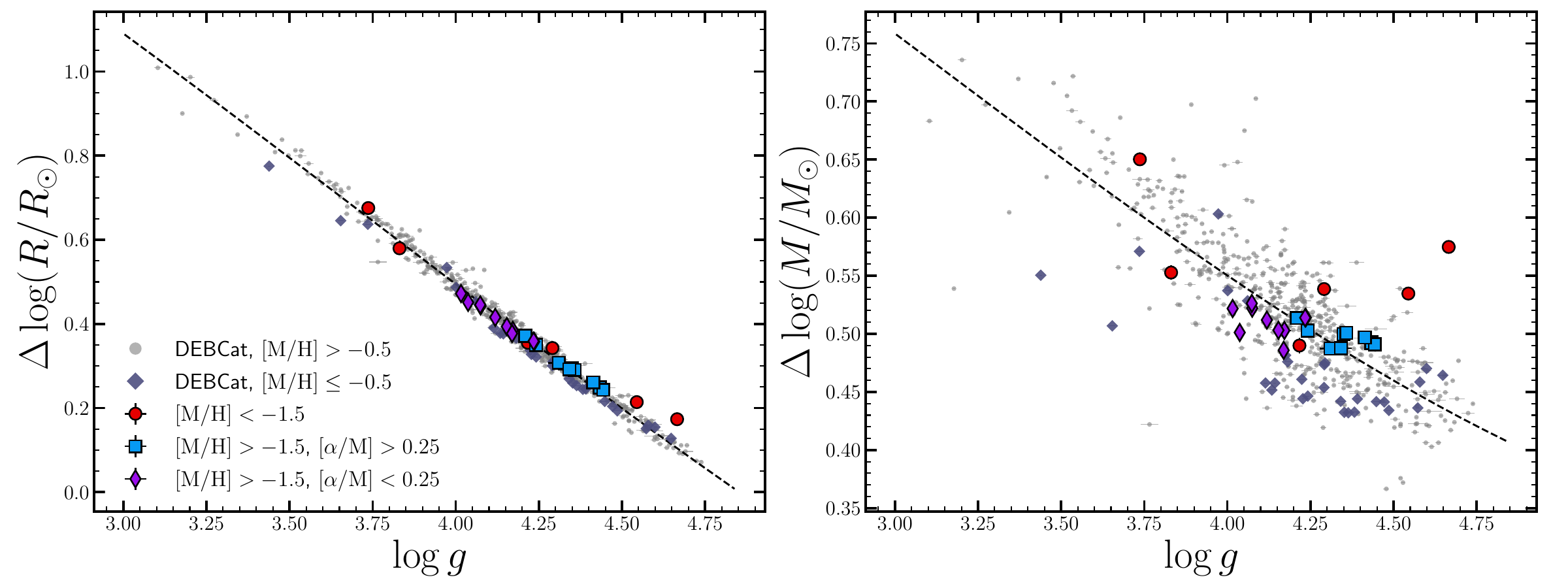}
    \caption{Empirical relations for $\log R$ (left) and $\log M$ (right) fit using Equation \ref{eqn:empirical_torres} with the effective temperature and metallicity contributions removed. The dashed curves show the remaining $\log g$ dependence of the relations, and the scatter is the residual of the fit.} 
    \label{fig:empirical_relations}
\end{figure*}

\begin{figure*}
    \centering
    \includegraphics[width=\linewidth]{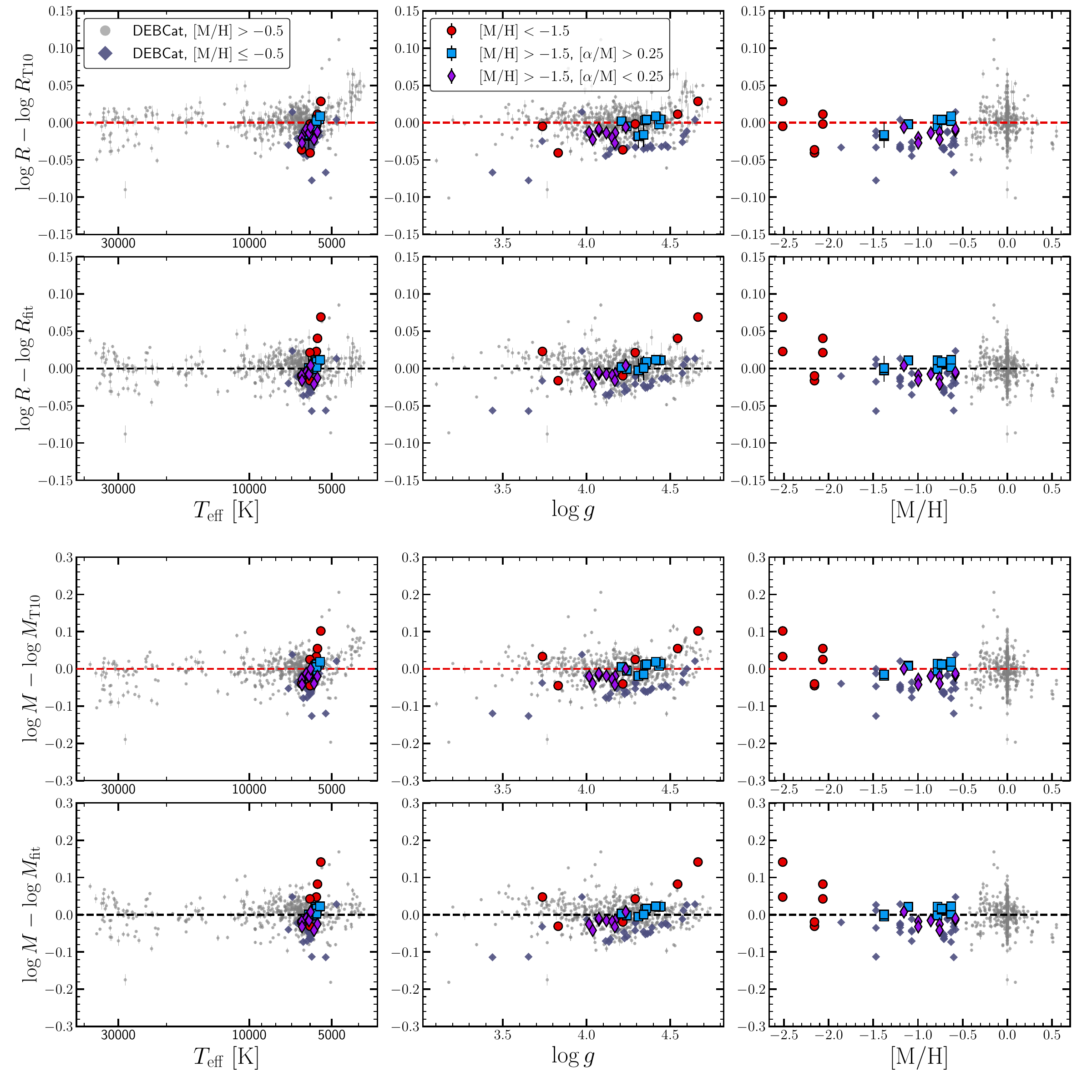}
    \caption{Residuals of the empirical relations for $\log R$ and $\log M$. The top two panels show the $\log R$ residuals using the coefficients from \citetalias{Torres10} and those we report in Table \ref{tab:empirical_relation}, respectively. The bottom two panels show the same but for $\log M$. The DEBCat targets are shown in gray and dark blue, and the binaries we characterize here are colored by their metallicity and $\alpha$-abundance. The spike in the residuals as a function of $\mh{}$ comes from systems without a measured metallicity where we assume $\mh{} = 0$.}
    \label{fig:empirical_relations_residuals}
\end{figure*}

Figure \ref{fig:empirical_relations} illustrates how well the new  empirical relations work in terms of the $\log g$ dependence of the mass and radius after removing the temperature and metallicity contributions and Figure \ref{fig:empirical_relations_residuals} shows the residuals of both models as a function of $T_{\rm{eff}}$, $\log g$ and $\mh{}$.  Not surprisingly, the new models do better at low metallicity. The overall scatters also improve, from $\sigma_{\log M} = 0.0385$ and $\sigma_{\log R} = 0.0203$ for the \citetalias{Torres10} coefficients when applied to the updated sample of 621 stars in EBs to $\sigma_{\log M} = 0.0357$ and $\sigma_{\log R} = 0.0177$ with the new fits. The dispersions of the new fits are modestly larger than those for the \citetalias{Torres10} model applied to their sample of 177 stars ($\sigma_{\log M} = 0.027$ and $\sigma_{\log R} = 0.014$).  This likely implies the need for a more complicated model given the large increase in the range of metallicities. When applied to the Sun, the \citetalias{Torres10} coefficients give $M = 1.051\ M_\odot$ and $R = 1.018\ R_\odot$. With our updated coefficients, we find $M = 1.036\ M_\odot$ and $R = 1.017\ R_\odot$. 

Figure \ref{fig:empirical_relations_residuals} shows that using the updated coefficients removes the correlated $T_{\rm{eff}}$ residuals at $T_{\rm{eff}} \lesssim  5000$~K. Metal-poor stars also show negative $\log R$ residuals using the \citetalias{Torres10} coefficients. We find a $\sigma_{\log R} = 0.0276$ for stars with $\mh{} < -0.5$ using the \citetalias{Torres10} coefficients, which improves to $\sigma_{\log R} = 0.0225$ using the coefficients we report in Table \ref{tab:empirical_relation}, a 19\% reduction. However, for the most metal-poor systems, the residuals get worse with our updated coefficients, which is also
likely evidence that the increased metallicity range requires a more complex model.

\renewcommand{\tablecaptiontext}{Fit coefficients of the empirical relation to $\log M$ and $\log R$ using effective temperature, surface gravity, and metallicity. We give the values reported in \citetalias{Torres10}, which were based on a sample of 177 stars with $\mh{} \gtrsim -0.5$, and the updated coefficients we find here using 621 stars. $\sigma_{\rm{int}}$ is the intrinsic scatter, $N_{\rm{fit}}$ is the number of stars used in the fit and $\sigma$ is the root-mean-squared scatter for the full set of 621 stars.\label{tab:empirical_relation}}
\begin{deluxetable*}{cccccc}[t]
\tablecaption{\tablecaptiontext}
\tablehead{
\colhead{$i$} & \colhead{term} & \colhead{$a_i$ $(\log M)$} & \colhead{$b_i$ $(\log R)$} & \colhead{$a_i$ $(\log M)$} & \colhead{$b_i$ $(\log R)$} \\
\colhead{} & \colhead{} & \colhead{Torres+2010} & \colhead{Torres+2010} & \colhead{This work} & \colhead{This work}
}
\startdata
1 & $1$ & $1.5689 \pm 0.058$ & $2.4427 \pm 0.038$ & $1.185 \pm 0.10$ & $2.1510 \pm 0.050$ \\
2 & $X$ & $1.3787 \pm 0.029$ & $0.6679 \pm 0.016$ & $1.4644 \pm 0.018$ & $0.73109 \pm 0.0091$ \\
3 & $X^2$ & $0.4243 \pm 0.029$ & $0.1771 \pm 0.027$ & $0.4220 \pm 0.035$ & $0.2088 \pm 0.018$ \\
4 & $X^3$ & $1.139 \pm 0.24$ & $0.705 \pm 0.13$ & $0.183 \pm 0.13$ & $0.1048 \pm 0.067$ \\
5 & $(\log g)^2$ & $-0.1425 \pm 0.011$ & $-0.21415 \pm 0.0075$ & $-0.0706 \pm 0.018$ & $-0.16073 \pm 0.0092$ \\
6 & $(\log g)^3$ & $0.01969 \pm 0.0019$ & $0.02306 \pm 0.0013$ & $0.00773 \pm 0.0030$ & $0.01430 \pm 0.0015$ \\
7 & ${\rm [M/H]}$ & $0.1010 \pm 0.014$ & $0.04173 \pm 0.0082$ & $0.10896 \pm 0.0040$ & $0.05424 \pm 0.0020$ \\
\hline
 & $\sigma_{\rm int}$ & \nodata & \nodata & $0.0351$ & $0.0173$ \\
 & $\sigma$ & $0.0385$ & $0.0203$ & $0.0357$ & $0.0177$ \\
 & $N_{\rm fit}$ & $177$ & $177$ & $621$ & $621$ \\
\enddata
\end{deluxetable*}

\subsection{Comparison with Gaia XP Metallicities}

\begin{figure}[t]
    \centering
    \includegraphics[width=\linewidth]{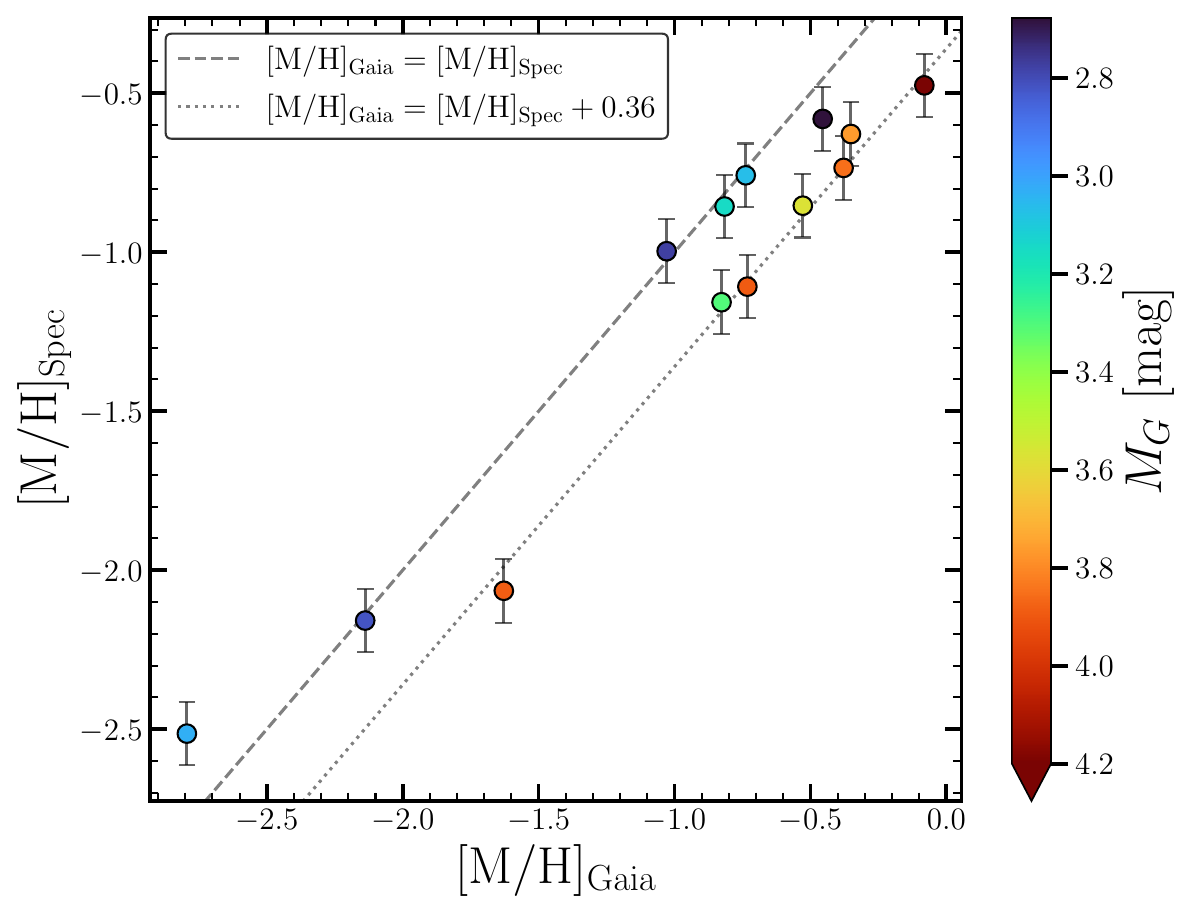}
    \caption{Comparison between the metallicities estimated from the \Gaia{} XP spectra \citep{Andrae23} and the metallicities derived here from the high-resolution spectra. The dashed line shows where the two agree, and the dotted line is $\mh{}_{\rm{Gaia}} - \mh{}_{\rm{spec}} = 0.36$. We find that the sample is split by absolute magnitude $M_G$, with more luminous systems having better agreement between the metallicity predicted by the XP spectra and the metallicity we measure from the high-resolution spectra.}
    \label{fig:delta_xp}
\end{figure}

Out of the 15 binaries investigated here, 13 have metallicities reported from the \Gaia{} XP spectra in \citet{Andrae23}. Figure \ref{fig:delta_xp} compares the metallicities we measure from the high-resolution spectra, $\mh{}_{\rm{spec}}$, and the predictions from the \Gaia{} XP catalog $\mh{}_{\rm{Gaia}}$. We find a bimodal distribution in the difference $\mh{}_{\rm{Gaia}} - \mh{}_{\rm{spec}}$, with approximately half of the systems having $\mh{}_{\rm{Gaia}} - \mh{}_{\rm{spec}} \approx 0.36$~dex. We find that these two groups are separated in absolute magnitude, $M_G$. We calculate absolute magnitudes using distances from \citet{BailerJones21} and extinctions from {\tt mwdust} \citep{Bovy16}. Binaries with $M_G < 3.2$ have mean $\mh{}_{\rm{Gaia}} - \mh{}_{\rm{spec}} = 0.04$~dex with scatter $0.06$~dex, excluding the very metal-poor system J0001$+$2156. The binaries with $M_G > 3.2$ instead have $\mh{}_{\rm{Gaia}} - \mh{}_{\rm{spec}} = 0.36 \pm 0.05$~dex. The two systems without \Gaia{} XP metallicities, J0350$+$0805 and J2352$-$0707, both have $M_G > 3.2$. 

This bimodal distribution of metallicity differences could be a result of the training sample. \citet{Andrae23} use a training sample selected from APOGEE DR17 \citep{Abdurrouf22} and the sample of metal-poor stars from \citet{Li22}. The training sample is particularly sparse for dwarfs with $\mh{} < -1$, though many of the systems with overpredicted \Gaia{} XP metallicities have $-1 < \mh{} < -0.5$. The limitations of the \Gaia{} XP metallicity catalog at the edges of the training set have been seen in other samples like the metal-rich stars and cool red giants observed in \citet{Saad26}. The \citet{Andrae23} model also uses the absolute magnitude as an input feature. Since many of our binaries are near-twins, $M_G$ is $\sim 0.7$~mag brighter than it would be for a single star. However, we do not observe a trend in $\mh{}_{\rm{Gaia}} - \mh{}_{\rm{spec}}$ with the spectroscopic luminosity ratio. Despite this offset, the \Gaia{} XP spectra remain useful for identifying metal-poor EB candidates. 

\subsection{Synchronization and Circularization}

\begin{figure}[!t]
    \centering
    \includegraphics[width=\linewidth]{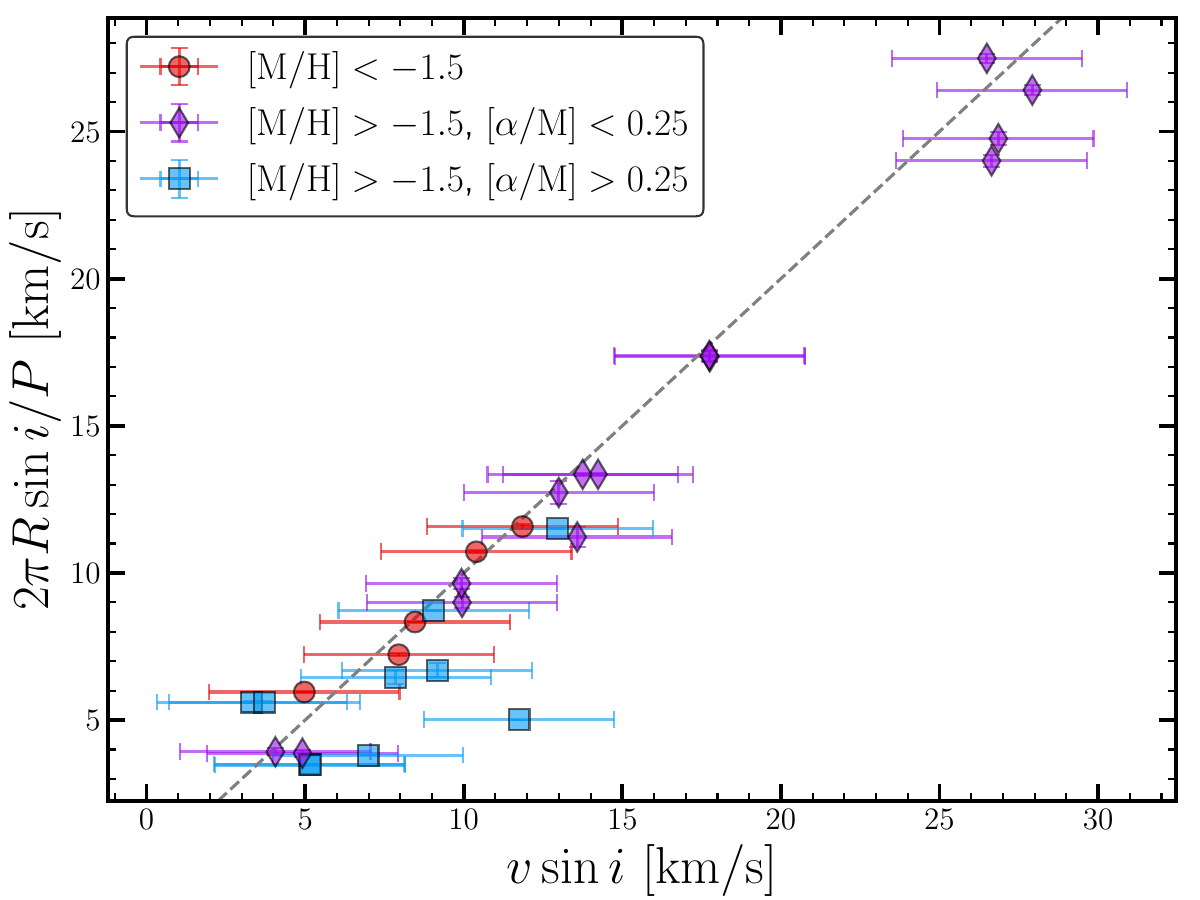}
    \caption{Comparison of the projected rotational velocity assuming tidal synchronization, $2\pi R \sin i /P$, to the measured spectroscopic $v \sin i$. The gray line shows $2\pi R \sin i /P = v \sin i$. We find all circular systems are consistent with being tidally synchronized and that the eccentric system J1047$+$4815 is pseudo-synchronized.}
    \label{fig:vsini}
\end{figure}

Since the binaries have short orbital periods, they are most likely tidally synchronized. Figure \ref{fig:vsini} compares the measured projected rotational velocities from the disentangled component spectra to the synchronous projected rotational velocity $v_{\rm{sync}} = 2\pi R \sin i /P$, which is determined from the joint light curve and RV orbit fit. This assumes that the binaries are in spin-orbit alignment, which may not always be the case for short-period binaries \citep[e.g.,][]{Wells25}. Future Rossiter-McLaughlin measurements could be used to directly measure the projected spin-orbit obliquity \citep[e.g.,][]{Albrecht07}. Assuming spin-orbit alignment, we find that all but one of the binaries are consistent with being tidally locked. 

The exception is J1047$+$4815, which has $e=\ecc{J1047+4815}$, the largest eccentricity in the sample. Using $v_{\rm{sync}} = 2\pi R \sin i /P$ assumes $e=0$, so for J1047$+$4815 we instead calculate the expected projected rotational velocity of a pseudo-synchronous orbit as
\begin{equation}
    v_{\rm{ps}} = v_{\rm{sync}} \frac{1+\frac{15}{2}e^2 + \frac{45}{8}e^4 + \frac{5}{16}e^6}{(1+3e^2+\frac{3}{8}e^4)(1-e^2)^{3/2}},
\end{equation}
\noindent from \citet{Hut81}. For J1047$+$4815, we find $v_{\rm{ps},1}=11.1$~km/s for the primary and $v_{\rm{ps},2}=8.5$~km/s for the secondary. These are consistent with our measured projected rotational velocities from the disentangled spectra (Table \ref{tab:spec_table}), so the components of J1047$+$4815 are likely pseudo-synchronized. 

We approximate the tidal synchronization and circularization timescales for stars with convective envelopes as:
\begin{equation}
    t_{\rm{sync}} \approx 10^4 \left[\frac{1+q}{2q}\right]^2 P^4\ \rm{years},
\end{equation}
\begin{equation}
    t_{\rm{circ}} \approx 10^6 q^{-1} \left[\frac{1+q}{2}\right]^{5/3} P^{16/3}\ \rm{years},
\end{equation}
\noindent where the orbital period $P$ is in days \citep{Zahn77, Zahn89}. We calculate the synchronization and circularization timescales for both components. We find that all systems have synchronization timescales $< 1$~Gyr, far shorter than all of our age estimates (Table \ref{tab:age_table}). J0807$+$6945, which is also the longest-period binary in the sample, has the longest synchronization timescale of the sample, $t_{\rm{sync}}\approx 530$~Myr. The agreement between the spectroscopic $v \sin i$ and the predicted $v \sin i$ from tidal synchronization then gives an additional cross-check on our disentangling and spectral synthesis results. Based on Figure \ref{fig:vsini}, it is likely that our assumed $\sigma_{v \sin i} = 3$~km/s is overestimated. 

%We take the circularization time of the binary as $1/t_{\rm{circ}} = 1/t_{\rm{circ},1} + 1/t_{\rm{circ},2}$. 

We take the circularization time of the binary as the shorter of the two component circularization times \citep[e.g.,][]{Sethi26}. We find five binaries have $t_{\rm{circ}} < 13.8$~Gyr. Of these five, the largest eccentricity is $e=\ecc{J0727+2018}$ for J0727$+$2018. The eccentric system J1047$+$4815 has a long circularization time $t_{\rm{circ}} \approx 830$~Gyr. This is surpassed only by the $P=15.2$~day binary J0807$+$6945, which has $t_{\rm{circ}} = 2000$~Gyr, approximately 140 times the age of the Universe. Despite this long circularization timescale, we measure $e=\ecc{J0807+6945}$, consistent with zero, suggesting that the low eccentricity reflects the system's eccentricity at formation. This system is similar to those in the ``cold-core'' of circular binaries with $10 \lesssim P \lesssim 20$~days in \citet{Zanazzi22}.

\subsection{Looking ahead to Gaia DR4}

\begin{figure*}
    \centering
    \includegraphics[width=\linewidth]{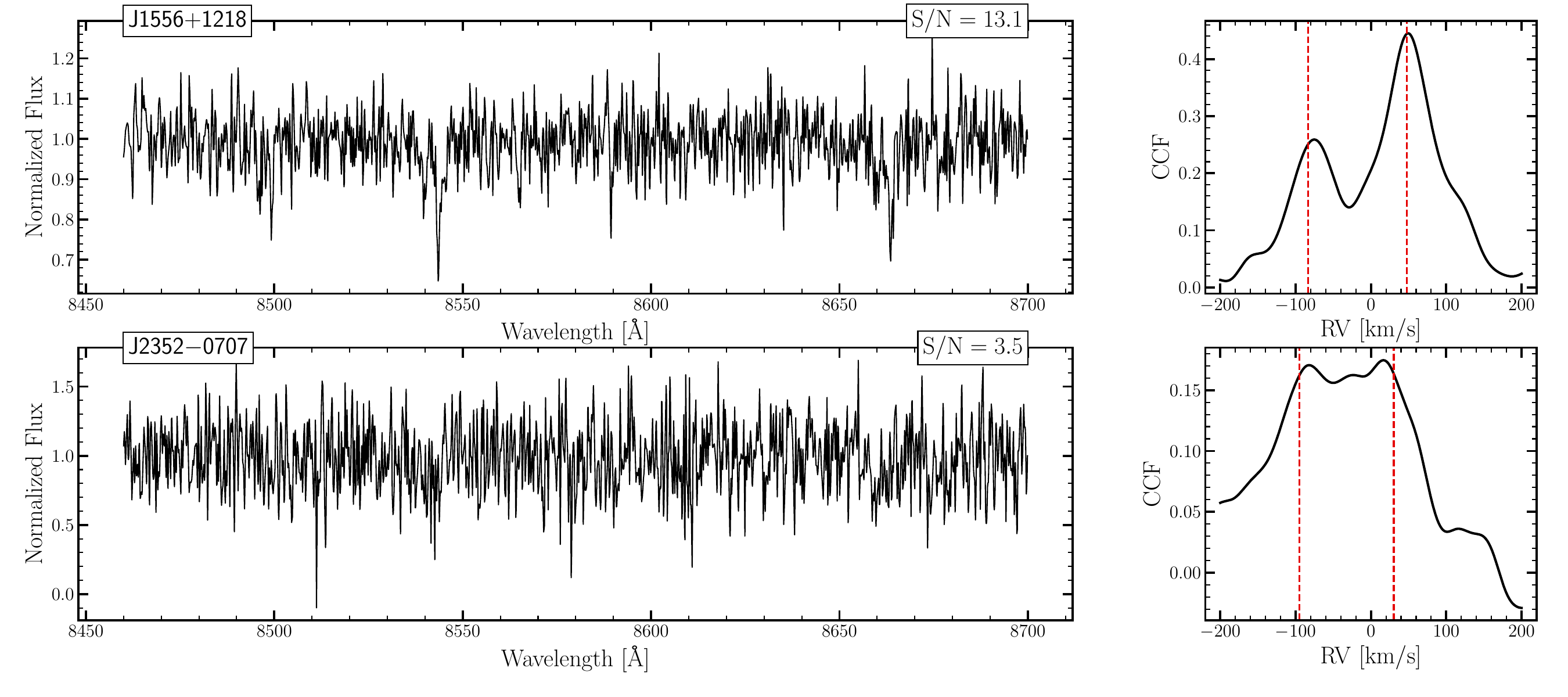}
    \caption{Simulated \Gaia{} RVS spectra for two binaries, chosen to have the highest and lowest predicted \Gaia{} RVS signal-to-noise in the sample. The left panels show the simulated spectra at RV quadrature. The right panel shows a cross-correlation function (CCF) using a template matching the photometric primary. The vertical red lines show the true RVs at the simulated epoch.}
    \label{fig:rvs_sim}
\end{figure*}

The sample of detached EBs with precise mass and radius measurements is primarily limited by resources for spectroscopic follow-up. Measuring RV orbits requires $\gtrsim 10$ multi-epoch high-resolution spectra. This landscape will change with the upcoming \Gaia{} DR4, which will include Radial Velocity Spectrometer \citep[RVS,][]{Cropper18} spectra taken during the full nominal mission (July 2014--January 2020). While the \Gaia{} RVS resolution is only $R\approx 11{,}500$, and the spectra cover just the calcium triplet region in the near-IR, radial velocities can be successfully derived for double-lined binaries. \Gaia{} DR3 included $\sim 5{,}000$ orbits for SB2s \citep{Arenou23}, many of which were also eclipsing binaries. \citet{Rowan23} used these published orbits alongside ASAS-SN light curves to measure masses and radii of 61 binaries with fractional uncertainties of $\sim 5$--10\%. 

In DR4, most targets will have $\sim 30$--40 RV transits. The primary challenge is that the short exposure times (4.4s) result in low spectral signal-to-noise in all but the brightest EBs. All of the targets in the \Gaia{} DR3 SB2 catalog have $G_{\rm{RVS}} < 10.7$, whereas the binaries in this sample all have $G_{\rm{RVS}} > 10.9$. We estimate that the EBs we characterize here will have \Gaia{} RVS spectra with $3.5 \leq S/N \leq 13.1$ based on the RVS performance\footnote{\url{https://www.cosmos.esa.int/web/gaia/science-performance}} in \Gaia{} DR3 \citep{Katz23}. Figure \ref{fig:rvs_sim} shows simulated \Gaia{} RVS spectra for the binaries with the highest and lowest predicted signal-to-noise. While precise RV determination will be challenging, especially for metal-poor stars like those presented here, RVs with $\sigma_{\rm{RV}} \sim 5$~km/s are still possible. Even measuring velocity semi-amplitudes with precisions of $\sigma_K \approx 3$~km/s can produce mass measurements with fractional uncertainties of $\lesssim 10\%$ for twin binaries with $P \sim 5$--10~days. Although this is an order of magnitude greater than the uncertainties that can be measured from dedicated spectroscopic follow-up, expanding the sample of benchmark detached EBs will allow for comparisons with stellar models to be done at scale, instead of on a case-by-case basis. 

\section{Conclusions} \label{sec:conclusions}

The evolutionary pathway of a star depends on its initial mass and chemical composition. Eclipsing binaries have long served as fundamental benchmarks for stellar models because they allow for a direct dynamical measurement of stellar masses. Here, we target low-metallicity eclipsing binaries as part of an effort to expand the sample of benchmark mass and radius measurements to the metal-poor regime. 

We identify metal-poor eclipsing binaries starting from large spectroscopic surveys, \Gaia{} XP spectra, and Galactic kinematics. We visually inspect \TESS{} light curves to identify detached eclipsing binaries. We use three spectrographs to monitor the RV orbits of 15 main-sequence binaries, and measure RVs using TODCOR (Figure \ref{fig:todcor_example}). We then jointly fit the \TESS{} light curves and the measured RVs with \PHOEBE{} (Figure \ref{fig:lc_rv}). 

As expected for a metal-poor population, many of our binaries are more evolved than those in existing catalogs of detached EBs (Figure \ref{fig:mass_radius}). We then use spectral disentangling to extract the component spectra (Figure \ref{fig:disentangling_example}), which we fit using spectral synthesis (Figure \ref{fig:disentangle_fit}) to measure metallicity and $\alpha$-abundance. We find three binaries with $\mh{} < -2.0$, all of which are $\alpha$-enhanced. The most metal-poor system, J0001$+$2156, has $\mh{} = \specmet{J0001+2156}$, making it the most metal-poor eclipsing binary with direct mass and radius measurements to date. Only one of the systems considered here has $\mh{} > -0.5$, J0345$-$0407, where we measure $\mh{} = \specmet{J0345-0407}$.

We find the degree of $\alpha$-enhancement generally correlates with Galactic kinematics: the stars with $\alpham{} > 0.25$ are all kinematically consistent with either the thick disk or halo, while most of the stars with $\alpham{} < 0.25$ are consistent with being part of the thin disk (Figure \ref{fig:alpha_iron_toomre}). 

We compare our measured masses and radii to stellar models with theoretical isochrones from PARSEC, MIST, and BaSTI. We estimate ages for both components independently and for a joint fit to both stars in the mass-radius plane, and use the $\Delta \chi^2$ to quantify the level of agreement with the models (Figure \ref{fig:age_example}). In most cases, we find that the predicted ages for both stars agree at the $1\sigma$ level. Unsurprisingly, many of our binaries are old, with the three most metal-poor systems predicted to have ages of $\age{J0001+2156}{padova}$, $\age{J1556+1218}{padova}$, and $\age{J1916+3828}{padova}$~Gyr from PARSEC isochrones. While these are not the oldest stars with precise age measurements, these systems do rival those characterized through other techniques, such as asteroseismology \citep[e.g.,][]{Lundkvist25}.

Empirical relations can be used to estimate stellar masses and radii from spectroscopic parameters. \citet{Torres10} presented empirical relations for $\log M$ and $\log R$ based on a sample of 177 stars in detached EBs. We update these relations to include binaries in DEBCat and the binaries we characterize here (Figure \ref{fig:empirical_relations}). We find that the coefficients from \citet{Torres10} give a scatter of $\sigma_{\log R} = 0.0203$ and $\sigma_{\log M} = 0.0385$ when applied to the DEBCat binaries and our systems. We update the coefficients (Table \ref{tab:empirical_relation}) and find an improved root-mean-squared scatter $\sigma_{\log R} = 0.0177$ and $\sigma_{\log M} = 0.0357$.

We use our measured metallicities to test the predictions from the \Gaia{} XP spectra (Figure \ref{fig:delta_xp}). We find that the agreement correlates with absolute magnitude, which could indicate a gap in the training sample in the \citet{Andrae23} catalog for metal-poor dwarfs. We also compare our measured projected rotational velocities to the expected rotation velocities if the binaries are tidally synchronized, finding that all of our circular binaries are synchronized and the eccentric binary J1047$+$4815 is pseudo-synchronized (Figure \ref{fig:vsini}). Finally, we look ahead to \Gaia{} DR4. While measuring precise SB2 RVs will be challenging with the low signal-to-noise RVS spectra, careful treatment could enable more than an order of magnitude increase in the sample of benchmark detached eclipsing binaries with mass and radius measurements.

\begin{acknowledgments}

We thank Jessica Lu, Jackie Blaum, Dan Weisz, Rhys Seeburger, and Todd Thompson for helpful discussions. DMR is supported by NASA Hubble Fellowship grant HST-HF2-51588.001-A awarded by the Space Telescope Science Institute, which is operated by the Association of Universities for Research in Astronomy, Inc., for NASA, under contract NAS5-26555. KZS and CSK are partially supported by NSF grants AST-2307385 and AST-2407206.

The LBT is an international collaboration among institutions in the United States and Europe. At the time data were acquired for this research, LBT Corporation Members were the University of Arizona on behalf of the Arizona Board of Regents; Istituto Nazionale di Astrofisica, Italy; and The Ohio State University, representing The Ohio State University, University of Notre Dame, University of Minnesota, and University of Virginia.  This research used the facilities of the Italian Center for Astronomical Archives (IA2) operated by INAF at the Astronomical Observatory of Trieste.  Observations have benefited from the use of ALTA Center (alta.arcetri.inaf.it) forecasts performed with the Astro-Meso-Nh model. Initialization data of the ALTA automatic forecast system come from the General Circulation Model (HRES) of the European Centre for Medium Range Weather Forecasts.

This work also made use of observations obtained with the Automated Planet Finder at Lick Observatory, operated by the University of California Observatories.

This research has used data from the CHIRON spectrograph on the
CTIO/SMARTS 1.5-m telescope, which is operated as part of the SMARTS
Consortium. We thank Todd Henry for his help and support over the past several years.

This work has made use of data from the European Space Agency (ESA) mission \textit{Gaia} (\url{https://www.cosmos.esa.int/gaia}), processed by the \textit{Gaia} Data Processing and Analysis Consortium (DPAC, \url{https://www.cosmos.esa.int/web/gaia/dpac/consortium}). Funding for the DPAC has been provided by national institutions, in particular the institutions participating in the \textit{Gaia} Multilateral Agreement.

\end{acknowledgments}

\appendix

\section{Light curve and RV orbit fits for all targets} 

Figure \ref{fig:lc_rv_appendix} shows the light curve, RV orbit models, and model residuals for all targets. J1556$+$1218 is shown in Figure \ref{fig:lc_rv}. For each target, the phase-folded light curve is shown on the left and the RVs on the right. The smaller bottom panels show the residuals.

\label{sec:appendix_lc_rvs}

\begin{figure*}
    \centering
    \includegraphics[trim={2cm 0 0 0},clip,width=\linewidth]{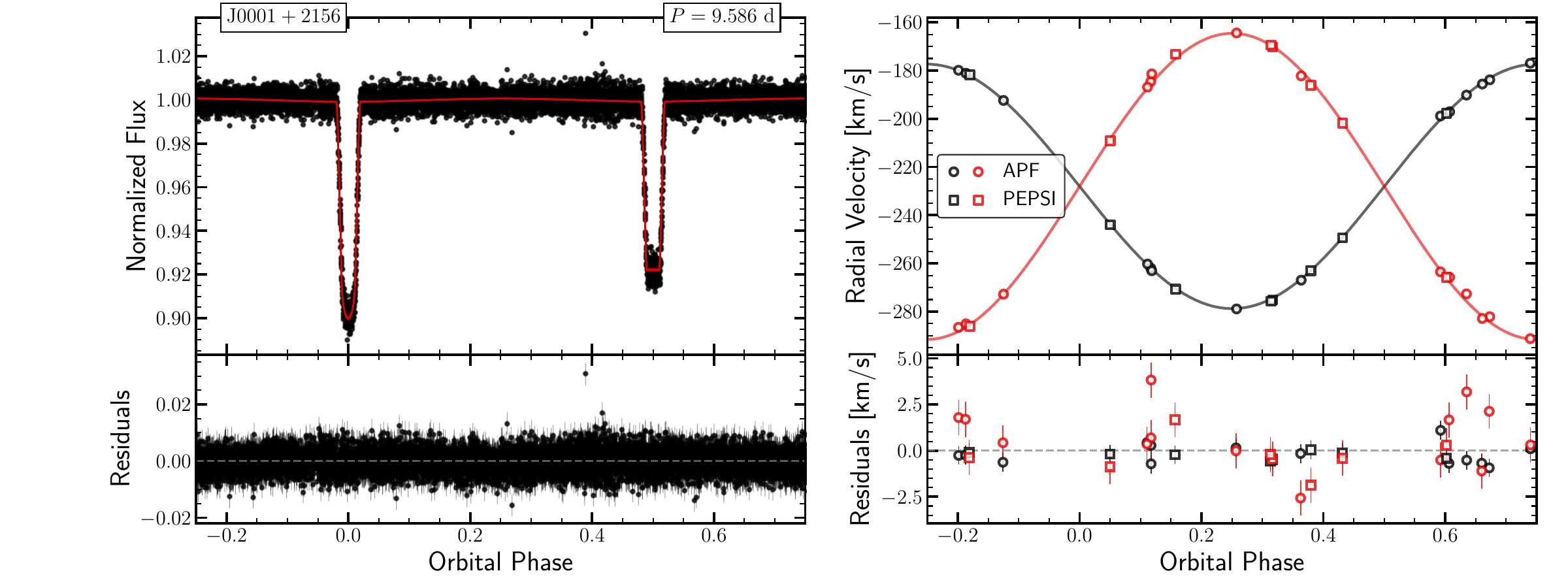}
    \includegraphics[trim={2cm 0 0 0},clip,width=\linewidth]{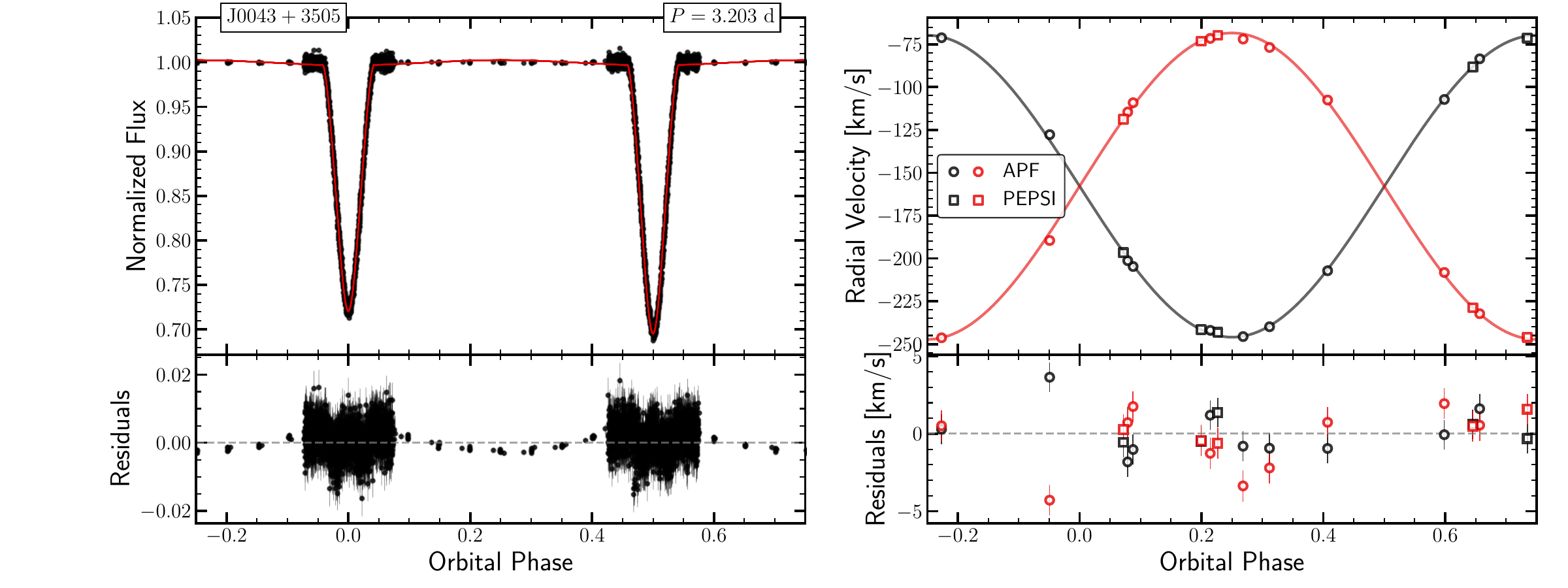}
    \includegraphics[trim={2cm 0 0 0},clip,width=\linewidth]{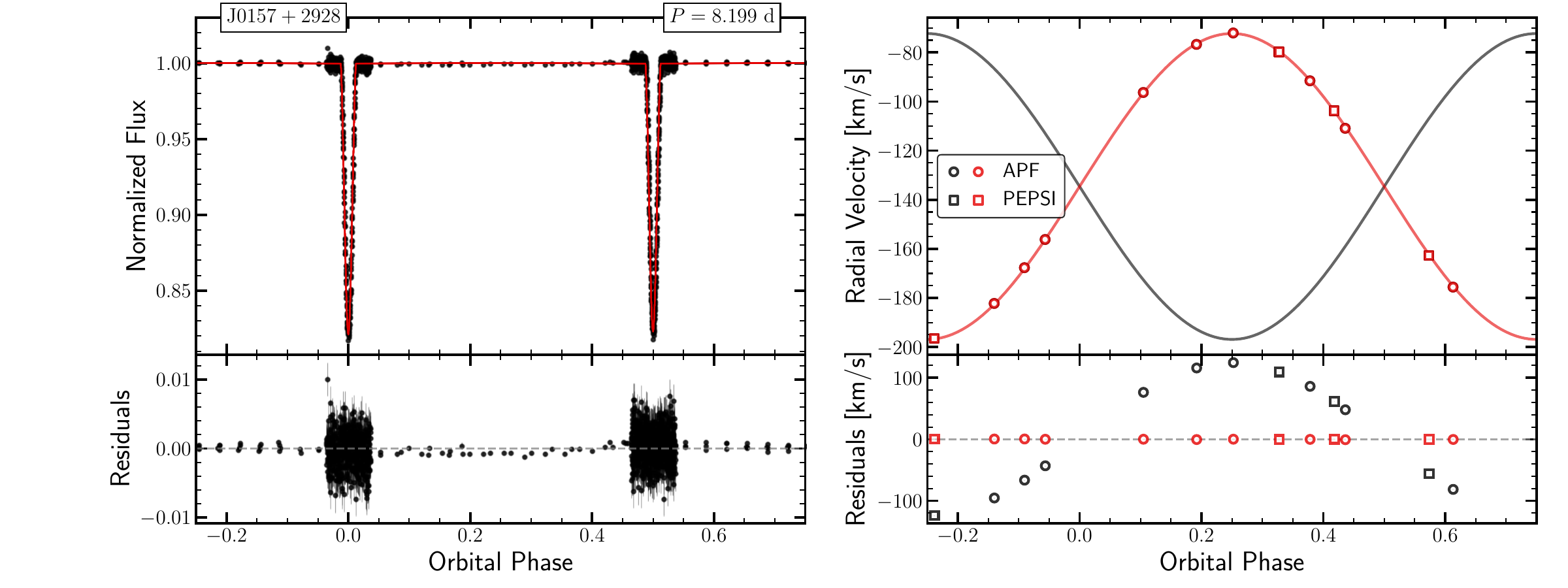}
    \caption{Light curve and RVs for all targets.}
    \label{fig:lc_rv_appendix}
\end{figure*}

\begin{figure*}
    \ContinuedFloat
    \centering
    \includegraphics[trim={2cm 0 0 0},clip,width=\linewidth]{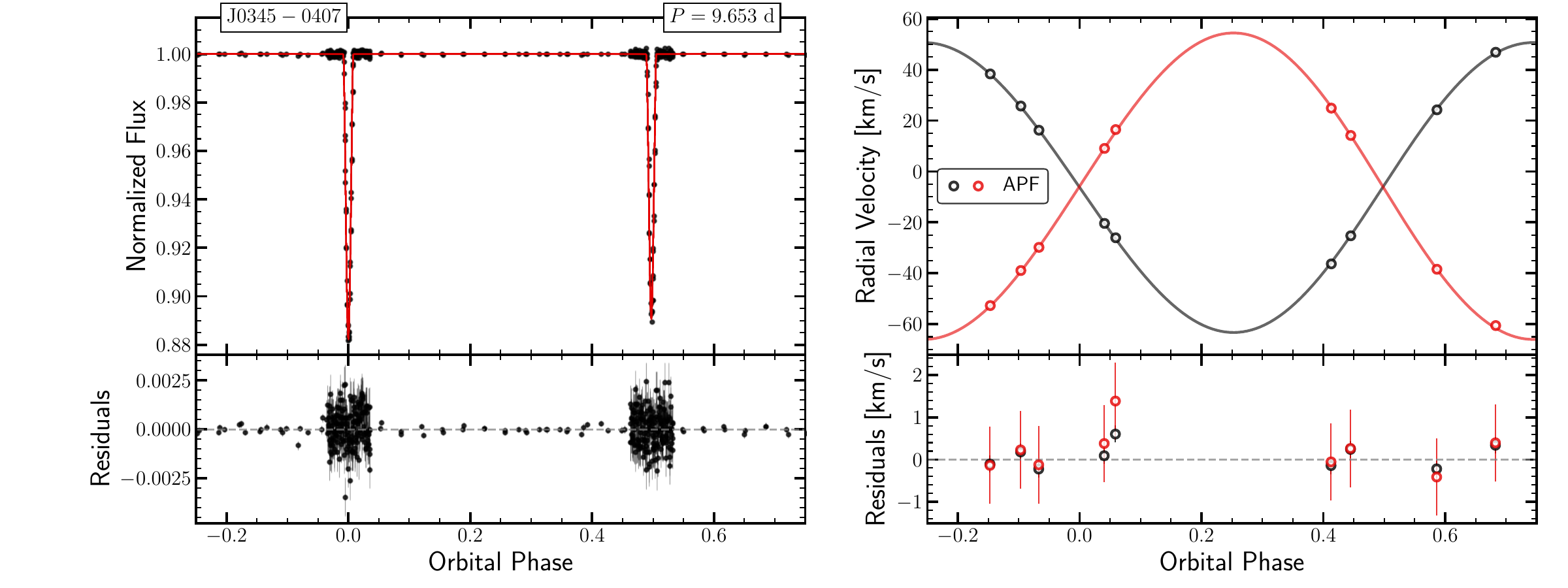}
    \includegraphics[trim={2cm 0 0 0},clip,width=\linewidth]{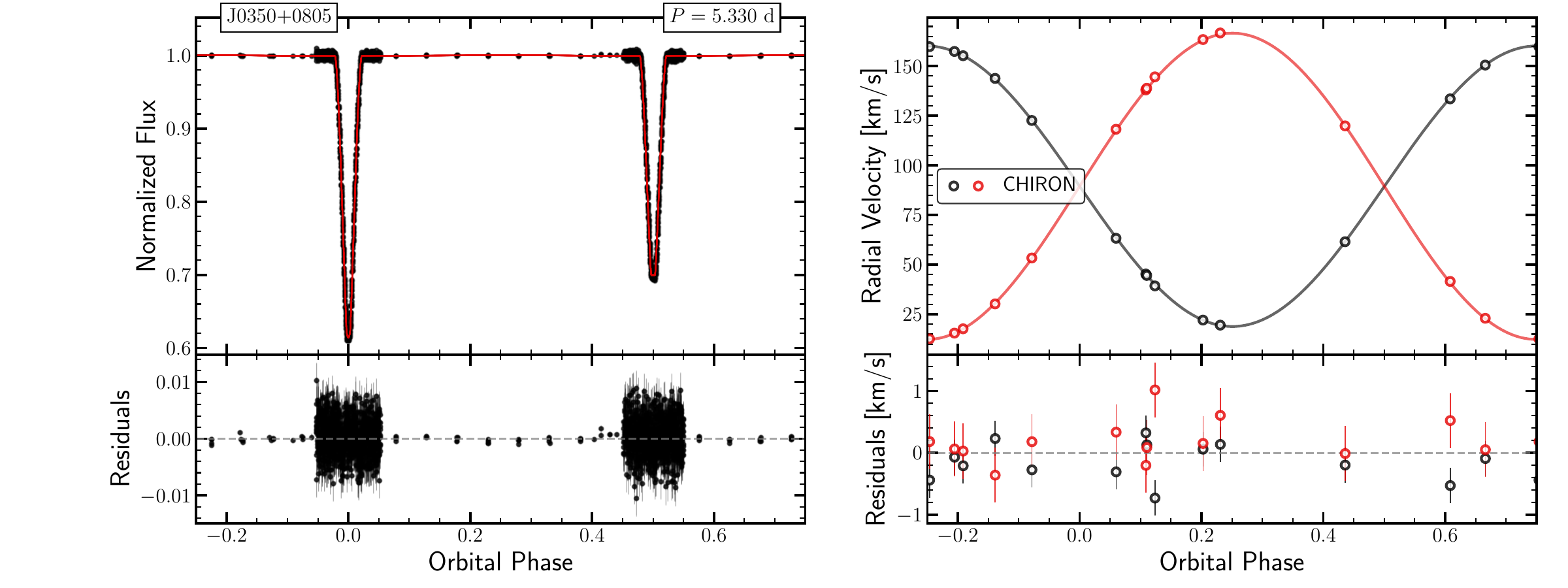}
    \includegraphics[trim={2cm 0 0 0},clip,width=\linewidth]{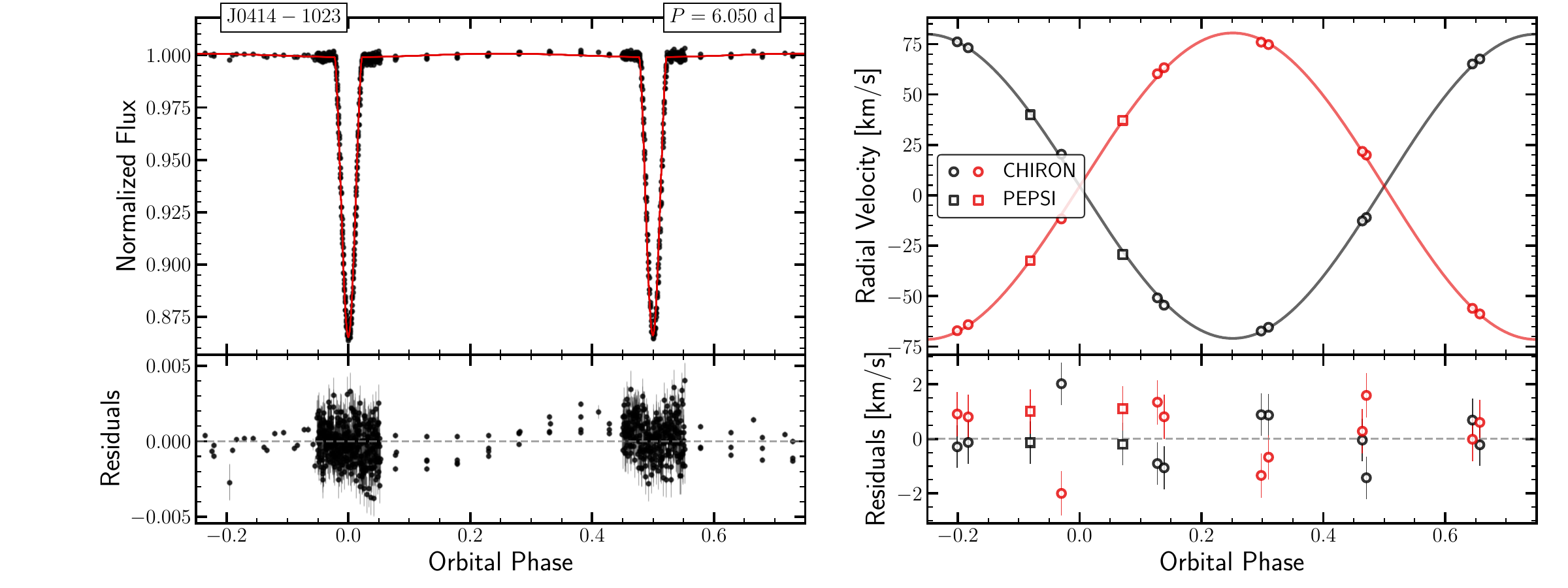}
\end{figure*}

\begin{figure*}
    \ContinuedFloat
    \centering
    \includegraphics[trim={2cm 0 0 0},clip,width=\linewidth]{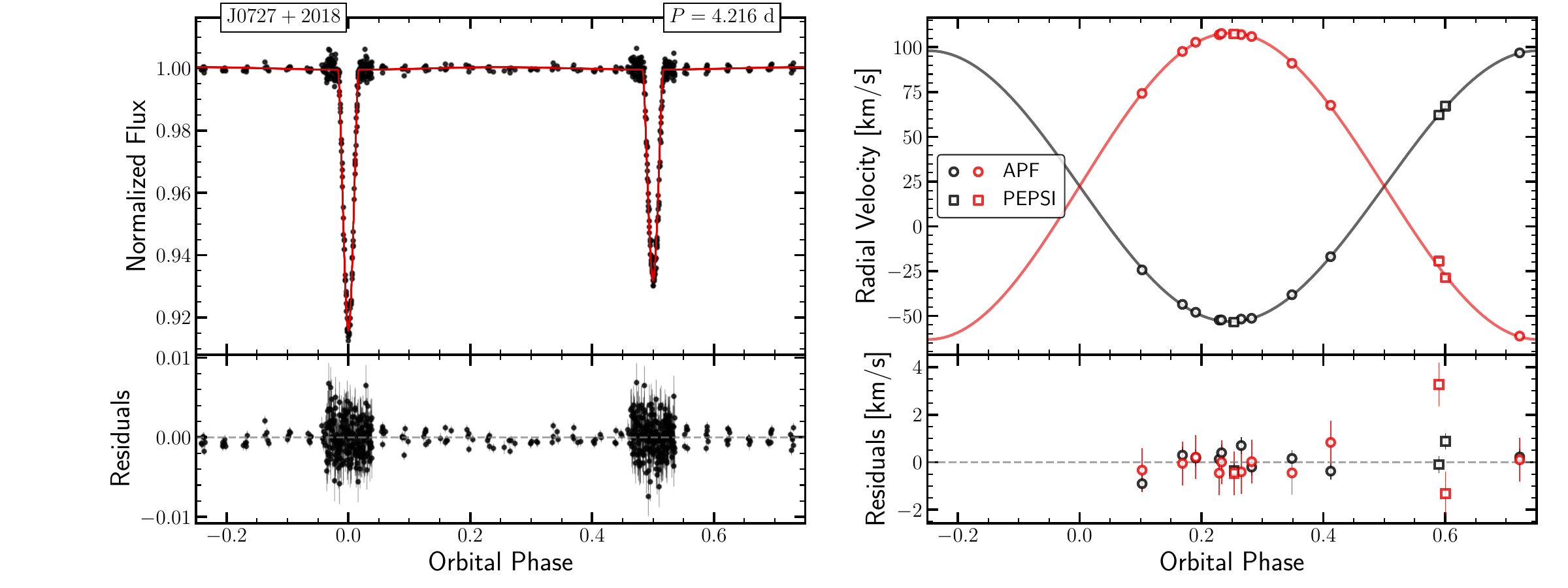}
    \includegraphics[trim={2cm 0 0 0},clip,width=\linewidth]{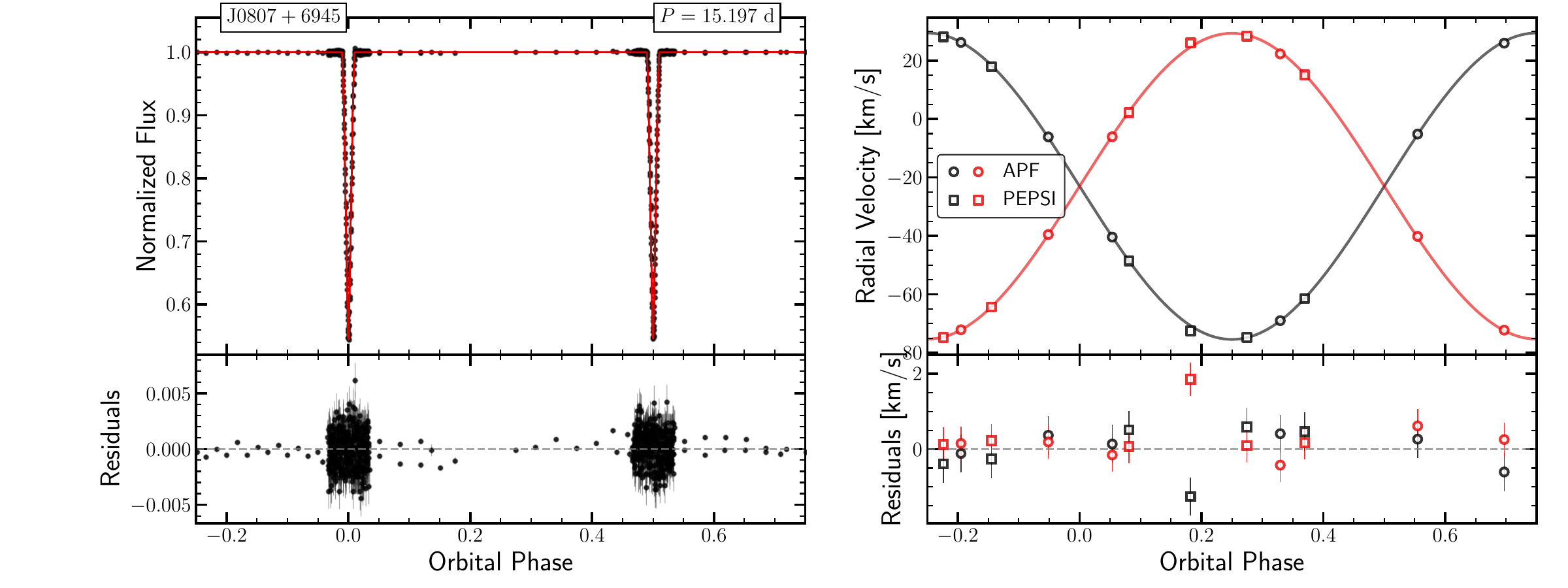}
    \includegraphics[trim={2cm 0 0 0},clip,width=\linewidth]{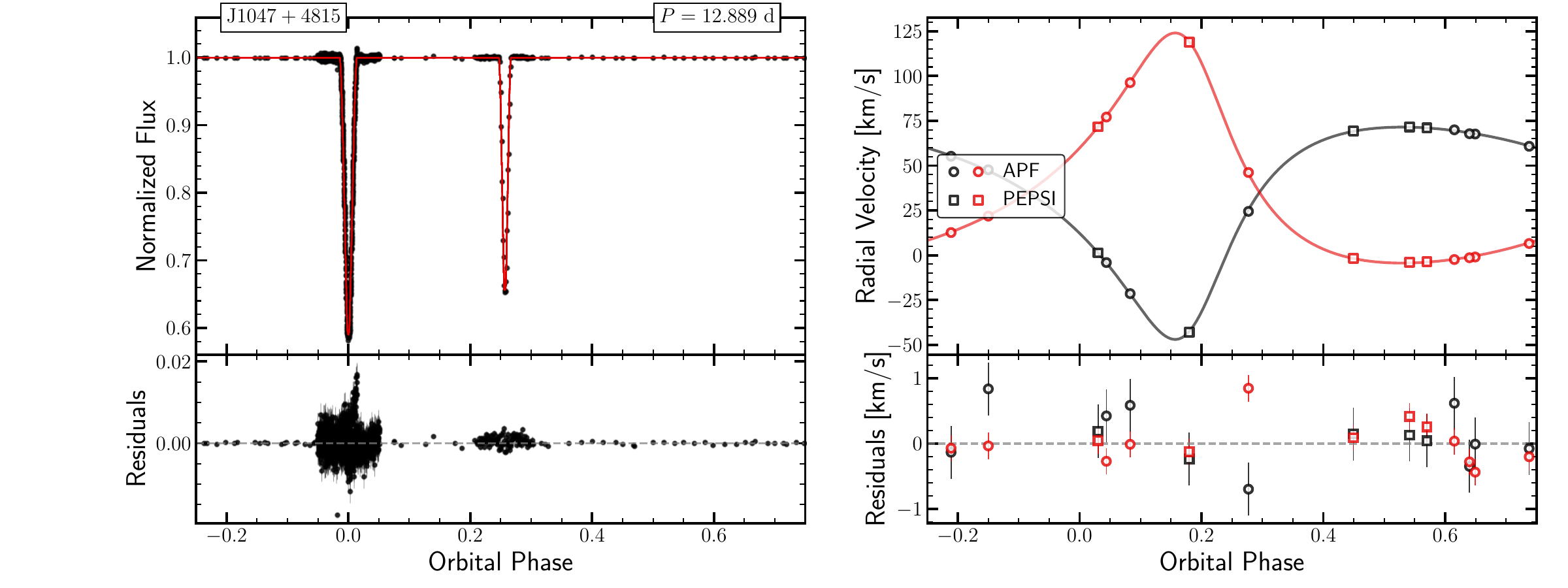}
\end{figure*}

\begin{figure*}
    \ContinuedFloat
    \centering
    \includegraphics[trim={2cm 0 0 0},clip,width=\linewidth]{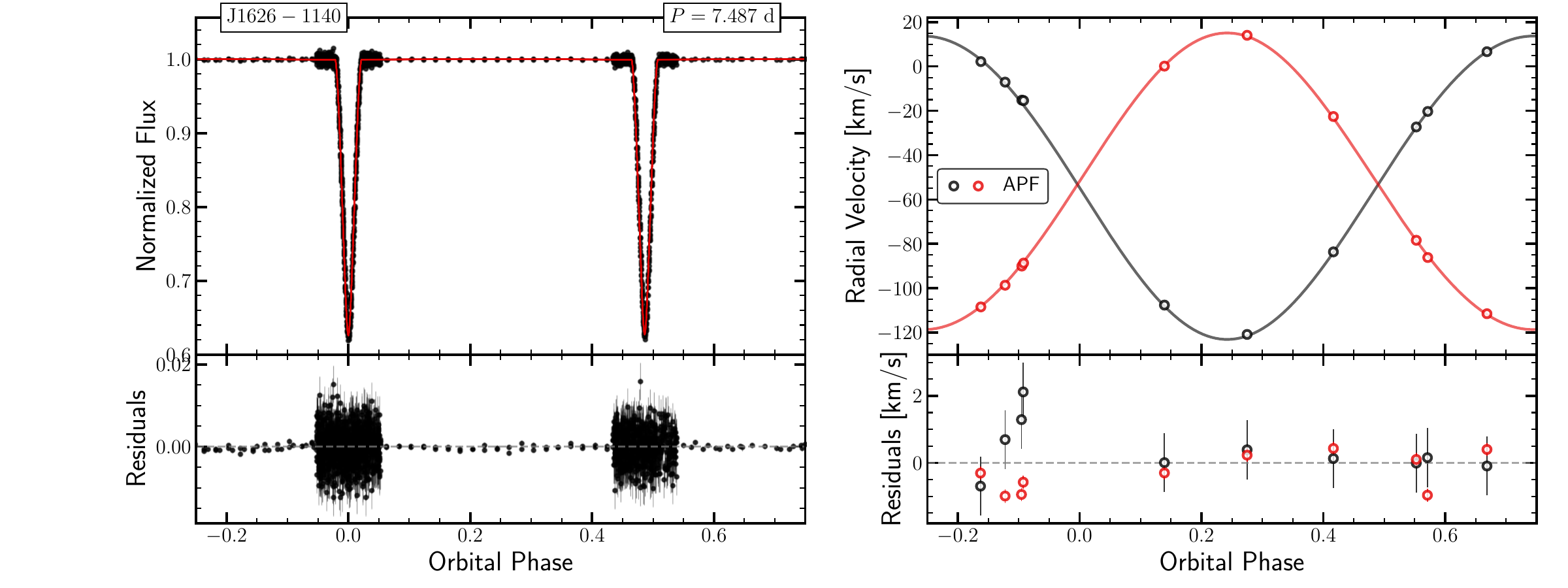}
    \includegraphics[trim={2cm 0 0 0},clip,width=\linewidth]{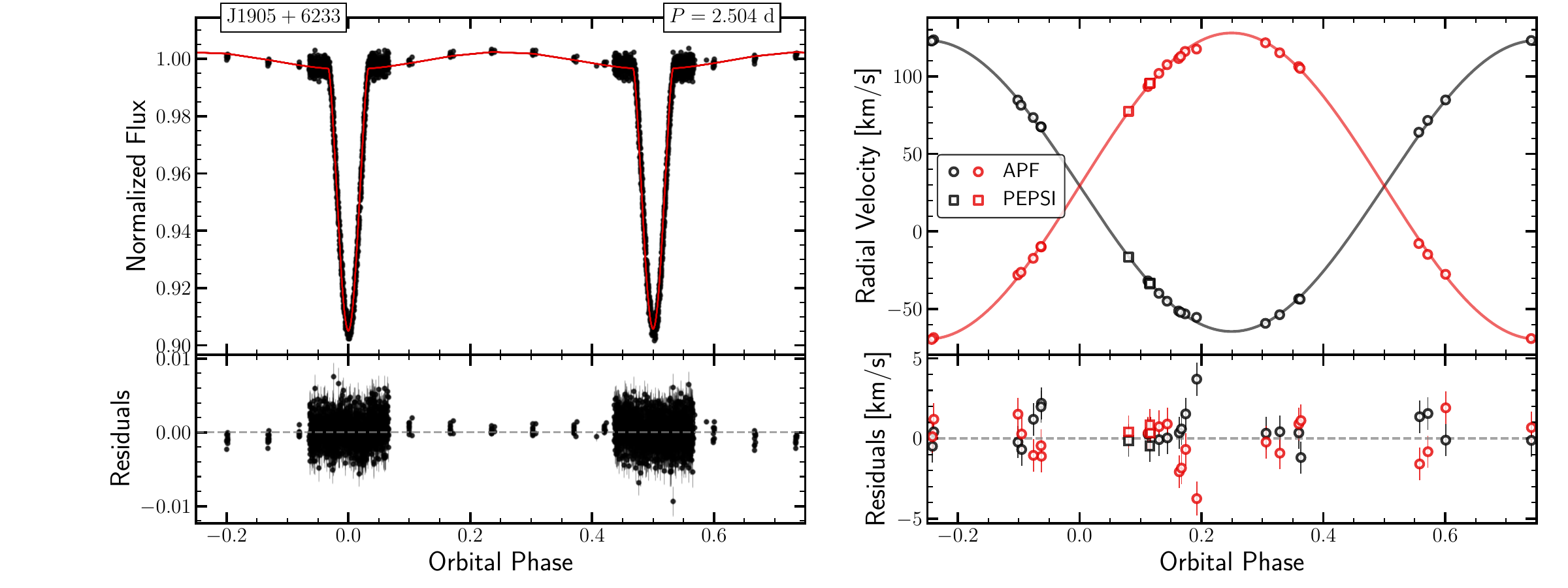}
    \includegraphics[trim={2cm 0 0 0},clip,width=\linewidth]{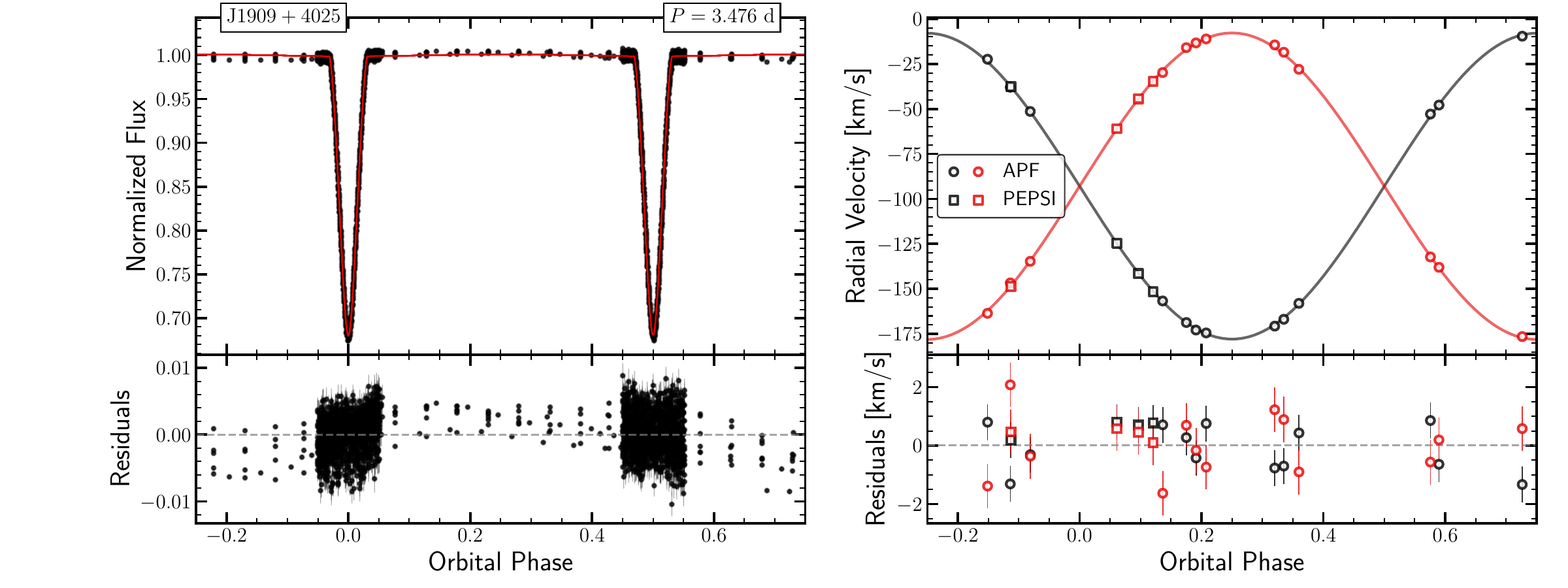}
\end{figure*}

\begin{figure*}
    \ContinuedFloat
    \centering
    \includegraphics[trim={2cm 0 0 0},clip,width=\linewidth]{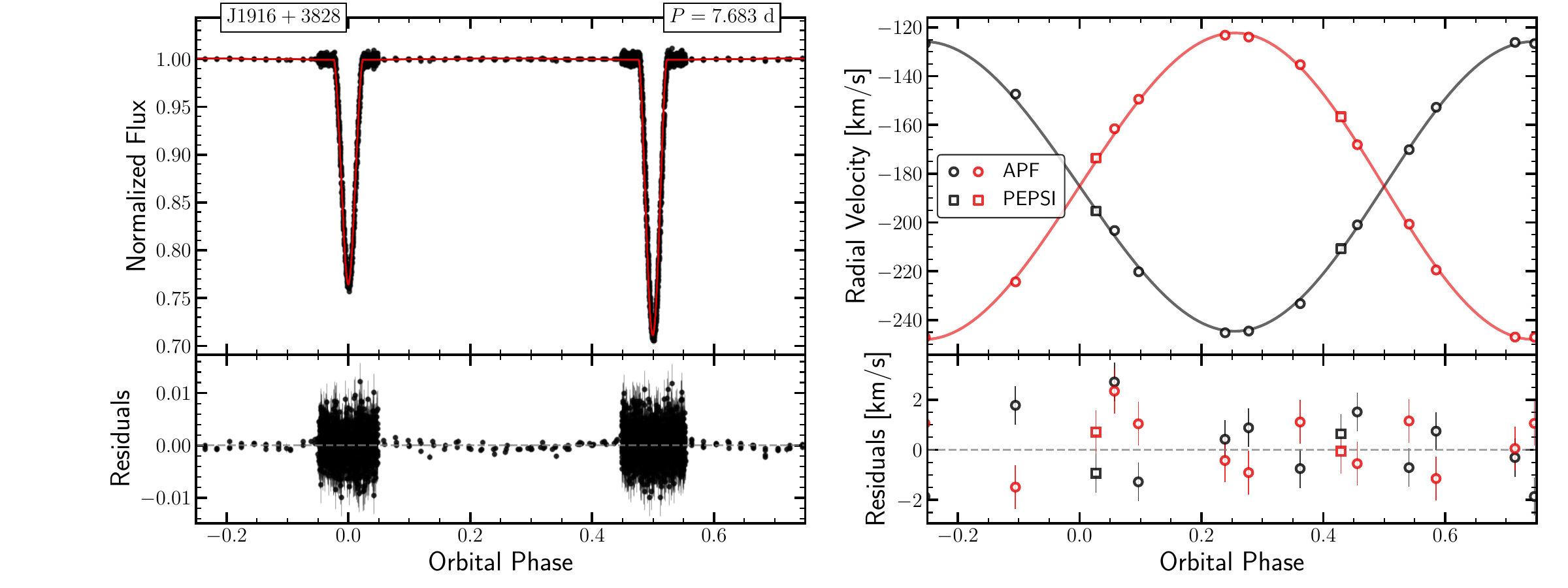}
    \includegraphics[trim={2cm 0 0 0},clip,width=\linewidth]{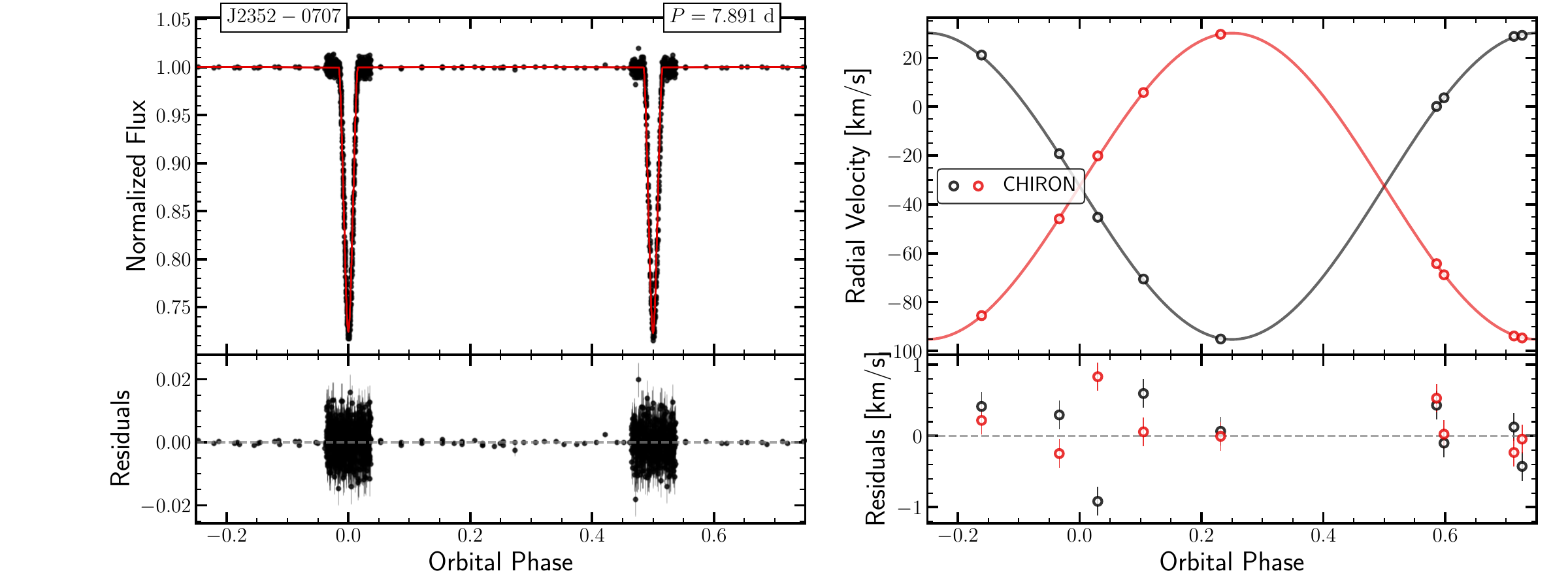}
\end{figure*}

\bibliography{main}{}
\bibliographystyle{aasjournalv7}

%% This command is needed to show the entire author+affiliation list when
%% the collaboration and author truncation commands are used.  It has to
%% go at the end of the manuscript.
%\allauthors

%% Include this line if you are using the \added, \replaced, \deleted
%% commands to see a summary list of all changes at the end of the article.
%\listofchanges

\end{document}